\documentstyle[pstricks,grp-my,epsf]{elsart}
\def\be{\begin{eqnarray}}
\def\ee{\end{eqnarray}}
\def\no{\nonumber}

\def\qb{\bar{q}}
\def\qbm{\bar{\mbox{q}}}
\def\la{\langle}
\newcommand{\gm}{\gamma}

\newcommand{\Sg}{\Sigma}
\newcommand{\dl}{\delta}
\newcommand{\SSg}{\tilde{\Sigma}}
\newcommand{\eq}{\begin{equation}}
\newcommand{\eqx}{\end{equation}}
\newcommand{\eqn}{\begin{eqnarray}}
\newcommand{\eqnx}{\end{eqnarray}}
\newcommand{\ben}{\begin{eqnaray}}
\newcommand{\een}{\end{eqnarray}}
\newcommand{\f}[2]{\frac{#1}{#2}}
\newcommand{\ra}{\longrightarrow}
\newcommand{\GG}{{\cal G}}
\renewcommand{\AA}{{\cal A}}

 \newcommand{\MM}{{\cal M}}
\newcommand{\BB}{{\cal B}}
\newcommand{\ZZ}{{\cal Z}}
\newcommand{\DD}{{\cal D}}
\newcommand{\HH}{{\cal H}}
\newcommand{\RR}{{\cal R}}
\newcommand{\GT}{{\cal G}_1 \otimes {\cal G}_2^T}

\newcommand{\zb}{\bar{z}}

\newcommand{\arr}[4]{
\left(\begin{array}{cc}
#1&#2\\
#3&#4
\end{array}\right)
}

\newcommand{\tr}{\mbox{\rm tr}\,}
\newcommand{\One}{\mbox{\bf 1}}

\newcommand{\corr}[1]{\la{#1}\rangle}
\newcommand{\br}[1]{\overline{#1}}
\newcommand{\phib}{\br{\phi}}
\newcommand{\psib}{\br{\psi}}
\newcommand{\lm}{\lambda}
\newcommand{\ksi}{\xi}

\newcommand{\Gb}{\br{G}}

\newcommand{\Gm}{G_{q\br{q}}}

\newcommand{\ggd}[2]{\GG_{#1}\otimes\GG^T_{#2}\Gamma}

\chardef\atcode=\catcode`\@
\catcode`\@=11  
 \@addtoreset{equation}{section}
 \@addtoreset{table}{section}
 \@addtoreset{figure}{section}

\catcode`\@=\atcode

\begin{fmffile}{ng}

\fmfcmd{%
	style_def ddbl expr p =
		begingroup
		save oldpen;
		pen oldpen;
		oldpen := currentpen;
		pickup oldpen scaled 6; 
		ccutdraw p;
		pickup oldpen scaled 4;
		cullit; undraw p; cullit; 
		pickup oldpen;
		endgroup;
	enddef;
}
\fmfset{dot_len}{1mm}

\begin{document}

\begin{frontmatter}

\title{Nonhermitean  Random Matrix Models }
\author[Jagel]{Romuald A. Janik},
\author[Jagel,GSI,TH]{Maciej A. Nowak},
\author[GSI,ELTE]{G\'{a}bor Papp} and
\author[SUNY]{Ismail Zahed}

\address[Jagel]{Department of Physics, Jagellonian University, 30-059
	Krak\'{o}w,Poland}
\address[GSI]{GSI, Plankstr. 1 , D-64291 Darmstadt, Germany}
\address[TH]{Inst\"{u}tut f\"{u}r Kernphysik, TH Darmstadt, D-64289
	Darmstadt, Germany}
\address[ELTE]{Institute for Theoretical Physics, E\"{o}tv\"{o}s
	University, Budapest, Hungary}
\address[SUNY]{Department of Physics, SUNY, Stony Brook, New York 11794, USA.}

\date{\today}

\begin{abstract}
We introduce an extension of the diagrammatic rules in random matrix 
theory and apply it to  nonhermitean random matrix models using 
the $1/N$ approximation. A number of one- and two-point functions are 
evaluated on their holomorphic and nonholomorphic 
supports to leading order in $1/N$. The one-point functions describe the
distribution of eigenvalues, while the two-point functions characterize their
macroscopic correlations. Generic form for the two-point functions are 
obtained, generalizing the concept of macroscopic universality to nonhermitean 
random matrices. We show that the holomorphic and nonholomorphic one- and 
two-point functions condition the behavior of pertinent partition functions 
to order ${\cal O}(1/N)$. We derive explicit conditions for the location and
distribution of their singularities.
Most of our analytical results are found to be in good agreement with
numerical calculations using large ensembles of complex matrices.
\\[4mm]
\end{abstract}

\end{frontmatter}


\eject
\newlength{\oldparskip}
\setlength{\oldparskip}{\parskip}
\addtolength{\parskip}{-4pt}
\tableofcontents
\vfil
\eject
\parskip \oldparskip

\section{Introduction}

Random matrix models provide an interesting setup for modeling a number of 
physical phenomena, where noise plays a prominent role. Applications range
from atomic physics to quantum gravity. In these phenomena, one may 
distinguish between the universal behavior of noise exhibited as fluctuations
at the mesoscopic scale (microscopic universality) and the macroscopic scale
(macroscopic universality) in bulk systems with or without absorption. Simple
physical realizations of these universalities are encountered in disordered 
mesoscopic systems \cite{MESO}.

In this work, we will be concerned with the generic behavior of two classes
of random matrix models: those without dissipation (hermitean) and those with 
dissipation (nonhermitean). Since conventional 
hermitean matrix ensembles have been 
considerably investigated \cite{GENERALHER} (For recent discussion on 
macroscopic correlators see e.g. \cite{ITOI}),
we will keep our discussion to 
the chiral unitary ensembles \cite{CHIRAL}, and we devote
most of our studies to nonhermitean ensembles, which  
are  less known although they play
an increasingly important role in quantum problems~\cite{GENERALNONHER,UPDATE}.

A number of methods have been devised to calculate with random matrix models. 
Most prominent perhaps  are the Schwinger-Dyson approach \cite{ZINNJUSTIN}
and the supersymmetric method \cite{EFETOV}. In this paper, we would like to 
elaborate on diagrammatic techniques for a variety of random matrix models, 
with special focus on the chiral and nonhermitean ensembles. 
The use of diagrammatic methods in random matrix models 
is not new, although those methods were not previously applied to 
nonhermitean matrices, and we refer the reader to \cite{DIAGRAMMOTHER} 
(and references therein). In this work we  will discuss
one- and two-point functions.  They 
carry information on the eigenvalue distributions
and their (macroscopic) correlations. They involve singular operators in the 
large $N$ limit, and in general break spontaneously holomorphic symmetry.
We show also that they condition in a simple way the character of pertinent
partition functions to order ${\cal O}(1/N)$, as in general expected. 

The organization of this paper is as follows: in section~2, we set up the 
notations for the hermitean case, and evaluate the resolvent in the $1/N$
approximation. We show how the result changes in the chiral case, and 
apply it to a schematic version of a Dirac operator. In section~3, we extend
our diagrammatic arguments to the nonhermitean case, using real and 
nonhermitean random matrix models as an example. The important difference 
between hermitean and nonhermitean matrix models, is the possibility of 
spontaneous breaking of holomorphic symmetry in the eigenvalue plane. 
In section~4, we work out the resolvents in the case of real and asymmetric 
matrices, nonhermitean chiral matrices and a random scattering model. In 
section~5, we evaluate a number of two-point correlators. We show that 
they all follow the general lore of macroscopic universality, thereby
generalizing the idea of Ambj{\o}rn, Jurkiewicz and Makeenko \cite{AMBJORN} and 
also Br\'{e}zin and Zee~\cite{BZ} to the nonhermitean case. All our results are
found in good agreement with large scale numerical calculations.
In section~6, we show that the one- and two-point functions may be used to
generate pertinent partition functions to order ${\cal O} (1/N)$ for both the
holomorphic as well as nonholomorphic case. We use the Ginibre-Girko
ensembles~\cite{GINIBRE},  as a way of illustration. A 
Straightforward generalization to other unitary
ensembles is briefly mentioned. Our conclusions are summarized in section~7.

\section{Hermitean Random Ensembles}

To illustrate the arguments and set up the notation for the diagrammatic 
arguments to follow, let us first consider the well known case of a 
deterministic plus random hermitean ensemble with Gaussian distribution.
It is convenient to use the diagrammatic notation introduced by \cite{BZ},
borrowing on the standard large $N$ diagrammatics for QCD \cite{THOOFT}. For 
the random part, the fermionic Lagrangian is simply
\eqn
{\cal L} =\bar{\psi}_a ( z {\bf 1}_a^b - H^b_a )\psi^b \,, 
\label{0plus0}
\eqnx
where $H$ is a hermitean random matrix with Gaussian weight. We will refer to 
$\psi$ as a ``quark" and to $H$ as a ``gluon". The ``Feynman 
graphs'' following from (\ref{0plus0}) allow only for the flow of  ``color'' 
(no momentum), since (\ref{0plus0}) defines a field-theory in $0+0$ 
dimensions. They are an efficient way of 
keeping track of ``color" factors.\footnote{The names ``quarks'',
``gluons'',
``color'' etc. are used here in a figurative sense, without any connection
to QCD. In order to avoid
any 
confusion, we always use these names in quotation marks.}   
The ``quark'' and ``gluon'' propagators (double line notation) are shown in 
Fig.~\ref{fig.rules}. For hermitean random matrices the Gaussian averaging 
preserves holomorphic symmetry in the large $N$ limit. This
is not the case for nonhermitean random matrices when singular operators 
are considered (see sections~3-6).

\begin{figure}[htbp]
\be
\parbox{20mm}{%
	\begin{fmfgraph}(20,3)
	\fmfleft{i1}
	\fmfright{o1}
	\fmf{quark}{o1,i1}
	\end{fmfgraph}
} & \mbox{\Large $\frac{1}{z}$} \no \\
\parbox{20mm}{%
	\begin{fmfgraph*}(10,10)
	\fmftop{i1,u1,u2,o1}
	\fmfbottom{i2,l1,l2,o2}
	\fmffreeze
    \fmfforce{(0,h)}{i1}
    \fmfforce{(.4w,h)}{u1}
    \fmfforce{(.6w,h)}{u2}
    \fmfforce{(w,h)}{o1}
    \fmfforce{(0,0)}{i2}
    \fmfforce{(.4w,0)}{l1}
    \fmfforce{(.6w,0)}{l2}
    \fmfforce{(w,0)}{o2}
	\fmf{quark}{u1,i1}
	\fmf{quark}{o1,u2}
	\fmf{quark}{i2,l1}
	\fmf{quark}{l2,o2}
	\fmf{plain}{u1,l1}
	\fmf{plain}{u2,l2}
	\fmfv{d.si=0,l=b,l.a=180,l.d=1mm}{i1}
	\fmfv{d.si=0,l=a,l.a=0,l.d=1mm}{o1}
	\fmfv{d.si=0,l=c,l.a=180,l.d=1mm}{i2}
	\fmfv{d.si=0,l=d,l.a=0,l.d=1mm}{o2}
	\end{fmfgraph*}
} & \hspace*{10mm}
  \mbox{\Large $\langle {\cal H}^a_b{\cal H}^c_d\rangle$ =
  $\frac{1}{N} \delta^a_d\delta^c_b $} 
  \no
\ee
\caption{Large $N$ ``Feynman" rules for ``quark'' and ``gluon" propagators.}
\label{fig.rules}
\end{figure}

\subsection{Pastur Equation}

The  pertinent resolvent  for (\ref{0plus0}) is given
 by \cite{BZ,PASTUR,WEGNER}
\eq
G(z):=\f{1}{N}<\tr \f{1}{z-D-\HH}> \,,
\label{1}
\eqx
where $\HH$ is an $N\times N$ hermitean random matrix,
and $D$ a deterministic part. The distribution of eigenvalues $\nu (z)$
of $H=D +\HH$ follows from (\ref{1}) through $\pi\nu (z) =\partial_{\zb} G$.

Throughout, the weight average is considered Gaussian for simplicity.  
However, all our arguments generalize to the non-Gaussian weight. 
In terms of the one-particle 
irreducible self energy $\SSg$, (\ref{1}) reads
\eq
G(z)=\f{1}{N}\tr \f{1}{z-D-\SSg} \,.
\label{2}
\eqx
We note that since the argument in (\ref{1}) displays poles on the real
axis only (hermitean matrices), it is safe to assume that after averaging
and in large $N$, the resolvent is holomorphic in the $z$-plane modulo possible
singularities on the real axis, for which $\nu(z)\neq 0$. This property is 
manifest in  (\ref{2}).

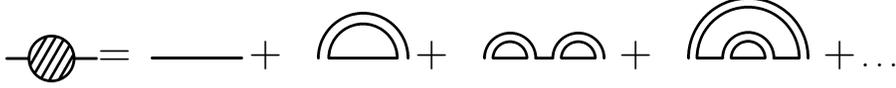
\begin{figure}[htbp]
\setlength{\unitlength}{0.8mm}
\be
 \parbox{15mm}{\raisebox{-3mm}{%
	\begin{fmfgraph}(15,8)
		\fmfleft{i1}
		\fmfright{o1}
		\fmf{plain}{i1,v,o1}
		\fmfv{d.sh=circle,d.f=.6,d.si=.5w}{v}
	\end{fmfgraph}
 }} \!\!\!&\mbox{\Large =}&\,
 \parbox{15mm}{\raisebox{-3mm}{%
	\begin{fmfgraph}(15,8)
		\fmfleft{i1}
		\fmfright{o1}
		\fmf{plain}{i1,o1}
	\end{fmfgraph}
 }} \!\mbox{\Large +} \quad\!
 \parbox{15mm}{\raisebox{-3mm}{%
	\begin{fmfgraph}(15,8)
		\fmfleft{i1}
		\fmfright{o1}
		\fmf{phantom}{i1,i2,v,o2,o1}
		\fmffreeze
		\fmfshift{(-0.13w,0)}{i2}
		\fmfshift{(0.13w,0)}{o2}
		\fmf{plain}{i2,o2}
		\fmf{plain,left}{i1,o1}
		\fmf{plain,left}{i2,o2}
	\end{fmfgraph}
 }} \!\mbox{\Large +} \quad\!
 \parbox{20mm}{\raisebox{-3mm}{%
	\begin{fmfgraph}(20,8)
		\fmfleft{i1}
		\fmfright{o1}
		\fmf{phantom}{i1,i2,vi1,vo1,vi2,vo2,o2,o1}
		\fmffreeze
		\fmfshift{(-0.07w,0)}{i2}
		\fmfshift{(0.07w,0)}{o2}
		\fmfshift{(0.07w,0)}{vi1}
		\fmfshift{(-0.07w,0)}{vo2}
		\fmf{plain}{i2,vi1}
		\fmf{plain}{vo1,vi2}
		\fmf{plain}{vo2,o2}
		\fmf{plain,left}{i2,vi1}
		\fmf{plain,left}{i1,vo1}
		\fmf{plain,left}{vo2,o2}
		\fmf{plain,left}{vi2,o1}
	\end{fmfgraph}
 }} \!\mbox{\Large +} \quad\!
 \parbox{20mm}{\raisebox{-3mm}{%
	\begin{fmfgraph}(20,8)
		\fmfleft{i1}
		\fmfright{o1}
		\fmf{phantom}{i1,i2,vi1,vo1,vi2,vo2,o2,o1}
		\fmffreeze
		\fmfshift{(-0.07w,0)}{i2}
		\fmfshift{(0.07w,0)}{o2}
		\fmfshift{(-0.07w,0)}{vo1}
		\fmfshift{(0.07w,0)}{vi2}
		\fmf{plain}{i2,vi1}
		\fmf{plain}{vo1,vi2}
		\fmf{plain}{vo2,o2}
		\fmf{plain,left}{i2,o2}
		\fmf{plain,left}{i1,o1}
		\fmf{plain,left}{vi1,vo2}
		\fmf{plain,left}{vo1,vi2}
	\end{fmfgraph}
 }} \!\mbox{\Large +} \ldots \no
\ee
\caption{Rainbow diagrams.}
\label{fig.rainbow}
\end{figure}

\begin{figure}[htbp]
\setlength{\unitlength}{1mm}
\be
 \parbox{15mm}{\raisebox{-3mm}{%
	\begin{fmfgraph*}(15,8)
		\fmfleft{i1}
		\fmfright{o1}
		\fmf{plain}{i1,v,o1}
		\fmfv{d.sh=circle,d.f=0,d.si=.5w,l=$\noexpand\Sigma$,l.d=0}{v}
	\end{fmfgraph*}
 }} \ &\mbox{\Large =}&\ 
 \parbox{20mm}{\raisebox{-3mm}{%
	\begin{fmfgraph}(20,8)
		\fmfleft{i1}
		\fmfright{o1}
		\fmf{phantom}{i1,i2,v,o2,o1}
		\fmffreeze
		\fmfshift{(-0.15w,0)}{i2}
		\fmfshift{(0.15w,0)}{o2}
		\fmf{plain}{i2,v,o2}
		\fmf{plain,left}{i2,o2}
		\fmf{plain,left}{i1,o1}
		\fmfv{d.sh=circle,d.f=.6,d.si=.3w}{v}
	\end{fmfgraph}
 }} \no
\ee
\caption{Pastur equation.}
\label{fig.pastur}
\end{figure}
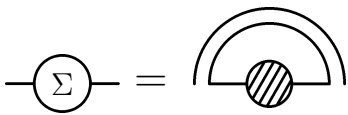

In the large $N$ limit the equation for the self energy $\SSg$ follows from 
resumming the rainbow diagrams of Fig.~\ref{fig.rainbow}.
All other diagrams (non-planar) are subleading in the large $N$ limit. 
The result is 
\eq
\left(\f{1}{z-D-\SSg}\right)_{bc} \DD_{ab,cd}= \SSg_{ad}
\label{e.hmat}
\eqx
where $\DD_{ab,cd}$ is the random matrix propagator given by
\eq
\DD_{ab,cd}=<\HH^a_b \HH^c_d>=\f{1}{N}\dl_{ad} \dl_{bc} \,.
\label{4}
\eqx
Inserting this into (\ref{e.hmat}) we find that $\SSg=\Sg\cdot\One$,
where $\One$ is the identity matrix, and $\Sg$ is a scalar satisfying
Pastur's equation (see Fig.~\ref{fig.pastur})
\eq
\f{1}{N}\tr\f{1}{z-D-\Sg}=\Sg \,.
\label{5}
\eqx
We note that $\Sg=G(z)$. 
It is convenient, for the generalization to the chiral case, to represent
the propagator as a tensor product of matrices
\eq
\DD=\One \otimes \One  \,.
\label{6}
\eqx
As a result equation (\ref{e.hmat}) becomes
\eq
\f{1}{N}\tr\left[ \f{1}{z-D-\SSg} \One \right] \otimes \One =\SSg \,.
\label{7}
\eqx
This result was derived by a number of authors \cite{BZ,PASTUR,WEGNER}.

\subsection{Chiral Pastur Equations}

In the chiral case the $0+0$ dimensional Lagrangian (\ref{0plus0}) becomes
\eq
{\cal L} = \bar{\psi}_a ( z \delta_a^b - {\MM}^b_a )\psi^b 
\label{chiralx}
\eqx
with the chiral  random matrix 
\eq
\MM=\arr{0}{\RR}{\RR^{\dagger}}{0} \,.
\label{8}
\eqx
Here, the ``quarks" are chiral in the sense that $\psi = (\psi_R, \psi_L)$,
following the block decomposition in (\ref{8}). The chirality operator is
\eq
\gamma_5 = \arr{{\bf 1}}{0}{0}{-{\bf 1}} \,.
\label{800}
\eqx
We may  now apply the results of the previous section but with a
different propagator. Namely
\eq
\DD_{ab,cd}=<\MM^a_b \MM^c_d>=\f{1}{N}\dl_{ad} \dl_{bc}\quad
\mbox{\rm $a>N,b<N$ or $a<N,b>N$}
\label{9}
\eqx
and zero otherwise. In the tensor product notation we get
\eq
\DD=\f{1}{N}\arr{\One}{0}{0}{0} \otimes \arr{0}{0}{0}{\One}+
\f{1}{N}\arr{0}{0}{0}{\One} \otimes \arr{\One}{0}{0}{0} \,.
\label{10}
\eqx
Again, the random averaging is carried using a Gaussian weight.
Since (\ref{chiralx}) is chirally symmetric for $z=0$, the question 
arises as to whether the averaging 
breaks spontaneously chiral symmetry in the
large $N$ limit. This is indeed the case for $z\rightarrow i0$ after 
$N\rightarrow\infty$. 
Inserting~(\ref{10}) into equation (\ref{e.hmat}) we obtain
\eqn
&&\phantom{+}\f{1}{N}\tr\left[ \f{1}{z-D-\SSg}\arr{\One}{0}{0}{0}  \right] \otimes
\arr{0}{0}{0}{\One}\nonumber\\
&&+\f{1}{N}
\tr\left[ \f{1}{z-D-\SSg}\arr{0}{0}{0}{\One}  \right] \otimes
\arr{\One}{0}{0}{0}=\SSg \,.
\label{11}
\eqnx
We see that $\SSg$ is diagonal
\eq
\SSg=\arr{\Sg_1\cdot\One}{0}{0}{\Sg_2\cdot\One}
\label{12}
\eqx
with $\Sg_1$ and $\Sg_2$ scalars and the Green's function is
$G(z)=\f{1}{2}(\Sg_1+\Sg_2)$. The chiral Pastur's equations now take
the form
\eqn
\f{1}{N} \tr\left[ \f{1}{z-D-\SSg}\arr{0}{0}{0}{\One}\right]
&=&\Sg_1\,,\\
\f{1}{N} \tr\left[ \f{1}{z-D-\SSg}\arr{\One}{0}{0}{0}\right] &=&\Sg_2
       \,.
\label{13}
\eqnx
As $z\rightarrow i0$, $\Sigma_1=\Sigma_2\neq 0$ implying that in the large $N$ 
limit, the Gaussian averaging breaks spontaneously chiral symmetry. This is 
particularly interesting as it shows the following: the mechanism of 
spontaneous symmetry breakdown requires minimally that the interaction is 
chirally symmetric and random, a point presently embodied in the instanton 
vacuum description~\cite{DIASHUR}. A related idea 
was first discussed in \cite{NEUBERGER}.

\subsection{Random Dirac Operators}

As an illustration of the above equations, consider the schematic case of 
chiral fermions with mass $m$.  In this case, the deterministic
$2N\times2N$ matrix is 
\eq
D=\arr{m}{0}{0}{-m} \,.
\label{14}
\eqx
Rearranging the rows and columns of the matrix $z-D-\SSg$ and inverting,
yield the following equations for $\Sg_1$ and $\Sg_2$
\eqn
\Sg_1 &=& \f{z-m-\Sg_1}{(z-m-\Sg_1)(z+m-\Sg_2)} \,,\nonumber \\
\Sg_2 &=& \f{z+m-\Sg_2}{(z-m-\Sg_1)(z+m-\Sg_2)} \,.
\label{16}
\eqnx
Using the fact that $\Sg_2/\Sg_1= (z+m)/(z-m)$ and $G=(\Sg_1+\Sg_2)/2$
we get the equation for the Green's function
\eq
(G^3-2zG^2)(m^2-z^2) + G z^2 (m^2- (z^2+1))
+z^3=0  \,.
\label{17}
\eqx
We note that when $m=0$ this gives exactly $z^3$ times the ordinary
-- hermitean -- Pastur's equation for this case. Solving for $\Sg_1$ gives
\eq
G=\f{z}{2}\left(1+\f{\sqrt{z^2-m^2-4}}{\sqrt{z^2-m^2}}\right)
\label{18}
\eqx
which is holomorphic in the $z$-plane modulo cuts on the real axis. 
The solutions to (\ref{16}) were used in~\cite{USRANDOM}.

\section{Nonhermitean Random Ensembles}

If we were to use nonhermitean matrices in the resolvent (\ref{1}), then 
configuration by configuration, the resolvent displays poles that are scattered
around $z=0$ (for $D=0$) in the complex $z$-plane. Their spread is given by the Gaussian 
distribution, which is of order one. In the large $N$ limit, the poles 
accumulate in general on finite surfaces (for unitary matrices on circles), 
over which the resolvent is no longer 
holomorphic. The (spontaneous) breaking of holomorphic symmetry follows from 
the large $N$ limit. As a result $\partial G/\partial\zb\neq 0$ on the 
nonholomorphic surface, with a finite eigenvalue distribution. In this section
we will set up the diagrammatic rules for investigating nonhermitean random 
matrix models.

For hermitean matrices, ``quarks" $\psi$ and ``conjugate-quarks" $\phi$
decouple in the ``thermodynamical'' limit ($N\to\infty$). Their 
respective resolvents follow from the $0+0$ dimensional Lagrangian 
\eq
{\cal L}_0 = \psib(z-\MM )\psi+\phib(\zb-\MM^{\dagger})\phi
\label{21}
\eqx
and do not `talk' to each other. They are holomorphic (antiholomorphic) functions modulo cuts on the
real  axis.  For nonhermitean matrices, this is not the case in the large 
$N$ limit. The spontaneous breaking of holomorphic symmetry in the large $N$ 
limit, may be probed in the $z$-plane by adding to (\ref{21})
\eq
{\cal L}_B = \lm(\psib\phi+\phib\psi)
\label{22}
\eqx
in the limit $\lm\ra 0$. The combination ${\cal L}_0+{\cal L}_B$ will be used 
below to investigate a number of random and nonhermitean ensembles
with or without deterministic parts.

{}From (\ref{21},\ref{22}) we define the matrix-valued resolvent through 
\be
\arr{{\cal G}_{qq}}{{\cal G}_{q\overline{q}}}{{\cal G}_{\overline{q}q}}
{{\cal G}_{\overline{q}\overline{q}}}
= \left\langle \arr{z-\MM}{\lambda}{\lambda}{\zb 
-\MM^{\dagger}}^{-1}\right\rangle
\label{19}
\ee
where the limit $N\rightarrow\infty$ is understood before $\lambda\rightarrow 
0$. For convenience, the ``quark'' field is organized in an isodoublet
$\ksi=(\psi,\phi) =(q, \overline{q})$.The ``quark'' spectral density follows 
from Gauss law,
\eq
\nu (z, \zb) = \frac 1{\pi} \partial_{\zb} \,\, G(z,\zb ) =
\frac 1{\pi N} \partial_{\zb} \,\, {\rm Tr}_N\,\,{\cal G}_{qq}
\label{20}
\eqx
which is the distribution of eigenvalues of $\MM$. 
For hermitean $\MM$, (\ref{20}) is valued on the real axis. As 
$\lambda\rightarrow 0$, the block-structure decouples, and we are left
with the original resolvent. For $z\rightarrow +i0$, the latter is just a 
measurement of the real eigenvalue distribution. For nonhermitean $\MM$,
(\ref{20}) is valued in the $z$-plane. As $\lambda\rightarrow 0$, the block
structure does not decouple, leading to a nonholomorphic resolvent for certain
parts of the $z$-plane. Holomorphic
symmetry is spontaneously broken in the large $N$ limit. The size of the 
nonholomorphic regions is conditioned by the divergence of the 
``quark-conjugate-quark'' two-point function (see below).

\subsection{Example}

To set up the diagrammatic rules we will use an example.
Consider the case of real and nonhermitean matrices with off-diagonal 
correlations. Specifically, let $M$ be a real and asymmetric $N\times N$
matrix so that
\eq
\corr{M_{ij}^2}=\f{1}{N} \quad\quad \corr{M_{ij}M_{ji}}=\f{\tau}{N} \,.
\label{23}
\eqx
The case $\tau=1$ reduces to symmetric real matrices. The random ensemble
described by (\ref{23}) occur naturally in neural network problems, and 
have been discussed in \cite{SOMMERS88} using the replica construction. 

\subsection{Diagrammatic rules}

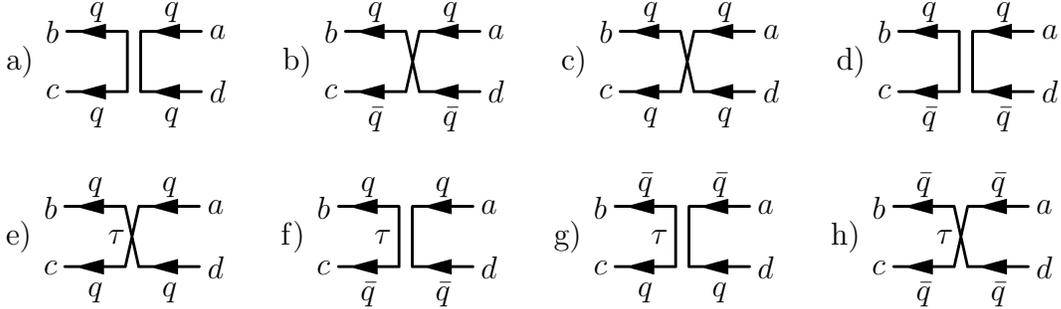
\begin{figure}[htbp]
\def\skel{%
	\fmftop{iu,u1,u2,ou}
	\fmfbottom{il,l1,l2,ol}
	\fmffreeze
	\fmfforce{(0,h)}{iu}
	\fmfforce{(.45w,h)}{u1}
	\fmfforce{(.55w,h)}{u2}
	\fmfforce{(w,h)}{ou}
	\fmfforce{(0,0)}{il}
	\fmfforce{(.45w,0)}{l1}
	\fmfforce{(.55w,0)}{l2}
	\fmfforce{(w,0)}{ol}
} 
a)\hspace*{3mm}
\parbox{18mm}{%
 \begin{fmfgraph*}(18,8)
	\skel
	\fmf{quark,label=$\noexpand q$}{u1,iu}
	\fmf{quark,label=$\noexpand q$}{ou,u2}
	\fmf{quark,label=$\noexpand q$,l.side=left}{l1,il}
	\fmf{quark,label=$\noexpand q$,l.side=left}{ol,l2}
	\fmf{plain}{u1,l1}
	\fmf{plain}{u2,l2}
	\fmfv{d.si=0,l=$b$,l.a=180,l.d=1mm}{iu}
	\fmfv{d.si=0,l=$c$,l.a=180,l.d=1mm}{il}
 	\fmfv{d.si=0,l=$a$,l.a=0,l.d=1mm}{ou}
	\fmfv{d.si=0,l=$d$,l.a=0,l.d=1mm}{ol}
\end{fmfgraph*}
}
\hspace*{8mm}
b)\hspace*{3mm}
\parbox{18mm}{%
 \begin{fmfgraph*}(18,8)
	\skel
	\fmf{quark,label=$\noexpand q$}{u1,iu}
	\fmf{quark,label=$\noexpand q$}{ou,u2}
	\fmf{quark,label=$\noexpand\qb$,l.side=left}{l1,il}
	\fmf{quark,label=$\noexpand\qb$,l.side=left}{ol,l2}
	\fmf{plain}{u1,l2}
	\fmf{plain}{u2,l1}
	\fmfv{d.si=0,l=$b$,l.a=180,l.d=1mm}{iu}
	\fmfv{d.si=0,l=$c$,l.a=180,l.d=1mm}{il}
 	\fmfv{d.si=0,l=$a$,l.a=0,l.d=1mm}{ou}
	\fmfv{d.si=0,l=$d$,l.a=0,l.d=1mm}{ol}
\end{fmfgraph*}
}
\hspace*{8mm}
c)\hspace*{3mm}
\parbox{18mm}{%
 \begin{fmfgraph*}(18,8)
	\skel
	\fmf{quark,label=$\noexpand\qb$}{u1,iu}
	\fmf{quark,label=$\noexpand\qb$}{ou,u2}
	\fmf{quark,label=$\noexpand q$,l.side=left}{l1,il}
	\fmf{quark,label=$\noexpand q$,l.side=left}{ol,l2}
	\fmf{plain}{u1,l2}
	\fmf{plain}{u2,l1}
	\fmfv{d.si=0,l=$b$,l.a=180,l.d=1mm}{iu}
	\fmfv{d.si=0,l=$c$,l.a=180,l.d=1mm}{il}
 	\fmfv{d.si=0,l=$a$,l.a=0,l.d=1mm}{ou}
	\fmfv{d.si=0,l=$d$,l.a=0,l.d=1mm}{ol}
\end{fmfgraph*}
}
\hspace*{8mm}
d)\hspace*{3mm}
\parbox{18mm}{%
 \begin{fmfgraph*}(18,8)
	\skel
	\fmf{quark,label=$\noexpand\qb$}{u1,iu}
	\fmf{quark,label=$\noexpand\qb$}{ou,u2}
	\fmf{quark,label=$\noexpand\qb$,l.side=left}{l1,il}
	\fmf{quark,label=$\noexpand\qb$,l.side=left}{ol,l2}
	\fmf{plain}{u1,l1}
	\fmf{plain}{u2,l2}
	\fmfv{d.si=0,l=$b$,l.a=180,l.d=1mm}{iu}
	\fmfv{d.si=0,l=$c$,l.a=180,l.d=1mm}{il}
 	\fmfv{d.si=0,l=$a$,l.a=0,l.d=1mm}{ou}
	\fmfv{d.si=0,l=$d$,l.a=0,l.d=1mm}{ol}
\end{fmfgraph*}
}

\vspace*{15mm}
e)\hspace*{3mm}
\parbox{18mm}{%
 \begin{fmfgraph*}(18,8)
	\skel
	\fmf{quark,label=$\noexpand q$}{u1,iu}
	\fmf{quark,label=$\noexpand q$}{ou,u2}
	\fmf{quark,label=$\noexpand q$,l.side=left}{l1,il}
	\fmf{quark,label=$\noexpand q$,l.side=left}{ol,l2}
	\fmf{plain}{u1,v,l2}
	\fmf{plain}{u2,v,l1}
	\fmfv{d.si=0,l=$b$,l.a=180,l.d=1mm}{iu}
	\fmfv{d.si=0,l=$c$,l.a=180,l.d=1mm}{il}
 	\fmfv{d.si=0,l=$a$,l.a=0,l.d=1mm}{ou}
	\fmfv{d.si=0,l=$d$,l.a=0,l.d=1mm}{ol}
	\fmfv{d.si=0,l=$\noexpand\tau$,l.a=180,l.d=1mm}{v}
\end{fmfgraph*}
}
\hspace*{8mm}
f)\hspace*{3mm}
\parbox{18mm}{%
 \begin{fmfgraph*}(18,8)
	\skel
	\fmf{quark,label=$\noexpand q$}{u1,iu}
	\fmf{quark,label=$\noexpand q$}{ou,u2}
	\fmf{quark,label=$\noexpand\qb$,l.side=left}{l1,il}
	\fmf{quark,label=$\noexpand\qb$,l.side=left}{ol,l2}
	\fmf{plain}{u1,v,l1}
	\fmf{plain}{u2,l2}
	\fmfv{d.si=0,l=$b$,l.a=180,l.d=1mm}{iu}
	\fmfv{d.si=0,l=$c$,l.a=180,l.d=1mm}{il}
 	\fmfv{d.si=0,l=$a$,l.a=0,l.d=1mm}{ou}
	\fmfv{d.si=0,l=$d$,l.a=0,l.d=1mm}{ol}
	\fmfv{d.si=0,l=$\noexpand\tau$,l.a=180,l.d=1mm}{v}
\end{fmfgraph*}
}
\hspace*{8mm}
g)\hspace*{3mm}
\parbox{18mm}{%
 \begin{fmfgraph*}(18,8)
	\skel
	\fmf{quark,label=$\noexpand\qb$}{u1,iu}
	\fmf{quark,label=$\noexpand\qb$}{ou,u2}
	\fmf{quark,label=$\noexpand q$,l.side=left}{l1,il}
	\fmf{quark,label=$\noexpand q$,l.side=left}{ol,l2}
	\fmf{plain}{u1,v,l1}
	\fmf{plain}{u2,l2}
	\fmfv{d.si=0,l=$b$,l.a=180,l.d=1mm}{iu}
	\fmfv{d.si=0,l=$c$,l.a=180,l.d=1mm}{il}
 	\fmfv{d.si=0,l=$a$,l.a=0,l.d=1mm}{ou}
	\fmfv{d.si=0,l=$d$,l.a=0,l.d=1mm}{ol}
	\fmfv{d.si=0,l=$\noexpand\tau$,l.a=180,l.d=1mm}{v}
\end{fmfgraph*}
}
\hspace*{8mm}
h)\hspace*{3mm}
\parbox{18mm}{%
 \begin{fmfgraph*}(18,8)
	\skel
	\fmf{quark,label=$\noexpand\qb$}{u1,iu}
	\fmf{quark,label=$\noexpand\qb$}{ou,u2}
	\fmf{quark,label=$\noexpand\qb$,l.side=left}{l1,il}
	\fmf{quark,label=$\noexpand\qb$,l.side=left}{ol,l2}
	\fmf{plain}{u1,v,l2}
	\fmf{plain}{u2,v,l1}
	\fmfv{d.si=0,l=$b$,l.a=180,l.d=1mm}{iu}
	\fmfv{d.si=0,l=$c$,l.a=180,l.d=1mm}{il}
 	\fmfv{d.si=0,l=$a$,l.a=0,l.d=1mm}{ou}
	\fmfv{d.si=0,l=$d$,l.a=0,l.d=1mm}{ol}
	\fmfv{d.si=0,l=$\noexpand\tau$,l.a=180,l.d=1mm}{v}
\end{fmfgraph*}
}
\vspace*{5mm}
\caption{All ``gluonic'' amplitudes for real nonhermitean matrices.}
\label{fig.eightglue}
\end{figure}

{}From (\ref{21},\ref{22}) we note that there are two kinds of ``quark''
propagators ($1/z$ for ``quarks'' $\psi$ and $1/\bar{z}$ for
``conjugate-quarks''$\phi$, where both can be incoming and outgoing) and 
four kinds of ``gluon'' propagators, associated to the following
``quark-conjugate-quark'' amplitudes (treating the ``quarks" as external
sources) 
\eqn
\langle\bar{\psi}_b M_a^b \psi^a \bar{\psi}_c  M_d^c \psi^d\rangle &=&
 \bar{\psi}_b \psi^a \bar{\psi}_c \psi^d \langle M^b_a M^c_d\rangle
	\nonumber \\ 
\langle\bar{\psi}_b M_a^b \psi^a \bar{\phi}_c  (M^T)_d^c \phi^d\rangle &=&
 \bar{\psi}_b \psi^a \bar{\phi}_c \phi^d \langle M^b_a (M^T)^c_d\rangle
	\nonumber \\ 
\langle\bar{\phi}_b (M^T)_a^b \phi^a \bar{\psi}_c  M_d^c \psi^d\rangle &=&
 \bar{\phi}_b \phi^a \bar{\psi}_c \psi^d \langle (M^T)^b_a M^c_d\rangle
	\\
\langle\bar{\phi}_b (M^T)_a^b \phi^a \bar{\phi}_c  (M^T)_d^c \phi^d\rangle &=&
 \bar{\phi}_b \phi^a \bar{\phi}_c \phi^d \langle (M^T)^b_a (M^T)^c_d\rangle
	\,. \nonumber
\label{fourgluons}
\eqnx
In each of these four cases, averaging is done with the weight
\eqn
<\ldots> =\int [dM] \exp \left[ -\frac{N}{2(1-\tau^2)}{\rm tr}(MM^T-\tau MM)
\right]
\label{GOEtwisted}
\eqnx
reproducing the correlations (\ref{23}). 
Indeed, 
\eqn 
<M^a_b M^d_c>=\frac{1}{N}(\delta_{bc}\delta^{ad} + \tau \delta^d_b \delta^a_c)
\label{twistprop}
\eqnx
where the second contribution corresponds to twisting the lines
with a ``penalty factor'' $\tau$. For $\tau=\pm 1$ this is equivalent
to symmetrizing or antisymmetrizing the double lines in the ``gluon''
propagator. All combinations arising from the averaging (\ref{GOEtwisted})
are shown in Fig.~\ref{fig.eightglue}.
The pairs (ae),(bf), (cg) and (dh) in Fig.~\ref{fig.eightglue}
correspond to four ``gluon'' propagators, evaluated with the weight
(\ref{GOEtwisted}). 
The lower row in Fig.~\ref{fig.eightglue} corresponds to terms with 
$\tau \neq 0$. The arrows denote incoming and outgoing particles, the 
``quarks'' and ``conjugate-quarks'' are distinguished by their labels.

\noindent

The generalization to the hermitean-antihermitean case, 
i.e. the case of complex matrices, corresponds to 
the substitution $M^T \rightarrow M^{\dagger}$ with 
the measure
\eqn
<\ldots> =\int [dM] \exp \left[ -\frac{N}{(1-\tau^2)}{\rm tr}(MM^{\dagger}
- \tau {\rm Re} MM)
\right] \,.
\label{GUEtwisted}
\eqnx
The relations (\ref{23}) should be replaced by 
\eqn
<|M_{ab}|^2>=\f{1}{N} \,\,\,\,\,\,\,\,\,\, <M_{ab} M_{ba}>=\f{\tau}{N}
\label{corrcomplex}
\eqnx
corresponding to hermitean ($\tau=1$), antihermitean ($\tau=-1$) or
general complex ($\tau=0$) matrix theory. In the latter
the relevant ``gluonic'' amplitudes correspond now to
Fig. \ref{fig.eightglue}a--\ref{fig.eightglue}d.

\section{One-Point Functions}

In this section we will explicitly calculate a number of one-point functions 
(resolvents) for several nonhermitean random matrix models, using and 
extending when needed the diagrammatic approach discussed above.

\subsection{Pure Complex Random Matrices}

\begin{figure}[htbp]
\be
 \parbox{13mm}{%
  \begin{fmfgraph*}(13,10)
	\fmfleft{i1}
	\fmfright{o1}
	\fmf{plain}{i1,v,o1}
	\fmfv{d.sh=circle,d.f=0,d.si=.5w,l=$\noexpand\Sigma_1$,l.d=0}{v}
  \end{fmfgraph*}
 } \ &\mbox{\Large =}& \quad
 \parbox{40mm}{\raisebox{10mm}{%
  \begin{fmfgraph*}(40,15)
	\fmfstraight
	\fmfbottom{i1,c,d,v,b,a,o1}
	\fmffreeze
	\fmfforce{(.22w,0)}{c}
	\fmfforce{(.28w,0)}{d}
	\fmfforce{(.72w,0)}{b}
	\fmfforce{(.78w,0)}{a}
	\fmf{quark,label=$q$,l.side=left}{c,i1}
	\fmf{quark,label=$q$,l.side=left}{v,d}
	\fmf{quark,label=$q$,l.side=left}{b,v}
	\fmf{quark,label=$q$,l.side=left}{o1,a}
	\fmf{plain,left,tension=0.2}{c,b}
	\fmf{plain,left,tension=0.2}{d,a}
	\fmfv{d.sh=circle,d.f=.5,d.si=.15w}{v}
  \end{fmfgraph*}
 }} \; \mbox{\Large +} \; \mbox{\Large $\tau$} \;
 \parbox{40mm}{\raisebox{10mm}{%
  \begin{fmfgraph*}(40,15)
	\fmfstraight
	\fmfbottom{i1,c,d,v,b,a,o1}
	\fmffreeze
	\fmfforce{(.22w,0)}{c}
	\fmfforce{(.28w,0)}{d}
	\fmfforce{(.72w,0)}{b}
	\fmfforce{(.78w,0)}{a}
	\fmf{quark,label=$q$,l.side=left}{c,i1}
	\fmf{quark,label=$q$,l.side=left}{v,d}
	\fmf{quark,label=$q$,l.side=left}{b,v}
	\fmf{quark,label=$q$,l.side=left}{o1,a}
	\fmf{plain,left,tension=0.2}{c,a}
	\fmf{plain,left,tension=0.2}{d,b}
	\fmfv{d.sh=circle,d.f=.5,d.si=.15w}{v}
  \end{fmfgraph*}
 }} \no
\ee
\vspace*{-8mm}
\caption{Self-energy equation for  real nonhermitean matrices.}
\label{fig.twistedpastur}
\end{figure}
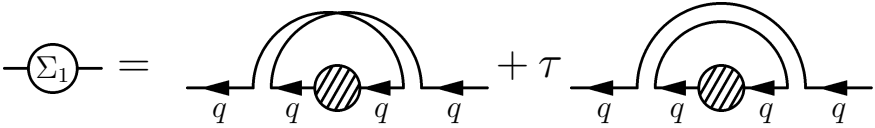

We seek the one-point functions for
$\corr{\bar{\ksi_i}\ksi_j}$ as a $2\times 2$ matrix. 
The equation for the one particle irreducible (1PI) self-energy  follows
from Figs.~\ref{fig.eightglue}--\ref{fig.twistedpastur} in the form
\eqn
\arr{\Sg_1}{\Sg_2}{\Sg_3}{\Sg_4}&=&
\f{1}{N}\tr_{N}\arr{\GG_{qq}}{\GG_{q\bar{q}}}{\GG_{\bar{q}q}}{\GG_{\bar{q}\bar{q}}}
\circ \arr{\tau}{1}{1}{\tau} \nonumber \\
&=&\f{1}{N}\tr_{N}
\arr{z-\Sg_1}{\lm-\Sg_2}{\lm-\Sg_3}{\zb-\Sg_4}^{-1}
\circ
\arr{\tau}{1}{1}{\tau}\,.
\label{24}
\eqnx
Here the trace is meant component-wise (block per block),
and the argument of the trace is the dressed 
propagator. The operation $\circ$ is {\em not} a matrix multiplication,
but a simple multiplication between the entries in the corresponding positions.
Here ${\rm tr}_N$ is short for the trace on the $N\times N$ block-matrices.

The first correlator in (\ref{23}) does not influence the
``quark-quark'' interaction - it corresponds to the double line with a twist
in the  corresponding Pastur equation - hence subleading. However, this
twist could be compensated by the twisted part of the propagator 
coming from the second correlator (\ref{23}), thereby 
explaining the factor $\tau$ in the upper left corner of (\ref{24}).
This is shown in Fig.~\ref{fig.twistedpastur},
where the first (subleading) graph comes from Fig.~\ref{fig.eightglue}a
and the second (leading) comes from Fig.~\ref{fig.eightglue}e. The other
entries in (\ref{24}) follow from Fig.~\ref{fig.eightglue} by inspection.

The ``quark'' one-point function is now
\eq
G(z , \zb )=\f{1}{N}{\rm tr_N }\,\, \GG_{qq}= ({\zb-\Sg_4})/{det} \,.
\label{25}
\eqx
It follows that $\Sg_2=\Sg_3$, with
 \eqn
det\,\,\Sg_1&=&{\tau(\zb-\Sg_4)} \label{e.s1}\\
det\,\,\Sg_4&=&{\tau(z-\Sg_1)} \label{e.s4}\\
det\,\, \Sg_2&=&\Sg_2-\lm \,,
\label{26}
\eqnx
where $det=(z-\Sg_1)(\zb-\Sg_4)-(\lm-\Sg_2)^2$. Substituting
$r=\Sg_2-\lm$ in the last relation in (\ref{26}) yields the
equation
\eq
((z-\Sg_1)(\zb-\Sg_4)-r^2)(r+\lm)=r \,.
\label{27}
\eqx
For $\lm=0$, the solution with $r=0$ is holomorphic while that with $r\neq 0$
is nonholomorphic. In the holomorphic case, $\Sigma_1 (z-\Sigma_1)=\tau$, and 
the resolvent is simply
\eq
G(z)=\f{z \mp\sqrt{z^2-4\tau}}{2\tau} \,.
\label{31}
\eqx
where the upper sign corresponds to the solution with the pertinent asymptotics. 
In the nonholomorphic case, $G(z, \zb) =\zb -\Sigma_4$, with
\eq
G(z , \zb)=\f{\zb-\tau z}{1-\tau^2}
\label{29}
\eqx
in agreement with \cite{SOMMERS88}. The boundary between the holomorphic and 
nonholomorphic solution follows from the condition $|G(z, \zb )|^2= 
|G(z)|^2=1$, that is
\eq
\f{x^2}{(1+\tau)^2}+\f{y^2}{(1-\tau)^2}=1
\label{34}
\eqx
which is an ellipse. Inside (\ref{34}) the solution is nonholomorphic and 
outside it is holomorphic. The case investigated by Ginibre and Girko 
\cite{GINIBRE} follows for $\tau=0$. 
The one-point function for the random matrix $(H_1+iH_2)/\sqrt{2}$ can be 
obtained similarly. The self-energy equation being the same as before but
with $\tau=0$.

\vspace*{-9mm}

\subsection{Real Asymmetric (Square) Matrices}

Following Khoruzhenko \cite{KHO} we define the random matrix ensemble
\eq
H_0+A
\label{35}
\eqx
where $H_0$ is a deterministic but otherwise arbitrary (e.g. complex) matrix
and $A$ is a real asymmetric random matrix with
\eq
\corr{a_{jk}a_{lm}}=\f{1}{N} v^2 \dl_{jl}\dl_{km} \,.
\label{36}
\eqx
The 1PI self energy equations are now
\eq
\arr{\Sg_1}{\Sg_2}{\Sg_3}{\Sg_4}=\tr_{N}
\arr{z-H_0-\Sg_1}{\lm-\Sg_2}{\lm-\Sg_3}{\zb-H_0^{\dagger}-\Sg_4}^{-1}
\circ
\arr{0}{v^2}{v^2}{0} \,.
\label{37}
\eqx
Care must be taken with the definition of the inverse matrix. The inverse
is taken only with respect to the ``isospin'' indices ($2\times 2$), so 
the determinant is now a matrix in N-space. We obtain
\eqn
\arr{\!\Sg_1\!}{\Sg_2\!}{\!\Sg_3\!}{\Sg_4\!}\!=\tr_{\!N}
\arr{\!\!\zb-H_0^{\dagger}-\Sg_4\!\!}{\Sg_2-\lm}{\Sg_3-\lm}{\!\!z-H_0-\Sg_1\!\!}
\circ \arr{0}{v^2 \, D_R^{-1}\!}{\!v^2\, D_L^{-1}}{0}
\label{38}
\eqnx
with $\Sg_2=\Sg_3=x$ for $\lm=0$ and 
\eqn
D_L&=&(\zb-H_0^{\dagger}-\Sg_4)(z-H_0-\Sg_1)-x^2 \,, \nonumber \\
D_R&=&(z-H_0-\Sg_1)(\zb-H_0^{\dagger}-\Sg_4)-x^2
\eqnx
since we do not impose that $H_0$ and $H_0^{\dagger}$ commute.
 We see that $\Sg_1=\Sg_4=0$ and 
the equation for $\Sg_2=x$ now
has the form
\eq
x=v^2 \cdot x \cdot\frac 1N \,tr_N [ 
(\zb-H_0^{\dagger})(z-H_0)-x^2]^{-1} \,.
\label{39}
\eqx
The solution with $x\neq 0$ corresponds to the non-analytical Green's function.
Its ``quark'' Green's function is given by
\eq
G(z , \zb )= \frac 1N \tr_N [(\zb-H_0^{\dagger})G_0(-x^2)]
\label{40}
\eqx
where $G_0(w)$ is the generalized resolvent
\eq
G_0(w)=\f{1}{ (\zb-H_0^{\dagger})(z-H_0)+w }
\label{41}
\eqx
and $x$ satisfies
\eq
\frac 1N \tr_N G_0(-x^2)=\f{1}{v^2} \,.
\label{42}
\eqx
These are the equations obtained by Khoruzhenko \cite{KHO}.

\subsection{Nonhermitean Chiral Matrices}
\label{s.chiral}

In this section we will reanalyze the chiral random matrix model
of section~2.2 in the presence of a nonhermitean part, $e.g.$
``chemical potential'' $\mu$. The motivation for that stems from the constant 
mode sector of the massless and chiral Dirac operator at finite chemical 
potential \cite{MULAT}. 
Here we will use Gaussian chiral and random matrices as 
described by (\ref{21}).

If we were to define
\eq
\gm=i\gamma_0 = \arr{0}{-1}{1}{0}\,\qquad{\rm and}\qquad
	\MM=\arr{0}{-\mu+\RR}{\mu+\RR^\dagger}{0}
\label{43}
\eqx
then the 1PI self-energy equations in the planar approximation are given by 
\eq
\arr{\Sg_1}{\Sg_2}{\Sg_3}{\Sg_4}=\frac 1N \tr_{N}
	\underbrace{\arr{z-\mu\gm-\Sg_1}{\lm-\Sg_2}{\lm-\Sg_3}%
	  {\zb+\mu\gm-\Sg_4}^{-1}}_{\mbox{\Large $\GG$}}
\circ
\arr{\DD}{\DD}{\DD}{\DD}
\label{44}
\eqx
where $\DD$ is the ``gluon'' propagator (\ref{9}), and 
$\Sg_i$ are diagonal $2N\times 2N$ matrices. Inverting in (\ref{44})
with respect to the ``isospin'' indices gives
\eq
\label{e.chse}
\arr{\Sg_1}{\Sg_2}{\Sg_3}{\Sg_4}=\tr_{N}
\arr{\zb+\mu\gm-\Sg_4}{\Sg_2-\lm}{\Sg_3-\lm}{z-\mu\gm-\Sg_1}
\cdot \Delta^{-1}
\circ
\arr{\DD}{\DD}{\DD}{\DD} \,.
\label{45}
\eqx
Here $\Delta$ is the determinant of the $2\times 2$ matrix with  $2N\times
2N$ entries. The equation for $\Sg_2\neq 0$ in (\ref{45}) gives
\eq
\label{e.master}
1=\f{1}{N}\tr \left( \Delta^{-1}\cdot \DD \right)
\eqx 
with
\eq
\Delta^{-1}=\arr{1}{-\mu (\zb-\Sg_4-z+\Sg_1)/{\alpha}}
{\mu (\zb-\Sg_4-z+\Sg_1)/{\alpha}}{1}
\label{47}
\eqx
and where $\alpha^{2N}$ is the determinant of the $2N\times 2N$ matrix $\Delta$.
Inserting (\ref{47}) into  (\ref{e.chse}) yields
\eqn
\Sg_4 &=& z-\Sg_1-\mu^2 \f{\zb-\Sg_4-z+\Sg_1}{\alpha}\\
\Sg_1 &=& \zb-\Sg_4+\mu^2 \f{\zb-\Sg_4-z+\Sg_1}{\alpha} \,.
\label{48}
\eqnx
In particular $\Sg_1+\Sg_4=\mbox{Re}\,z\equiv x$.
The resolvent for the ``quark-quark'' part is
simply $\Sg_1$. Solving (\ref{48}) in conjunction with (\ref{e.master}) gives
\eq
\alpha^2-\alpha+\mu^2(\zb-\Sg_4-z+\Sg_1)^2=0 \,.
\label{49}
\eqx
Simple algebra for the ``quark-quark'' resolvent gives
\eq
G (z, \zb) = \f{1}{2N} \tr_{N}\ \GG_{qq} =\f{x}{2}-iy-\f{1}{2}\f{iy}{y^2-\mu^2}
\label{50}
\eqx
a  result first derived in \cite{STEPHANOV} using different arguments.

For $\Sigma_2=0$ we recover the holomorphic
solution~\cite{STEPHANOV,USMUX}, $\Sigma_1(z)=G(z)\One$,
$\Sigma_4=\Sigma_1^\dagger$, with $G(z)$ fulfilling
the cubic Pastur equation
\eq
G^3(z) -2zG^2(z) + (z^2 +\mu^2 +1)G(z)-z=0.
\label{cubicpasturmux}
\eqx

\subsection{Random Scattering Model}

As a final application of the techniques introduced above, we consider the
random matrix model discussed by Mahaux and Weidenm\"{u}ller \cite{MAHAUX}
in the context of chaotic resonance scattering. The model is generically 
described by a nonhermitean Hamiltonian of the form
\eq
H-i\gm VV^{\dagger}
\label{51}
\eqx
where $H$ is random Gaussian (orthogonal) and $N\times N$ matrix-valued,
while $V$ is $N\times M$ matrix-valued \cite{UPDATE,HAAKE}.
 The form (\ref{51}) 
is dictated by unitarity. The $V$ propagator is given by
\eq
\corr{V^a_k V^b_l}=\f{1}{N}\dl^{ab}\dl_{kl} \,.
\label{52}
\eqx
\begin{figure}[htbp]
\be
  \parbox{15mm}{%
	\begin{fmfgraph*}(15,10)
	\fmfleft{a,c,k,v}
	\fmfright{l,b,d}
	\fmffreeze
	\fmfforce{(0,0.7h)}{a}
	\fmfforce{(w,0.7h)}{b}
	\fmfforce{(0.33w,0)}{c}
	\fmfforce{(0.67w,0)}{d}
	\fmfforce{(.17w,0.7h)}{k}
	\fmfforce{(.83w,0.7h)}{l}
	\fmfforce{(.5w,0)}{v}
	\fmfv{d.sh=circle,d.f=1,d.si=0,l=$a$,l.d=2mm,l.a=90}{a}
	\fmfv{d.sh=circle,d.f=1,d.si=0,l=$b$,l.d=2mm,l.a=90}{b}
	\fmfv{d.sh=circle,d.f=1,d.si=0,l=$k$,l.d=2mm,l.a=90}{k}
	\fmfv{d.sh=circle,d.f=1,d.si=0,l=$l$,l.d=2mm,l.a=90}{l}
	\fmfv{d.sh=circle,d.f=1,d.si=0,l=q,l.d=2mm,l.a=-110}{c}
	\fmfv{d.sh=circle,d.f=1,d.si=0,l=q,l.d=2mm,l.a=-70}{d}
	\fmf{plain}{a,c}
	\fmf{plain}{k,v,l}
	\fmf{plain}{b,d}
 \end{fmfgraph*}
}\; i\gamma V^a_k V^b_l \;, \qquad
  \parbox{15mm}{%
	\begin{fmfgraph*}(15,10)
	\fmfleft{a,c,k,v}
	\fmfright{l,b,d}
	\fmffreeze
	\fmfforce{(0,0.7h)}{a}
	\fmfforce{(w,0.7h)}{b}
	\fmfforce{(0.33w,0)}{c}
	\fmfforce{(0.67w,0)}{d}
	\fmfforce{(.17w,0.7h)}{k}
	\fmfforce{(.83w,0.7h)}{l}
	\fmfforce{(.5w,0)}{v}
	\fmfv{d.sh=circle,d.f=1,d.si=0,l=$a$,l.d=2mm,l.a=90}{a}
	\fmfv{d.sh=circle,d.f=1,d.si=0,l=$b$,l.d=2mm,l.a=90}{b}
	\fmfv{d.sh=circle,d.f=1,d.si=0,l=$k$,l.d=2mm,l.a=90}{k}
	\fmfv{d.sh=circle,d.f=1,d.si=0,l=$l$,l.d=2mm,l.a=90}{l}
	\fmfv{d.sh=circle,d.f=1,d.si=0,l=$\noexpand\qbm$,l.d=1.5mm,l.a=-110}{c}
	\fmfv{d.sh=circle,d.f=1,d.si=0,l=$\noexpand\qbm$,l.d=1.5mm,l.a=-70}{d}
	\fmf{plain}{a,c}
	\fmf{plain}{k,v,l}
	\fmf{plain}{b,d}
 \end{fmfgraph*}
} 
\; -i\gamma V^a_k V^b_l \quad 
\left\{ \begin{array}{c} a,b=1,\ldots,N \cr k,l=1,\ldots,M \end{array}
  \right. \no
\ee
\caption{Vertex corresponding to the $VV^{\dagger}$ terms in the Hamiltonian.}
\label{fig.vv}
\end{figure}

\noindent
In the following we will use the notation $\dl=i\gm$ and $m=M/N$. 
The quadratic appearance of $V$ in (\ref{52}) calls for additional
changes in the preceding Feynman rules. In particular, 
the 1PI self-energy has two new contributions: 

\begin{enumerate}
\item a channel-vertex correction:\\
 $\dl m$ for ``quark'', $-\dl m$ for
``conjugate-quark'', and 0 for mixed.
\item a sum of rainbow diagrams
\end{enumerate}

\noindent each of which is shown in Fig.~\ref{fig.self}.
Here the extra diagrammatic complications stem from the fact
that two $V$ ``gluon'' lines couple to either the ``quark''
or the ``conjugate-quark'' lines through the new vertex (see
Fig.~\ref{fig.vv}). 

\begin{figure}[htbp]
\be
\mbox{\Large $\Sigma_1$} \!\! &\mbox{\Large =}& 
\underbrace{%
 \parbox{15mm}{\raisebox{-12mm}{%
	\begin{fmfgraph*}(15,8)
		\fmfleft{i1}
		\fmfright{o1}
		\fmf{phantom}{i1,i2,v,o2,o1}
		\fmffreeze
		\fmfshift{(-0.13w,0)}{i2}
		\fmfshift{(0.13w,0)}{o2}
		\fmf{plain,label=$q$,l.d=1mm}{i2,v}
		\fmf{plain,label=$q$,l.d=1mm}{v,o2}
		\fmfv{d.sh=circle,d.f=.6,d.si=.2w}{v}
		\fmf{plain,left}{i1,o1}
		\fmf{plain,left}{i2,o2}
	\end{fmfgraph*}
 }} \quad \mbox{\Large +} \quad \hspace*{-12mm}
}_{\mbox{\Large $G$}}
\underbrace{%
 \parbox{25mm}{\raisebox{4mm}{%
	\begin{fmfgraph*}(25,8)
		\fmfstraight
		\fmfbottom{i1,v,o1}
		\fmf{ddbl,right=2.5}{v,v}
		\fmffreeze
		\fmf{phantom}{i1,v,o1}
		\fmfv{d.si=0,l=q q,l.a=-90,l.d=1mm}{v}
	\end{fmfgraph*}
 }} \hspace*{-12mm} \quad \mbox{\Large +} \quad
 \parbox{15mm}{\raisebox{-4mm}{%
	\begin{fmfgraph}(15,8)
		\fmfstraight
		\fmfbottom{i1,i2,i3,v,o3,o2,o1}
		\fmffreeze
		\fmfshift{(-.09w,0)}{i2}
		\fmfshift{(.09w,0)}{o2}
		\fmfshift{(-.15w,0)}{i3}
		\fmfshift{(.15w,0)}{o3}
		\fmf{plain}{i3,v,o3}
		\fmfv{d.sh=circle,d.f=.6,d.si=.2w}{v}
		\fmf{plain,left}{i1,o1}
		\fmf{plain,left}{i2,o2}
		\fmf{plain,left=0.55}{i2,o2}
		\fmf{plain,left=0.5}{i3,o3}
	\end{fmfgraph}
 }} \quad \mbox{\Large +} \quad
 \parbox{20mm}{\raisebox{4mm}{%
	\begin{fmfgraph}(20,8)
		\fmfstraight
		\fmfbottom{i1,i2,i3,v1,vc1,vc2,vc3,v2,o3,o2,o1}
		\fmffreeze
		\fmfshift{(-.04w,0)}{i2}
		\fmfshift{(.04w,0)}{o2}
		\fmfshift{(-.08w,0)}{i3}
		\fmfshift{(.08w,0)}{o3}
		\fmfshift{(.04w,0)}{vc1}
		\fmfshift{(-.04w,0)}{vc3}
		\fmfshift{(-.02w,0)}{v1}
		\fmfshift{(.02w,0)}{v2}
		\fmf{plain}{i3,v1,vc1}
		\fmf{plain}{vc3,v2,o3}
		\fmfv{d.sh=circle,d.f=.6,d.si=.15w}{v1}
		\fmfv{d.sh=circle,d.f=.6,d.si=.15w}{v2}
		\fmf{plain,left}{i1,o1}
		\fmf{plain,left}{i2,o2}
		\fmf{plain,left}{i2,vc2}
		\fmf{plain,left}{i3,vc1}
		\fmf{plain,left}{vc2,o2}
		\fmf{plain,left}{vc3,o3}
	\end{fmfgraph}
 }}
}_{\mbox{\Large $m F_{qq}$}}
 \quad \mbox{\Large +} \ldots \no \\
%
%
 &\mbox{\Large $\equiv$} &
 \parbox{15mm}{\raisebox{-12mm}{%
	\begin{fmfgraph*}(15,8)
		\fmfleft{i1}
		\fmfright{o1}
		\fmf{phantom}{i1,i2,v,o2,o1}
		\fmffreeze
		\fmfshift{(-0.13w,0)}{i2}
		\fmfshift{(0.13w,0)}{o2}
		\fmf{plain,label=$q$,l.d=1mm}{i2,v}
		\fmf{plain,label=$q$,l.d=1mm}{v,o2}
		\fmfv{d.sh=circle,d.f=.6,d.si=.2w}{v}
		\fmf{plain,left}{i1,o1}
		\fmf{plain,left}{i2,o2}
	\end{fmfgraph*}
 }} \quad \mbox{\Large +} \quad
 \parbox{15mm}{\raisebox{5mm}{%
	\begin{fmfgraph*}(15,10)
		\fmfbottom{i1,i2,o2,o1}
		\fmfforce{(.1w,0)}{i2}
		\fmfforce{(.9w,0)}{o2}
		\fmf{plain,wi=2thick}{i2,o2}
		\fmf{plain,left}{i2,o2}
		\fmf{plain,left}{i1,o1}
		\fmfv{d.si=0,l=$q$,l.a=-100,l.d=2mm}{i2}
		\fmfv{d.si=0,l=$q$,l.a=-80,l.d=2mm}{o2}
	\end{fmfgraph*}
  }} 
\no \\
%
%
\mbox{\Large $\Sigma_4$} \!\! &\mbox{\Large =}& 
 \parbox{15mm}{\raisebox{-12mm}{%
	\begin{fmfgraph*}(15,8)
		\fmfleft{i1}
		\fmfright{o1}
		\fmf{phantom}{i1,i2,v,o2,o1}
		\fmffreeze
		\fmfshift{(-0.13w,0)}{i2}
		\fmfshift{(0.13w,0)}{o2}
		\fmf{plain,label=$\noexpand\qb$,l.d=1mm}{i2,v}
		\fmf{plain,label=$\noexpand\qb$,l.d=1mm}{v,o2}
		\fmfv{d.sh=circle,d.f=.6,d.si=.2w}{v}
		\fmf{plain,left}{i1,o1}
		\fmf{plain,left}{i2,o2}
	\end{fmfgraph*}
 }} \quad \mbox{\Large +} \quad
 \parbox{15mm}{\raisebox{5mm}{%
	\begin{fmfgraph*}(15,10)
		\fmfbottom{i1,i2,o2,o1}
		\fmfforce{(.1w,0)}{i2}
		\fmfforce{(.9w,0)}{o2}
		\fmf{plain,wi=2thick}{i2,o2}
		\fmf{plain,left}{i2,o2}
		\fmf{plain,left}{i1,o1}
		\fmfv{d.si=0,l=$\noexpand\qb$,l.a=-100,l.d=2mm}{i2}
		\fmfv{d.si=0,l=$\noexpand\qb$,l.a=-80,l.d=2mm}{o2}
	\end{fmfgraph*}
  }} \no
\\
\mbox{\Large $\Sigma_2$} \!\! &\mbox{\Large =}& 
 \parbox{15mm}{\raisebox{-12mm}{%
	\begin{fmfgraph*}(15,8)
		\fmfleft{i1}
		\fmfright{o1}
		\fmf{phantom}{i1,i2,v,o2,o1}
		\fmffreeze
		\fmfshift{(-0.13w,0)}{i2}
		\fmfshift{(0.13w,0)}{o2}
		\fmf{plain,label=$q$,l.d=1.5mm}{i2,v}
		\fmf{plain,label=$\noexpand\qb$,l.d=1mm}{v,o2}
		\fmfv{d.sh=circle,d.f=.6,d.si=.2w}{v}
		\fmf{plain,left}{i1,o1}
		\fmf{plain,left}{i2,o2}
	\end{fmfgraph*}
 }} 
\quad \mbox{\Large +} \quad 
 \parbox{15mm}{\raisebox{5mm}{%
	\begin{fmfgraph*}(15,10)
		\fmfbottom{i1,i2,o2,o1}
		\fmfforce{(.1w,0)}{i2}
		\fmfforce{(.9w,0)}{o2}
		\fmf{plain,wi=2thick}{i2,o2}
		\fmf{plain,left}{i2,o2}
		\fmf{plain,left}{i1,o1}
		\fmfv{d.si=0,l=$q$,l.a=-100,l.d=2mm}{i2}
		\fmfv{d.si=0,l=$\noexpand\qb$,l.a=-80,l.d=1.5mm}{o2}
	\end{fmfgraph*}
  }} \no
\ee
\vspace*{-6mm}
\caption{Self-energy equations for the random scattering model.}
\label{fig.self}
\end{figure}

The self-energy equations are explicitly given by ($G=G_{qq}$,
$\Gb=G_{\qb\qb}$)
\eqn
\Sg_1&=&G+mF_{qq} \,, \nonumber \\
\Sg_4&=&\Gb+mF_{\bar{q}\bar{q}} \,, \nonumber \\
\Sg_2&=&\Gm+mF_{q\bar{q}}\,, 
\label{53} \\
\Sg_3&=&G_{\bar{q}q} + mF_{\bar{q}q} \nonumber
\eqnx
where $G=(\zb-\Sg_4)/det$, $\Gm=(\Sg_2-\lm)/det$, etc., with
$det=(z-\Sg_1)(\zb-\Sg_4)-(\lm-\Sg_2)^2$. 
The equations for the $F$'s are just the generating (``Lippmann-Schwinger") 
equations for the rainbow graphs, that is (see Fig.~\ref{fig.LS})
\eqn
F_{qq}      &=&   \delta +\dl G F_{qq}+\dl \Gm F_{q\bar{q}} \,, \nonumber \\
F_{\bar{q}\bar{q}} &=& -\delta   -\dl \Gb F_{\bar{q}\bar{q}}-\dl\Gm
	F_{q\bar{q}} \,, \nonumber\\ 
F_{q \bar{q}} &=& -\dl \Gb F_{q \bar{q}} -\dl \Gm F_{qq} \,,
\label{54} \\
F_{\bar{q} q} &=& \dl G_{\bar{q}q} F_{\bar{q}\bar{q}} + \dl G
	F_{\bar{q}q} \,. \nonumber
\eqnx

\noindent
One may solve for $F_{q\bar{q}}$ to obtain
\eq
F_{q\bar{q}}=\f{-\dl^2  \Gm}{denom} \,.
\label{55}
\eqx
The equation for the mixed self-energy $\Sg_2\neq 0$ now gives
\eq
\label{e.sommaster}
det=1-\f{\dl^2 m}{denom}
\label{56}
\eqx
where $denom$ is
\eq
denom=1+\dl(\Gb-G)-\f{\dl^2}{det} \,.
\label{57}
\eqx

\begin{figure}[htbp]
\be
\mbox{\Large $F_{qq}$} \!\! &\mbox{\Large =}& \!\!
 \parbox{15mm}{\raisebox{-10mm}{%
	\begin{fmfgraph*}(15,8)
		\fmfleft{i1}
		\fmfright{o1}
		\fmffreeze
		\fmfshift{(0.15w,0)}{i1}
		\fmfshift{(-0.15w,0)}{o1}
		\fmffixed{(0.15w,0)}{i1,i2}
		\fmffixed{(-0.15w,0)}{o1,o2}
		\fmffixed{(-.15w,.2w)}{i1,v1}
		\fmffixed{(-.15w,.2w)}{i2,v2}
		\fmffixed{(.15w,.2w)}{o1,v3}
		\fmffixed{(.15w,.2w)}{o2,v4}
		\fmfv{d.sh=circle,d.f=1,d.si=0,l=$q$,l.d=2mm,l.a=-90}{i2}
		\fmfv{d.sh=circle,d.f=1,d.si=0,l=$q$,l.d=2mm,l.a=-90}{o2}
		\fmf{plain,w=thin}{i1,v1}
		\fmf{plain,w=thin}{i2,v2}
		\fmf{plain,w=thin}{o1,v3}
		\fmf{plain,w=thin}{o2,v4}
		\fmf{plain,width=2thick}{i2,o2}
	\end{fmfgraph*}
 }} \; \mbox{\Large =} \;
  \parbox{10mm}{\raisebox{2mm}{%
	\begin{fmfgraph*}(10,5)
	\fmfleft{a,c,k,v}
	\fmfright{l,b,d}
	\fmffreeze
	\fmfforce{(0,0.7h)}{a}
	\fmfforce{(w,0.7h)}{b}
	\fmfforce{(0.33w,0)}{c}
	\fmfforce{(0.67w,0)}{d}
	\fmfforce{(.17w,0.7h)}{k}
	\fmfforce{(.83w,0.7h)}{l}
	\fmfforce{(.5w,0)}{v}
	\fmf{plain}{a,c}
	\fmf{plain}{k,v,l}
	\fmf{plain}{b,d}
	\fmfv{d.sh=circle,d.f=1,d.si=0,l=q,l.d=2mm,l.a=-110}{c}
	\fmfv{d.sh=circle,d.f=1,d.si=0,l=q,l.d=2mm,l.a=-70}{d}
 \end{fmfgraph*}
}} \; \mbox{\Large +} \quad
 \parbox{25mm}{\raisebox{12mm}{%
	\begin{fmfgraph*}(25,14)
		\fmfbottom{i1,o1}
		\fmffreeze
		\fmfshift{(0.08w,0)}{i1}
		\fmfshift{(-0.08w,0)}{o1}
		\fmffixed{(0.08w,0)}{i1,i2}
		\fmffixed{(-0.08w,0)}{o1,o2}
		\fmffixed{(-0.08w,0)}{o2,o3}
		\fmffixed{(-.08w,.1w)}{i1,v1}
		\fmffixed{(-.08w,.1w)}{i2,v2}
		\fmffixed{(.08w,.1w)}{o1,v3}
		\fmffixed{(.08w,.1w)}{o2,v4}
		\fmfforce{(.46w,0)}{vc1}
		\fmfforce{(.53w,0)}{vc2}
		\fmf{plain,w=thin}{vc2,v,o3}
		\fmfv{d.sh=circle,d.f=1,d.si=0,l=$q$,l.d=2mm,l.a=-90}{o3}
		\fmfv{d.sh=circle,d.f=1,d.si=0,l=$q$,l.d=2mm,l.a=-80}{vc2}
		\fmfv{d.sh=circle,d.f=.6,d.si=.1w}{v}
		\fmf{plain,w=2thick}{i2,vc1}
		\fmfv{d.sh=circle,d.f=1,d.si=0,l=$q$,l.d=2mm,l.a=-90}{i2}
		\fmfv{d.sh=circle,d.f=1,d.si=0,l=$q$,l.d=2mm,l.a=-100}{vc1}
		\fmf{plain}{i1,v1}
		\fmf{plain}{i2,v2}
		\fmf{plain,w=thin}{o1,v3}
		\fmf{plain,w=thin}{o2,v4}
		\fmf{plain,left}{vc1,o2}
		\fmf{plain,left}{vc2,o3}
		\fmfv{d.sh=circle,d.f=1,d.si=0,l=q,l.d=2mm,l.a=-70}{o1}
	\end{fmfgraph*}
 }}
	\; \mbox{\Large +} \quad
 \parbox{25mm}{\raisebox{12mm}{%
	\begin{fmfgraph*}(25,14)
		\fmfbottom{i1,o1}
		\fmffreeze
		\fmfshift{(0.08w,0)}{i1}
		\fmfshift{(-0.08w,0)}{o1}
		\fmffixed{(0.08w,0)}{i1,i2}
		\fmffixed{(-0.08w,0)}{o1,o2}
		\fmffixed{(-0.08w,0)}{o2,o3}
		\fmffixed{(-.08w,.1w)}{i1,v1}
		\fmffixed{(-.08w,.1w)}{i2,v2}
		\fmffixed{(.08w,.1w)}{o1,v3}
		\fmffixed{(.08w,.1w)}{o2,v4}
		\fmfforce{(.46w,0)}{vc1}
		\fmfforce{(.53w,0)}{vc2}
		\fmf{plain,w=thin}{vc2,v,o3}
		\fmfv{d.sh=circle,d.f=1,d.si=0,l=$q$,l.d=2.5mm,l.a=-90}{o3}
		\fmfv{d.sh=circle,d.f=1,d.si=0,l=$\noexpand\qb$,l.d=2mm,l.a=-80}{vc2}
		\fmfv{d.sh=circle,d.f=.6,d.si=.1w}{v}
		\fmf{plain,w=2thick}{i2,vc1}
		\fmfv{d.sh=circle,d.f=1,d.si=0,l=$q$,l.d=2.5mm,l.a=-90}{i2}
		\fmfv{d.sh=circle,d.f=1,d.si=0,l=$\noexpand\qb$,l.d=2mm,l.a=-100}{vc1}
		\fmf{plain}{i1,v1}
		\fmf{plain}{i2,v2}
		\fmf{plain,w=thin}{o1,v3}
		\fmf{plain,w=thin}{o2,v4}
		\fmf{plain,left}{vc1,o2}
		\fmf{plain,left}{vc2,o3}
		\fmfv{d.sh=circle,d.f=1,d.si=0,l=q,l.d=2.5mm,l.a=-70}{o1}
	\end{fmfgraph*}
 }} \no \\[-12mm]
%
%
\mbox{\Large $F_{\qb\qb}$} \!\! &\mbox{\Large =}& \!\!
 \parbox{15mm}{\raisebox{-10mm}{%
	\begin{fmfgraph*}(15,8)
		\fmfleft{i1}
		\fmfright{o1}
		\fmffreeze
		\fmfshift{(0.15w,0)}{i1}
		\fmfshift{(-0.15w,0)}{o1}
		\fmffixed{(0.15w,0)}{i1,i2}
		\fmffixed{(-0.15w,0)}{o1,o2}
		\fmffixed{(-.15w,.2w)}{i1,v1}
		\fmffixed{(-.15w,.2w)}{i2,v2}
		\fmffixed{(.15w,.2w)}{o1,v3}
		\fmffixed{(.15w,.2w)}{o2,v4}
		\fmfv{d.sh=circle,d.f=1,d.si=0,l=$\noexpand\qb$,l.d=2mm,l.a=-90}{i2}
		\fmfv{d.sh=circle,d.f=1,d.si=0,l=$\noexpand\qb$,l.d=2mm,l.a=-90}{o2}
		\fmf{plain,w=thin}{i1,v1}
		\fmf{plain,w=thin}{i2,v2}
		\fmf{plain,w=thin}{o1,v3}
		\fmf{plain,w=thin}{o2,v4}
		\fmf{plain,width=2thick}{i2,o2}
	\end{fmfgraph*}
 }} \; \mbox{\Large =} \; 
  \parbox{10mm}{\raisebox{2mm}{%
	\begin{fmfgraph*}(10,5)
	\fmfleft{a,c,k,v}
	\fmfright{l,b,d}
	\fmffreeze
	\fmfforce{(0,0.7h)}{a}
	\fmfforce{(w,0.7h)}{b}
	\fmfforce{(0.33w,0)}{c}
	\fmfforce{(0.67w,0)}{d}
	\fmfforce{(.17w,0.7h)}{k}
	\fmfforce{(.83w,0.7h)}{l}
	\fmfforce{(.5w,0)}{v}
	\fmf{plain}{a,c}
	\fmf{plain}{k,v,l}
	\fmf{plain}{b,d}
	\fmfv{d.sh=circle,d.f=1,d.si=0,l=$\noexpand\qbm$,l.d=2mm,l.a=-110}{c}
	\fmfv{d.sh=circle,d.f=1,d.si=0,l=$\noexpand\qbm$,l.d=2mm,l.a=-70}{d}
 \end{fmfgraph*}
}} \; \mbox{\Large +} \quad
 \parbox{25mm}{\raisebox{12mm}{%
	\begin{fmfgraph*}(25,14)
		\fmfbottom{i1,o1}
		\fmffreeze
		\fmfshift{(0.08w,0)}{i1}
		\fmfshift{(-0.08w,0)}{o1}
		\fmffixed{(0.08w,0)}{i1,i2}
		\fmffixed{(-0.08w,0)}{o1,o2}
		\fmffixed{(-0.08w,0)}{o2,o3}
		\fmffixed{(-.08w,.1w)}{i1,v1}
		\fmffixed{(-.08w,.1w)}{i2,v2}
		\fmffixed{(.08w,.1w)}{o1,v3}
		\fmffixed{(.08w,.1w)}{o2,v4}
		\fmfforce{(.46w,0)}{vc1}
		\fmfforce{(.53w,0)}{vc2}
		\fmf{plain,w=thin}{vc2,v,o3}
		\fmfv{d.sh=circle,d.f=1,d.si=0,l=$\noexpand\qb$,l.d=2mm,l.a=-90}{o3}
		\fmfv{d.sh=circle,d.f=1,d.si=0,l=$\noexpand\qb$,l.d=2mm,l.a=-80}{vc2}
		\fmfv{d.sh=circle,d.f=.6,d.si=.1w}{v}
		\fmf{plain,w=2thick}{i2,vc1}
		\fmfv{d.sh=circle,d.f=1,d.si=0,l=$\noexpand\qb$,l.d=2mm,l.a=-90}{i2}
		\fmfv{d.sh=circle,d.f=1,d.si=0,l=$\noexpand\qb$,l.d=2mm,l.a=-100}{vc1}
		\fmf{plain}{i1,v1}
		\fmf{plain}{i2,v2}
		\fmf{plain,w=thin}{o1,v3}
		\fmf{plain,w=thin}{o2,v4}
		\fmf{plain,left}{vc1,o2}
		\fmf{plain,left}{vc2,o3}
		\fmfv{d.sh=circle,d.f=1,d.si=0,l=$\noexpand\qbm$,l.d=2mm,l.a=-70}{o1}
	\end{fmfgraph*}
 }}
	\; \mbox{\Large +} \quad
 \parbox{25mm}{\raisebox{12mm}{%
	\begin{fmfgraph*}(25,14)
		\fmfbottom{i1,o1}
		\fmffreeze
		\fmfshift{(0.08w,0)}{i1}
		\fmfshift{(-0.08w,0)}{o1}
		\fmffixed{(0.08w,0)}{i1,i2}
		\fmffixed{(-0.08w,0)}{o1,o2}
		\fmffixed{(-0.08w,0)}{o2,o3}
		\fmffixed{(-.08w,.1w)}{i1,v1}
		\fmffixed{(-.08w,.1w)}{i2,v2}
		\fmffixed{(.08w,.1w)}{o1,v3}
		\fmffixed{(.08w,.1w)}{o2,v4}
		\fmfforce{(.46w,0)}{vc1}
		\fmfforce{(.53w,0)}{vc2}
		\fmf{plain,w=thin}{vc2,v,o3}
		\fmfv{d.sh=circle,d.f=1,d.si=0,l=$\noexpand\qb$,l.d=2mm,l.a=-90}{o3}
		\fmfv{d.sh=circle,d.f=1,d.si=0,l=$q$,l.d=2.5mm,l.a=-80}{vc2}
		\fmfv{d.sh=circle,d.f=.6,d.si=.1w}{v}
		\fmf{plain,w=2thick}{i2,vc1}
		\fmfv{d.sh=circle,d.f=1,d.si=0,l=$\noexpand\qb$,l.d=2mm,l.a=-90}{i2}
		\fmfv{d.sh=circle,d.f=1,d.si=0,l=$q$,l.d=2.5mm,l.a=-100}{vc1}
		\fmf{plain}{i1,v1}
		\fmf{plain}{i2,v2}
		\fmf{plain,w=thin}{o1,v3}
		\fmf{plain,w=thin}{o2,v4}
		\fmf{plain,left}{vc1,o2}
		\fmf{plain,left}{vc2,o3}
		\fmfv{d.sh=circle,d.f=1,d.si=0,l=$\noexpand\qbm$,l.d=2mm,l.a=-70}{o1}
	\end{fmfgraph*}
 }}  \no \\[-12mm]
%
%
\mbox{\Large $F_{q\qb}$} \!\! &\mbox{\Large =}& \!\!
 \parbox{15mm}{\raisebox{-10mm}{%
	\begin{fmfgraph*}(15,8)
		\fmfleft{i1}
		\fmfright{o1}
		\fmffreeze
		\fmfshift{(0.15w,0)}{i1}
		\fmfshift{(-0.15w,0)}{o1}
		\fmffixed{(0.15w,0)}{i1,i2}
		\fmffixed{(-0.15w,0)}{o1,o2}
		\fmffixed{(-.15w,.2w)}{i1,v1}
		\fmffixed{(-.15w,.2w)}{i2,v2}
		\fmffixed{(.15w,.2w)}{o1,v3}
		\fmffixed{(.15w,.2w)}{o2,v4}
		\fmfv{d.sh=circle,d.f=1,d.si=0,l=$q$,l.d=2mm,l.a=-90}{i2}
		\fmfv{d.sh=circle,d.f=1,d.si=0,l=$\noexpand\qb$,l.d=1.5mm,l.a=-90}{o2}
		\fmf{plain,w=thin}{i1,v1}
		\fmf{plain,w=thin}{i2,v2}
		\fmf{plain,w=thin}{o1,v3}
		\fmf{plain,w=thin}{o2,v4}
		\fmf{plain,width=2thick}{i2,o2}
	\end{fmfgraph*}
 }} \; \mbox{\Large =} \; \mbox{\Large +}\;
 \parbox{25mm}{\raisebox{12mm}{%
	\begin{fmfgraph*}(25,14)
		\fmfbottom{i1,o1}
		\fmffreeze
		\fmfshift{(0.08w,0)}{i1}
		\fmfshift{(-0.08w,0)}{o1}
		\fmffixed{(0.08w,0)}{i1,i2}
		\fmffixed{(-0.08w,0)}{o1,o2}
		\fmffixed{(-0.08w,0)}{o2,o3}
		\fmffixed{(-.08w,.1w)}{i1,v1}
		\fmffixed{(-.08w,.1w)}{i2,v2}
		\fmffixed{(.08w,.1w)}{o1,v3}
		\fmffixed{(.08w,.1w)}{o2,v4}
		\fmfforce{(.46w,0)}{vc1}
		\fmfforce{(.53w,0)}{vc2}
		\fmf{plain,w=thin}{vc2,v,o3}
		\fmfv{d.sh=circle,d.f=1,d.si=0,l=$\noexpand\qb$,l.d=2mm,l.a=-90}{o3}
		\fmfv{d.sh=circle,d.f=1,d.si=0,l=$\noexpand\qb$,l.d=2mm,l.a=-80}{vc2}
		\fmfv{d.sh=circle,d.f=.6,d.si=.1w}{v}
		\fmf{plain,w=2thick}{i2,vc1}
		\fmfv{d.sh=circle,d.f=1,d.si=0,l=$q$,l.d=2.5mm,l.a=-90}{i2}
		\fmfv{d.sh=circle,d.f=1,d.si=0,l=$\noexpand\qb$,l.d=2mm,l.a=-100}{vc1}
		\fmf{plain}{i1,v1}
		\fmf{plain}{i2,v2}
		\fmf{plain,w=thin}{o1,v3}
		\fmf{plain,w=thin}{o2,v4}
		\fmf{plain,left}{vc1,o2}
		\fmf{plain,left}{vc2,o3}
		\fmfv{d.sh=circle,d.f=1,d.si=0,l=$\noexpand\qbm$,l.d=2mm,l.a=-70}{o1}
	\end{fmfgraph*}
 }}
	\; \mbox{\Large +} \quad
 \parbox{25mm}{\raisebox{12mm}{%
	\begin{fmfgraph*}(25,14)
		\fmfbottom{i1,o1}
		\fmffreeze
		\fmfshift{(0.08w,0)}{i1}
		\fmfshift{(-0.08w,0)}{o1}
		\fmffixed{(0.08w,0)}{i1,i2}
		\fmffixed{(-0.08w,0)}{o1,o2}
		\fmffixed{(-0.08w,0)}{o2,o3}
		\fmffixed{(-.08w,.1w)}{i1,v1}
		\fmffixed{(-.08w,.1w)}{i2,v2}
		\fmffixed{(.08w,.1w)}{o1,v3}
		\fmffixed{(.08w,.1w)}{o2,v4}
		\fmfforce{(.46w,0)}{vc1}
		\fmfforce{(.53w,0)}{vc2}
		\fmf{plain,w=thin}{vc2,v,o3}
		\fmfv{d.sh=circle,d.f=1,d.si=0,l=$\noexpand\qb$,l.d=1.5mm,l.a=-90}{o3}
		\fmfv{d.sh=circle,d.f=1,d.si=0,l=$q$,l.d=2mm,l.a=-80}{vc2}
		\fmfv{d.sh=circle,d.f=.6,d.si=.1w}{v}
		\fmf{plain,w=2thick}{i2,vc1}
		\fmfv{d.sh=circle,d.f=1,d.si=0,l=$q$,l.d=2mm,l.a=-90}{i2}
		\fmfv{d.sh=circle,d.f=1,d.si=0,l=$q$,l.d=2mm,l.a=-100}{vc1}
		\fmf{plain}{i1,v1}
		\fmf{plain}{i2,v2}
		\fmf{plain,w=thin}{o1,v3}
		\fmf{plain,w=thin}{o2,v4}
		\fmf{plain,left}{vc1,o2}
		\fmf{plain,left}{vc2,o3}
		\fmfv{d.sh=circle,d.f=1,d.si=0,l=$\noexpand\qbm$,l.d=1.5mm,l.a=-70}{o1}
	\end{fmfgraph*}
 }} \no
\ee
\vspace*{-12mm}
\caption{``Lippmann-Schwinger" equations for $F$'s.}
\label{fig.LS}
\end{figure}

\noindent
The way to proceed now is as follows. One can write an equivalent 
generating equations for
$F_{\bar{q}q}$ by adding the graphs on the ``quark'' side. This shows that 
$F_{q\bar{q}}=F_{\bar{q}q}$. 
This in conjunction with
(\ref{e.sommaster}) gives the relation
\eq
F_{qq} +F_{\bar{q}\bar{q}}= \frac{1}{m}(G+\Gb)(1-det) \,. 
\label{58}
\eqx
Inserting this into the self-energy equations one obtains
\eq
G+\Gb=x
\label{59}
\eqx
Re-expressing $F_{\bar{q}\bar{q}}$ by $F_{q\bar{q}}$ in the appropriate 
self-energy equations and using
(\ref{e.sommaster}) one obtains
\eq
\zb- G\,det+\dl\zb\Gb=\dl(1-m)+\Gb(1+\dl x)-{\dl}/{det} \,.
\label{60}
\eqx
Eliminating $\Gb$, one solves this in conjunction with (\ref{e.sommaster})
treated as an equation for the determinant $det$:
\eq
[1+\dl(x-2G)-{\dl^2}/{det}]\,det=1+\dl(x-2G)-{\dl^2}/{det}-\dl^2 m \,.
\label{61}
\eqx
The final result (after changing $\dl$ into $-i\gm$) is 
\eq
G (z, \zb) =\f{x}{2}+\f{i}{2}\left[ \f{1}{\gm}+\f{m}{y}+
\f{\gm}{1-\gm y}\right] \,.
\label{62}
\eqx
This result was first derived by Haake,
Sommers and coworkers using the replica method \cite{HAAKE}, and the
supersymmetric method \cite{SOMMERS95}.

Before closing this section we would like to mention briefly the connection
of our diagrammatical approach to the mathematical concepts of free random 
variables~\cite{VOICULESCU,SPEICHER}.
 For  applications 
of free random variables to various physical problems 
see e.g. \cite{DOUGLAS,GROSS,ENGELHARDT,ZEE,USBLUE}. 
With the compact notations
\eqn
\bf{\Sigma}&=&\arr{\Sg_1}{\Sg_2}{\Sg_3}{\Sg_4}\,\,\,\,\,
\bf{F}=\arr{F_{qq}}{F_{q\bar{q}}}{F_{\bar{q}q}}{F_{\bar{q}\bar{q}}}\nonumber
\\
\GG&=&\arr{G_{qq}}{G_{q\bar{q}}}{G_{\bar{q}q}}{G_{\bar{q}\bar{q}}}\,\,\,\,
\, \bf{\hat{\delta}}=\arr{\delta}{0}{0}{-\delta}
\label{boldnot}
\eqnx
the equations (\ref{53}) and (\ref{54}) could be rewritten in the form
\eqn
\bf{\Sigma}&=& \GG +m {\bf F} \nonumber \,, \\
\bf{F}&=&\bf{\hat{\delta}}+\bf{\hat{\delta}}\GG \bf{F} \,.
\label{boldDS}
\eqnx
Solving for the matrix $\bf{\Sg}$ we get
\eqn
{\bf{\Sg}} = \GG + m(1-\bf{\hat{\delta}} \GG)^{-1} {\bf{\hat{\delta}}}
\eqnx
Introducing the 
{\em generalized} Blue's function~\cite{JNPWZ} (an extension
of Zee's approach~\cite{ZEE} to nonhermitean ensembles), 
as $2\times 2$ matrix valued function defined by 
\eq
\BB(\GG)=\ZZ=\arr{z}{\lm}{\lm}{\zb}
\label{genblue}
\eqx
with $\lm\rightarrow 0$, we see that
 $\bf{\Sg}=\BB(\GG)-\GG^{-1}$. 
Therefore
\eqn
\BB(\GG)=\GG +\GG^{-1}+ m(1-\bf{\hat{\delta}} \GG)^{-1} \bf{\hat{\delta}}
\label{blueadd}
\eqnx
in which we recognize the generalized addition 
law postulated in~\cite{JNPWZ},
with  generalized Gaussian Blue's function 
\eqn
\BB_{gauss}(\AA)=\frac{1}{\AA} +\AA
\eqnx
and generalized Blue's function for the  random scattering 
part of the Hamiltonian  
form
\eq
   \BB_{-i \gamma VV^{\dagger} }(\AA)= m(1-\bf{\hat{\delta}}
 \AA)^{-1}\bf{\hat{\delta}}
 +\frac{1}{\AA}  
\label{blueivv}
\eqx
with  
\eq
\bf{\hat{\delta}}= \left( 
\begin{array}{cc} i\gamma & 0 \\ 0 & -i\gamma \end{array} \right) \,.
\eqx
This completes the diagrammatic proof of the result announced already 
in our earlier work~\cite{JNPWZ}.

\section{Two-Point Functions}

To probe the character of the correlations between the eigenvalues of 
nonhermitean random matrices, either on their holomorphic or nonholomorphic 
supports, it is relevant to investigate two-point functions. A measure of
the breaking of holomorphic symmetry in the eigenvalue distribution 
is given by the connected two-point function or correlator
\eq
N^2G_c(z,\bar{z})=\corr{\left| \tr \f{1}{z-D-\HH}\right|^2}_c
\label{63}
\eqx
where the $z$ and $\zb$ content of the averaging is probed simultaneously.
In \cite{USMUX} the correlation function (\ref{63}) was shown to diverge
precisely on the nonholomorphic support of the eigenvalue distribution,
indicating an accumulation in the eigenvalue density.
In the conventional language of ``quarks'' and ``gluons'', 
(\ref{63}) is just the correlation function between ``quarks" and their
``conjugates". A divergence in (\ref{63}) in the $z$-plane reflects 
large fluctuations between the eigenvalues of the nonhermitean operators
on finite $z$-supports, hence their ``condensation". 

It was shown in \cite{AMBJORN} and \cite{BREZIN} that for hermitean
matrices (with $\zb\to w$) the fluctuations in connected and smoothed
two-point functions satisfy the general lore of macroscopic
universality. 
This means that all smoothed correlation functions are universal and 
could be classified by the support of the spectral densities, 
independently of the specifics of the random ensemble and genera in the 
topological expansion (see \cite{AKEMANN} for a recent discussion).  In 
this section we will evaluate a number of two-point functions for random
but nonhermitean ensembles on both their holomorphic and nonholomorphic
supports. We confirm that the general lore of macroscopic universality 
extends to the nonhermitean case as well.

The concept of smoothed correlation functions will become manifest
when comparing to  numerical calculations. Indeed, when correlating 
$N\times N$ matrices at finite eigenvalue-separations, oscillations are
expected due to the occurrence of a large number of eigenvalues. The 
correlation functions produced in the $1/N$ analysis are smoothed. 
Unsmoothed correlation functions are of interest for the studies
of spectral form factors in the crossover region \cite{BREHI}. They will not
be discussed here.

\subsection{Notations}

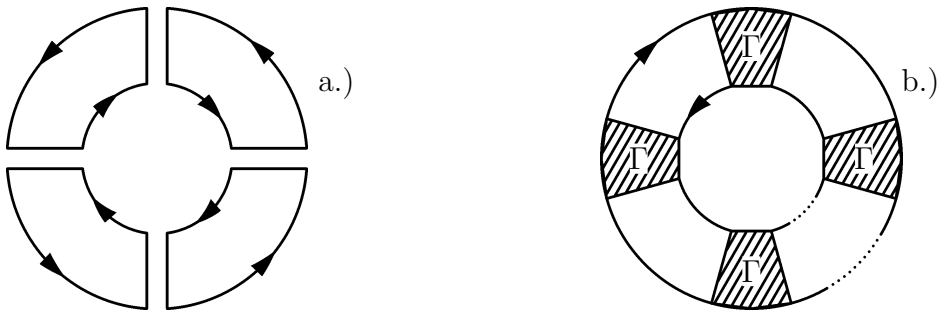
\begin{figure}[htbp]
%
\centerline{%
\parbox{40mm}{%
  \begin{fmfgraph*}(40,40)
	\fmfleft{l1}
	\fmfright{l2}
	\fmf{phantom}{l1,s1,g1,g2,g3,g4,s2,l2}
	\fmffreeze
	\fmfforce{(.25w,.5h)}{s1}
	\fmfforce{(.75w,.5h)}{s2}
	\fmfforce{(.125w,.5h)}{g1}
	\fmfforce{(.875w,.5h)}{g3}
	\fmfforce{(.5w,.875h)}{g2}
	\fmfforce{(.5w,.125h)}{g4}
	\fmf{phantom,left,tension=0.2,tag=1}{l1,l2}
	\fmf{phantom,left,tension=0.2,tag=2}{l2,l1}
	\fmf{phantom,left,tension=0.2,tag=3}{s1,s2}
	\fmf{phantom,left,tension=0.2,tag=4}{s2,s1}
	\fmfposition
	\fmfipath{p[]}
	\fmfiset{p1}{vpath1(__l1,__l2)}
	\fmfiset{p2}{vpath2(__l2,__l1)}
	\fmfiset{p3}{vpath3(__s1,__s2)}
	\fmfiset{p4}{vpath4(__s2,__s1)}
	\fmfi{quark}{subpath (23length(p1)/48,length(p1)/48) of p1}
	\fmfi{quark}{subpath (47length(p1)/48,25length(p1)/48) of p1}
	\fmfi{quark}{subpath (23length(p2)/48,length(p2)/48) of p2}
	\fmfi{quark}{subpath (47length(p2)/48,25length(p2)/48) of p2}
	\fmfi{quark}{subpath (length(p3)/24,11length(p3)/24) of p3}
	\fmfi{quark}{subpath (13length(p3)/24,23length(p3)/24) of p3}
	\fmfi{quark}{subpath (length(p4)/24,11length(p4)/24) of p4}
	\fmfi{quark}{subpath (13length(p4)/24,23length(p4)/24) of p4}
	\fmfi{plain}{point length(p1)/48 of p1 -- point length(p3)/24 of p3}
	\fmfi{plain}{point 23length(p1)/48 of p1 -- point 11length(p3)/24 of p3}
	\fmfi{plain}{point 25length(p1)/48 of p1 -- point 13length(p3)/24 of p3}
	\fmfi{plain}{point 47length(p1)/48 of p1 -- point 23length(p3)/24 of p3}
	\fmfi{plain}{point length(p2)/48 of p2 -- point length(p4)/24 of p4}
	\fmfi{plain}{point 23length(p2)/48 of p2 -- point 11length(p4)/24 of p4}
	\fmfi{plain}{point 25length(p2)/48 of p2 -- point 13length(p4)/24 of p4}
	\fmfi{plain}{point 47length(p2)/48 of p2 -- point 23length(p4)/24 of p4}
 \end{fmfgraph*}
} \raisebox{10mm}{a.)}
\hspace*{30mm}
%
%
\parbox{40mm}{%
  \begin{fmfgraph*}(40,40)
	\fmfleft{l1}
	\fmfright{l2}
	\fmf{phantom}{l1,s1,g1,g2,g3,g4,s2,l2}
	\fmffreeze
	\fmfforce{(.25w,.5h)}{s1}
	\fmfforce{(.75w,.5h)}{s2}
	\fmfforce{(.125w,.5h)}{g1}
	\fmfforce{(.875w,.5h)}{g3}
	\fmfforce{(.5w,.875h)}{g2}
	\fmfforce{(.5w,.125h)}{g4}
	\fmf{phantom,left,tension=0.2,tag=1}{l1,l2}
	\fmf{phantom,left,tension=0.2,tag=2}{l2,l1}
	\fmf{phantom,left,tension=0.2,tag=3}{s1,s2}
	\fmf{phantom,left,tension=0.2,tag=4}{s2,s1}
	\fmfposition
	\fmfipath{p[]}
	\fmfiset{p1}{vpath1(__l1,__l2)}
	\fmfiset{p2}{vpath2(__l2,__l1)}
	\fmfiset{p3}{vpath3(__s1,__s2)}
	\fmfiset{p4}{vpath4(__s2,__s1)}
	\fmfi{quark}{subpath (0,length(p1)/2) of p1}
	\fmfi{plain}{subpath (length(p1)/2,length(p1)) of p1}
	\fmfi{plain}{subpath (0,length(p2)/6) of p2}
	\fmfi{dots}{subpath (length(p2)/6,length(p2)/3) of p2}
 	\fmfi{plain}{subpath (length(p2)/3,length(p2)) of p2}
	\fmfi{quark}{subpath (length(p3)/2,0) of p3}
	\fmfi{plain}{subpath (length(p3)/2,length(p3)) of p3}
	\fmfi{plain}{subpath (0,length(p4)/6) of p4}
	\fmfi{dots}{subpath (length(p4)/6,length(p4)/3) of p4}
 	\fmfi{plain}{subpath (length(p4)/3,length(p4)) of p4}
	\fmfcmd{shadedraw (point 0 of p1 -- point length(p1)/12 of p1 --
  point length(p3)/12 of p3 -- point 11length(p4)/12 of p4 -- point
  11length(p2)/12 of p2) -- cycle;}
	\fmfcmd{shadedraw (point 5length(p1)/12 of p1 -- point length(p1)/2
  of p1 -- point 7length(p1)/12 of p1 -- point 7length(p3)/12 of p3 -- point
  5length(p3)/12 of p3) -- cycle;}
	\fmfcmd{shadedraw (point 5length(p2)/12 of p2 -- point length(p2)/2
  of p2 -- point 7length(p2)/12 of p2 -- point 7length(p4)/12 of p4 -- point
  5length(p4)/12 of p4) -- cycle;}
	\fmfcmd{shadedraw (point length(p1) of p1 -- point 11length(p1)/12
  of p1 -- point 11length(p3)/12 of p3 -- point length(p4)/12 of p4 -- point
  length(p2)/12 of p2) -- cycle;}
	\fmfcmd{unfill (vloc(__g1)+(-1.2mm,2mm))--(vloc(__g1)+(-1.2mm,-2mm))--(vloc(__g1)+(1.2mm,-2mm))--(vloc(__g1)+(1.2mm,2mm))--cycle;}
	\fmfcmd{unfill (vloc(__g2)+(-1.2mm,2mm))--(vloc(__g2)+(-1.2mm,-2mm))--(vloc(__g2)+(1.2mm,-2mm))--(vloc(__g2)+(1.2mm,2mm))--cycle;}
	\fmfcmd{unfill (vloc(__g3)+(-1.2mm,2mm))--(vloc(__g3)+(-1.2mm,-2mm))--(vloc(__g3)+(1.2mm,-2mm))--(vloc(__g3)+(1.2mm,2mm))--cycle;}
	\fmfcmd{unfill (vloc(__g4)+(-1.2mm,2mm))--(vloc(__g4)+(-1.2mm,-2mm))--(vloc(__g4)+(1.2mm,-2mm))--(vloc(__g4)+(1.2mm,2mm))--cycle;}
	\fmfv{l=$\noexpand\Gamma$,l.d=0}{g1}
	\fmfv{l=$\noexpand\Gamma$,l.d=0}{g2}
	\fmfv{l=$\noexpand\Gamma$,l.d=0}{g3}
	\fmfv{l=$\noexpand\Gamma$,l.d=0}{g4}
 \end{fmfgraph*}
}\raisebox{10mm}{b.)}
}
\caption{Sample wheel diagram ($n=4$) for the two-point correlators (GUE), 
(a) and generalized wheel diagram with kernel $\Gamma$ (b).}
\label{fig.wheel}
\end{figure}

To establish the notations for the two-point correlators, we follow 
\cite{BREZIN} and define the two-point connected (c) correlator for GUE,
\eqn
G_c(z,w)&=&\left< \frac{1}{N} {\rm tr}\frac{1}{z-\HH}
\frac{1}{N} {\rm tr}\frac{1}{w-\HH}\right>_c \no \\
&=&\partial_z \partial_w \left< \frac{1}{N} {\rm tr} \log (z-\HH)
\frac{1}{N} {\rm tr} \log (w-\HH) \right>_c \,.
\label{gausscor}
\eqnx
Expansion of the logarithms yields 
\eqn
G_c(z,w)=\partial_z \partial_w
\sum_{n=1}^{\infty} \sum_{m=1}^{\infty}\frac{1}{z^n w^m}\left<
\frac{1}{Nn} {\rm tr} \HH^n \frac{1}{Nm} {\rm tr} \HH^m \right>_c  \,.
\eqnx
Let us first consider $n=m$. The graphs in this case are represented 
by the wheel diagram (Fig.~\ref{fig.wheel}a), where the rungs of the wheel
are  given by the ``gluon" propagator from Fig.~\ref{fig.rules}. 
The diagram splits into $n$ disconnected sectors, corresponding to the
fact that
$<{\rm tr}\HH^n {\rm tr} \HH^n> =n$. Resumming the diagonal terms give
to 
\eq
N^2 G_c(z,w) = -\partial_z \partial_w \log (1-\frac{1}{zw}) \,.
\eqx
All the remaining (planar) diagrams
are rainbow-like (see Fig.~\ref{fig.rainbow}) and correspond 
to dressing the bare ``quark" propagators, i.e. $1/z \rightarrow G(z)$.
The result is
\eq
N^2 G_c(z,w) = -\partial_z \partial_w \log [1-G(z)G(w)] \,.
\label{XAX}
\eqx
The two-point correlator follows functionally from the one-point correlator. 
Such a relation for the case of Gaussian ensemble was already 
noted in \cite{FRENCHMELLO} prior to the universality arguments.

This
result can be further reduced \cite{AMBJORN,BREZIN}, to show that the
two-point 
correlator depends only on the end-points of the spectral density irrespective
of the choice of the weight. Since the smoothed correlation function between 
the eigenvalues follow from (\ref{XAX}), it reflects on 
correlations between eigenvalues separated by $N$ in the spectrum. Its generic
form reflects on macroscopic universality in the eigenvalue correlations. This
is to be contrasted with microscopic universality which is a statement about 
the correlations between eigenvalues one-level spacing apart in the spectrum
\cite{GENERALHER}.

\subsection{Ginibre-Girko Correlator}

In this section we show how to generalize (\ref{XAX}) to the nonholomorphic
region of complex and random matrices. We begin our discussion with the 
simplest case  $(H_1 + i H_2)/\sqrt{2}$, where $ H_1$ and $H_2$ are random
hermitean Gaussians.
 The essential difference between this case and the case discussed
above is that the ``quark'' propagators do not commute, and care must be taken 
with respect to their ordering.

The sectors of the wheel (kernel, denoted by  $\Gamma$) 
correspond to  the ``two-quark" irreducible
scattering amplitude. In our case it is not 1 (like
in the hermitean Gaussian case) but a tensor 
product in ``isospin'' space
\eq
\Gamma^{f_2f_3;g_2g_3}=\arr{\One}{0}{0}{0}_{f_2f_3} \otimes
\arr{0}{0}{0}{\One}_{g_2g_3} +
\arr{0}{0}{0}{\One}_{f_2f_3} \otimes \arr{\One}{0}{0}{0}_{g_2g_3}\!\!.
\eqx
The two terms correspond to the ``gluon" amplitudes of
Fig.~\ref{fig.eightglue}b and \ref{fig.eightglue}c,  respectively.
This particular form becomes transparent in Fig.~\ref{fig.sector}.

\begin{figure}[htbp]
\be
 \parbox{30mm}{\raisebox{-7mm}{%
	\begin{fmfgraph*}(30,12)
		\fmfstraight
		\fmfbottom{i1,vl,l1,vl1,vl2,o1}
		\fmftop{i2,vu,vu1,l2,vu2,o2}
		\fmfright{v}
		\fmffreeze
		\fmfforce{(.15w,0)}{vl}
		\fmfforce{(.15w,h)}{vu}
		\fmfforce{(.5w,0)}{vl1}
		\fmfforce{(.5w,h)}{vu1}
		\fmfforce{(.65w,.5h)}{v}
		\fmfforce{(.8w,0)}{vl2}
		\fmfforce{(.8w,h)}{vu2}
		\fmfforce{(.33w,0)}{l1}
		\fmfforce{(.33w,h)}{l2}
		\fmffreeze
		\fmf{plain}{i1,vl}
		\fmf{quark}{vl,vl1}
		\fmf{quark}{vu1,vu}
		\fmf{plain}{i2,vu}
		\fmf{quark}{vl2,o1}
		\fmf{quark}{o2,vu2}
		\fmfv{d.sh=circle,d.f=.5,d.si=.12w}{vl}
		\fmfv{d.sh=circle,d.f=.5,d.si=.12w}{vu}
		\fmfv{d.si=0,l=$\noexpand g_1$,l.a=180,l.d=2mm}{i1}
		\fmfv{d.si=0,l=$\noexpand f_1$,l.a=180,l.d=2mm}{i2}
		\fmfv{d.si=0,l=$\noexpand g_2$,l.a=90,l.d=1mm}{l1}
		\fmfv{d.si=0,l=$\noexpand f_2$,l.a=90,l.d=1mm}{l2}
		\fmfv{d.si=0,l=$\noexpand f_3$,l.a=0,l.d=1mm}{o2}
		\fmfv{d.si=0,l=$\noexpand g_3$,l.a=0,l.d=1mm}{o1}
		\fmfcmd{shadedraw (vloc(__vu1))--(vloc(__vl1))--(vloc(__vl2))--(vloc(__vu2))--cycle;}
		\fmfcmd{unfill (vloc(__v)+(-1.2mm,2mm))--(vloc(__v)+(-1.2mm,-2mm))--(vloc(__v)+(1.2mm,-2mm))--(vloc(__v)+(1.2mm,2mm))--cycle;}
		\fmfv{d.si=0,l=$\noexpand\Gamma$,l.d=0}{v}
	\end{fmfgraph*}
 }} \quad \mbox{\Large =} \quad
	\mbox{\Large ${\cal G}_{f_1f_2} \otimes {\cal G}_{g_1g_2}^T\ 
	\Gamma^{f_2f_3;g_2g_3}$} \no
\ee
\caption{Two-point kernel with $f,g = q, \bar{q}$.}
\label{fig.sector}
\end{figure}

 The factors 
\eqn
\frac{1}{z w}
\eqnx
get transformed (after ``dressing'') into
\eqn
\GG(z) \otimes {\GG}^T(w) \,.
\label{otimes}
\eqnx
Here $\GG(z)$ and $\GG(w)$ are the resolvents discussed
in section~4.1 with $\tau=0$. Transposition keeps track of the ``flow''
 of indices. 

Another generalization deals with the treatment of the derivatives. Now
the logarithm is not a holomorphic functions so we must treat $\partial_z$
as
\eq
\partial_z=\f{1}{2}(\partial_x-i\partial_y)
\label{68}
\eqx
where $z=x+iy$ and
\eq
\partial_{\br{w}}=\f{1}{2}(\partial_u+i\partial_v)
\label{69}
\eqx
where $w=u+iv$. Note that in order to obtain $G(z,\zb)$ we {\em cannot}
write the derivatives as $\partial_z\partial_{\zb}$ but rather $\partial_z
\partial_{\br{w}}$ and set $\br{w}=\zb$ at the end of the calculation.

The choice of ``isospin" in the lower fermion
loop is done by choosing the appropriate derivative $\partial_w$ for the 
``quark" and $\partial_{\br{w}}$ for the ``conjugate-quark".
As in \cite{BZ}, two distinct contributions may be identified. 
First, when the derivatives (bullets) act on the exposed or hidden
$\Gamma$ lines within the same sector of the wheel, as shown in 
Fig.~\ref{fig.secsam}. Second, when the derivatives act on  different
sectors as shown in Fig.~\ref{fig.secdif}.

\begin{figure}[htbp]
\be
\partial_1\partial_2 \left( \;\;\;
  \parbox{12mm}{\raisebox{-3mm}{%
	\begin{fmfgraph*}(12,10)
	\fmfbottom{i1,vl1,vl2,o1}
	\fmftop{i2,vu1,vu2,o2}
	\fmffreeze
	\fmfforce{(.4w,0)}{vl1}
	\fmfforce{(.4w,h)}{vu1}
	\fmfforce{(.7w,0)}{vl2}
	\fmfforce{(.7w,h)}{vu2}
	\fmffreeze
	\fmf{plain}{i1,vl1}
	\fmf{plain}{i2,vu1}
	\fmf{dots}{vl2,o1}
	\fmf{dots}{vu2,o2}
	\fmfv{d.si=0,l=$1$,l.a=180,l.d=1mm}{i2}
	\fmfv{d.si=0,l=$2$,l.a=180,l.d=1mm}{i1}
	\fmffreeze
	\fmfcmd{shadedraw (vloc(__vu1))--(vloc(__vl1))--(vloc(__vl2))--(vloc(__vu2))--cycle;}
	\end{fmfgraph*}
  }} \right) \mbox{\Large =} \;
  \parbox{12mm}{\raisebox{-3mm}{%
	\begin{fmfgraph}(12,10)
	\fmfbottom{i1,v1,vl1,vl2,o1}
	\fmftop{i2,v2,vu1,vu2,o2}
	\fmffreeze
	\fmfforce{(.2w,0)}{v1}
	\fmfforce{(.2w,h)}{v2}
	\fmfforce{(.4w,0)}{vl1}
	\fmfforce{(.4w,h)}{vu1}
	\fmfforce{(.7w,0)}{vl2}
	\fmfforce{(.7w,h)}{vu2}
	\fmffreeze
	\fmf{plain}{i1,vl1}
	\fmf{plain}{i2,vu1}
	\fmf{dots}{vl2,o1}
	\fmf{dots}{vu2,o2}
	\fmfv{d.sh=circle,d.f=1,d.si=.1w}{v1}
	\fmfv{d.sh=circle,d.f=1,d.si=.1w}{v2}
	\fmffreeze
	\fmfcmd{shadedraw (vloc(__vu1))--(vloc(__vl1))--(vloc(__vl2))--(vloc(__vu2))--cycle;}
	\end{fmfgraph}
  }} \; \mbox{\Large +} \;
  \parbox{12mm}{\raisebox{-3mm}{%
	\begin{fmfgraph}(12,10)
	\fmfbottom{i1,vl1,v1,vl2,o1}
	\fmftop{i2,v2,vu1,vu2,o2}
	\fmfforce{(.4w,0)}{vl1}
	\fmfforce{(.4w,h)}{vu1}
	\fmfforce{(.7w,0)}{vl2}
	\fmfforce{(.7w,h)}{vu2}
	\fmfforce{(.2w,h)}{v2}
	\fmfforce{(.55w,0)}{v1}
	\fmf{plain}{i1,vl1}
	\fmf{plain}{i2,vu1}
	\fmf{dots}{vl2,o1}
	\fmf{dots}{vu2,o2}
	\fmfv{d.sh=circle,d.f=1,d.si=.1w}{v1}
	\fmfv{d.sh=circle,d.f=1,d.si=.1w}{v2}
	\fmfcmd{shadedraw (vloc(__vu1))--(vloc(__vl1))--(vloc(__vl2))--(vloc(__vu2))--cycle;}
	\end{fmfgraph}
  }} \; \mbox{\Large +} \;
  \parbox{12mm}{\raisebox{-3mm}{%
	\begin{fmfgraph}(12,10)
	\fmfbottom{i1,v1,vl1,vl2,o1}
	\fmftop{i2,vu1,v2,vu2,o2}
	\fmfforce{(.4w,0)}{vl1}
	\fmfforce{(.4w,h)}{vu1}
	\fmfforce{(.7w,0)}{vl2}
	\fmfforce{(.2w,0)}{v1}
	\fmfforce{(.55w,h)}{v2}
	\fmfforce{(.7w,h)}{vu2}
	\fmf{plain}{i1,vl1}
	\fmf{plain}{i2,vu1}
	\fmf{dots}{vl2,o1}
	\fmf{dots}{vu2,o2}
	\fmfv{d.sh=circle,d.f=1,d.si=.1w}{v1}
	\fmfv{d.sh=circle,d.f=1,d.si=.1w}{v2}
	\fmfcmd{shadedraw (vloc(__vu1))--(vloc(__vl1))--(vloc(__vl2))--(vloc(__vu2))--cycle;}
	\end{fmfgraph}
  }} \; \mbox{\Large +} \;
  \parbox{12mm}{\raisebox{-3mm}{%
	\begin{fmfgraph}(12,10)
	\fmfbottom{i1,vl1,v1,vl2,o1}
	\fmftop{i2,vu1,v2,vu2,o2}
	\fmfforce{(.4w,0)}{vl1}
	\fmfforce{(.4w,h)}{vu1}
	\fmfforce{(.55w,0)}{v1}
	\fmfforce{(.55w,h)}{v2}
	\fmfforce{(.7w,0)}{vl2}
	\fmfforce{(.7w,h)}{vu2}
	\fmf{plain}{i1,vl1}
	\fmf{plain}{i2,vu1}
	\fmf{dots}{vl2,o1}
	\fmf{dots}{vu2,o2}
	\fmfv{d.sh=circle,d.f=1,d.si=.1w}{v1}
	\fmfv{d.sh=circle,d.f=1,d.si=.1w}{v2}
	\fmfcmd{shadedraw (vloc(__vu1))--(vloc(__vl1))--(vloc(__vl2))--(vloc(__vu2))--cycle;}
	\end{fmfgraph}
  }} \; \mbox{\Large =} \ 
\partial_1\partial_2 \left(\GG_1 \otimes \GG_2^T \Gamma \right) \no
\ee	
\vspace*{-3mm}
\caption{Derivatives acting within the same sector of $\Gamma$ in the 
wheel.}
\label{fig.secsam}
\end{figure}

\setlength{\unitlength}{0.8mm}
\begin{figure}[htbp]
\be
\partial_1\partial_2 \left( \;\;\;
  \parbox{16mm}{\raisebox{-3mm}{%
	\begin{fmfgraph*}(20,10)
	\fmfbottom{i1,vl1,vl2,m1,vl3,vl4,o1}
	\fmftop{i2,vu1,vu2,m2,vu3,vu4,o2}
	\fmffreeze
	\fmfforce{(.2w,0)}{vl1}
	\fmfforce{(.2w,h)}{vu1}
	\fmfforce{(.35w,0)}{vl2}
	\fmfforce{(.35w,h)}{vu2}
	\fmfforce{(.5w,0)}{m1}
	\fmfforce{(.5w,h)}{m2}
	\fmfforce{(.7w,0)}{vl3}
	\fmfforce{(.7w,h)}{vu3}
	\fmfforce{(.85w,0)}{vl4}
	\fmfforce{(.85w,h)}{vu4}
	\fmffreeze
	\fmf{plain}{i1,vl1}
	\fmf{plain}{i2,vu1}
	\fmf{dots}{vl2,m1}
	\fmf{dots}{vu2,m2}
	\fmf{plain}{m1,vl3}
	\fmf{plain}{m2,vu3}
	\fmf{dots}{vl4,o1}
	\fmf{dots}{vu4,o2}
	\fmfv{d.si=0,l=$1$,l.a=180,l.d=1mm}{i2}
	\fmfv{d.si=0,l=$2$,l.a=180,l.d=1mm}{i1}
	\fmffreeze
	\fmfcmd{shadedraw (vloc(__vu1))--(vloc(__vl1))--(vloc(__vl2))--(vloc(__vu2))--cycle;}
	\fmfcmd{shadedraw (vloc(__vu3))--(vloc(__vl3))--(vloc(__vl4))--(vloc(__vu4))--cycle;}
	\end{fmfgraph*}
  }} \; \right) \hspace*{0mm} &\mbox{\Large =}& \hspace*{0mm} \left( \;
  \parbox{16mm}{\raisebox{-3mm}{%
	\begin{fmfgraph}(20,10)
	\fmfbottom{i1,v,vl1,vl2,m1,vl3,vl4,o1}
	\fmftop{i2,vu1,vu2,m2,vu3,vu4,o2}
	\fmffreeze
	\fmfforce{(.2w,0)}{vl1}
	\fmfforce{(.2w,h)}{vu1}
	\fmfforce{(.1w,0)}{v}
	\fmfforce{(.35w,0)}{vl2}
	\fmfforce{(.35w,h)}{vu2}
	\fmfforce{(.5w,0)}{m1}
	\fmfforce{(.5w,h)}{m2}
	\fmfforce{(.7w,0)}{vl3}
	\fmfforce{(.7w,h)}{vu3}
	\fmfforce{(.85w,0)}{vl4}
	\fmfforce{(.85w,h)}{vu4}
	\fmffreeze
	\fmf{plain}{i1,vl1}
	\fmf{plain}{i2,vu1}
	\fmfv{d.sh=circle,d.f=1,d.si=.075w}{v}
	\fmfcmd{shadedraw (vloc(__vu1))--(vloc(__vl1))--(vloc(__vl2))--(vloc(__vu2))--cycle;}
	\fmf{dots}{vl2,m1}
	\fmf{dots}{vu2,m2}
	\fmf{plain}{m1,vl3}
	\fmf{plain}{m2,vu3}
	\fmfcmd{shadedraw (vloc(__vu3))--(vloc(__vl3))--(vloc(__vl4))--(vloc(__vu4))--cycle;}
	\fmf{dots}{vl4,o1}
	\fmf{dots}{vu4,o2}
	\end{fmfgraph}
  }} \ \mbox{\Large +} \;
  \parbox{16mm}{\raisebox{-3mm}{%
	\begin{fmfgraph}(20,10)
	\fmfbottom{i1,vl1,vl2,m1,vl3,v,vl4,o1}
	\fmftop{i2,vu1,vu2,m2,vu3,vu4,o2}
	\fmffreeze
	\fmfforce{(.2w,0)}{vl1}
	\fmfforce{(.2w,h)}{vu1}
	\fmfforce{(.35w,0)}{vl2}
	\fmfforce{(.35w,h)}{vu2}
	\fmfforce{(.5w,0)}{m1}
	\fmfforce{(.5w,h)}{m2}
	\fmfforce{(.7w,0)}{vl3}
	\fmfforce{(.7w,h)}{vu3}
	\fmfforce{(.775w,0)}{v}
	\fmfforce{(.85w,0)}{vl4}
	\fmfforce{(.85w,h)}{vu4}
	\fmffreeze
	\fmf{plain}{i1,vl1}
	\fmf{plain}{i2,vu1}
	\fmfcmd{shadedraw (vloc(__vu1))--(vloc(__vl1))--(vloc(__vl2))--(vloc(__vu2))--cycle;}
	\fmf{dots}{vl2,m1}
	\fmf{dots}{vu2,m2}
	\fmf{plain}{m1,vl3}
	\fmf{plain}{m2,vu3}
	\fmfcmd{shadedraw (vloc(__vu3))--(vloc(__vl3))--(vloc(__vl4))--(vloc(__vu4))--cycle;}
	\fmfv{d.sh=circle,d.f=1,d.si=.075w}{v}
	\fmf{dots}{vl4,o1}
	\fmf{dots}{vu4,o2}
	\end{fmfgraph}
  }} \ \right) \mbox{\Large $\cdot$} \left( \;
  \parbox{16mm}{\raisebox{-3mm}{%
	\begin{fmfgraph}(20,10)
	\fmfbottom{i1,vl1,vl2,m1,vl3,vl4,o1}
	\fmftop{i2,v,vu1,vu2,m2,vu3,vu4,o2}
	\fmffreeze
	\fmfforce{(.2w,0)}{vl1}
	\fmfforce{(.2w,h)}{vu1}
	\fmfforce{(.1w,h)}{v}
	\fmfforce{(.35w,0)}{vl2}
	\fmfforce{(.35w,h)}{vu2}
	\fmfforce{(.5w,0)}{m1}
	\fmfforce{(.5w,h)}{m2}
	\fmfforce{(.7w,0)}{vl3}
	\fmfforce{(.7w,h)}{vu3}
	\fmfforce{(.85w,0)}{vl4}
	\fmfforce{(.85w,h)}{vu4}
	\fmffreeze
	\fmf{plain}{i1,vl1}
	\fmf{plain}{i2,vu1}
	\fmfv{d.sh=circle,d.f=1,d.si=.075w}{v}
	\fmfcmd{shadedraw (vloc(__vu1))--(vloc(__vl1))--(vloc(__vl2))--(vloc(__vu2))--cycle;}
	\fmf{dots}{vl2,m1}
	\fmf{dots}{vu2,m2}
	\fmf{plain}{m1,vl3}
	\fmf{plain}{m2,vu3}
	\fmfcmd{shadedraw (vloc(__vu3))--(vloc(__vl3))--(vloc(__vl4))--(vloc(__vu4))--cycle;}
	\fmf{dots}{vl4,o1}
	\fmf{dots}{vu4,o2}
	\end{fmfgraph}
  }} \ \mbox{\Large +} \;
  \parbox{16mm}{\raisebox{-3mm}{%
	\begin{fmfgraph}(20,10)
	\fmfbottom{i1,vl1,vl2,m1,vl3,vl4,o1}
	\fmftop{i2,vu1,vu2,m2,vu3,v,vu4,o2}
	\fmffreeze
	\fmfforce{(.2w,0)}{vl1}
	\fmfforce{(.2w,h)}{vu1}
	\fmfforce{(.35w,0)}{vl2}
	\fmfforce{(.35w,h)}{vu2}
	\fmfforce{(.5w,0)}{m1}
	\fmfforce{(.5w,h)}{m2}
	\fmfforce{(.7w,0)}{vl3}
	\fmfforce{(.7w,h)}{vu3}
	\fmfforce{(.775w,h)}{v}
	\fmfforce{(.85w,0)}{vl4}
	\fmfforce{(.85w,h)}{vu4}
	\fmffreeze
	\fmf{plain}{i1,vl1}
	\fmf{plain}{i2,vu1}
	\fmfcmd{shadedraw (vloc(__vu1))--(vloc(__vl1))--(vloc(__vl2))--(vloc(__vu2))--cycle;}
	\fmf{dots}{vl2,m1}
	\fmf{dots}{vu2,m2}
	\fmf{plain}{m1,vl3}
	\fmf{plain}{m2,vu3}
	\fmfcmd{shadedraw (vloc(__vu3))--(vloc(__vl3))--(vloc(__vl4))--(vloc(__vu4))--cycle;}
	\fmfv{d.sh=circle,d.f=1,d.si=.075w}{v}
	\fmf{dots}{vl4,o1}
	\fmf{dots}{vu4,o2}
	\end{fmfgraph}
  }} \; \right) \no \\[3mm]
 &\mbox{\Large =} & 
	\ldots \partial_1 \left( \GG_1 \otimes
	\GG_2^T \Gamma \right) \ldots \partial_2 \left(\GG_1 \otimes
	\GG_2^T \Gamma \right) \no
\ee	
\vspace*{-6mm}
\caption{Derivatives (bullets) acting on two different sectors of $\Gamma$.}
\label{fig.secdif}
\end{figure}
\setlength{\unitlength}{1mm}

Finally, we have
\eqn
N^2G_c(z,w) ={\rm tr}_{q\bar{q}} 
&& \left( \f{1}{1-\GT\Gamma}\partial_z\partial_w
\left(\GT\Gamma\right)  \right.\nonumber\\
 &&\left. +\f{1}{1-\GT\Gamma}\partial_z \left(\GT\Gamma\right) \right.
\nonumber\\
 && \left.\times \f{1}{1-\GT\Gamma}\partial_w \left(\GT\Gamma\right) \right)
\eqnx
as an apparent generalization of the Br\'{e}zin-Hikami-Zee formalism.
The denominators in the above formula come from the fact that the dotted 
lines in Figs.~\ref{fig.secsam},~\ref{fig.secdif} 
could represent $0,1, 2,..\infty$ 
sectors $\Gamma$ of the wheel (see Fig.~\ref{fig.wheel}b) 
and have to be resummed.   
The last result  may be rewritten as 
\eq
N^2G_c(z,w)=-\partial_z\partial_w {\rm tr}_{q\bar{q}} \log\left(
1-\GT\Gamma\right) \,.
\label{logxx}
\eqx
Here the logarithm is understood as a power series expansion.
The operator $\GT\Gamma$ is a tensor product of $2\times 2$ matrices. One can
rewrite it as an ordinary $4\times 4$ matrix. The trace $\tr_{q\bar{q}}$
 becomes
an ordinary matrix trace. This allows the following reformulation
\eq
\label{l.form}
N^2G_c(z,w)=-\partial_z\partial_w \log \det(1-\GT\Gamma) \,.
\eqx

\noindent
Using the explicit expressions for the Ginibre-Girko resolvents~(\ref{24})
\eqn
\GG_1 &=& \left( \begin{array}{cc} \zb &g_z\\g_z&z \end{array}\right) \,, \\
\GG_2 &=&
\left( \begin{array}{cc} \bar{w} &g_w\\g_w&w \end{array}\right)
\eqnx
where $g_z^2=|z|^2-1$,  the operator $\GT\Gamma$ can be evaluated to 
\eq
\left[\arr{\zb}{0}{g_z}{0}\otimes\arr{0}{g_w}{0}{w}+
\arr{0}{g_z}{0}{z}\otimes\arr{\bar{w}}{0}{g_w}{0}\right] \,.
\eqx
In a 4 by 4 matrix notation $\GT\Gamma$ is
\eq
\GT\Gamma = \left(\begin{array}{cccc}
0& \zb g_w & g_z \bar{w} & 0 \\
0& \zb w  & g_zg_w  & 0 \\
0& g_z g_w   &z \bar{w}& 0 \\
0& g_z w & z g_w& 0
\end{array}\right) \,.
\eqx
The determinant of $1-\GT\Gamma$ gives $|z-w|^2$, so 
\eqn
N^2 G_{qq}(z,w)&=&\f{-1}{(w-z)^2} \label{93} \,.\\
N^2 G_{q\qb}(z,w) &=& 0 \,.
\label{64}
\eqnx

\subsection{Elliptic Correlator}
%

We consider now  the general case  of subsection~4.1.
In this case, there is an extra contribution to the Ginibre-Girko 
diagrams, due to the $\tau$ terms in the ``gluonic" propagator. 
In terms of graphs, another diagram may contribute to $\Gamma$ 
with a penalty factor
$\tau$ (see Fig.~\ref{fig.eightglue}e and \ref{fig.eightglue}h).
As a result,
\eqn
\label{e.gamma}
\Gamma&=&\hphantom{\tau} \arr{\One}{0}{0}{0} \otimes \arr{0}{0}{0}{\One}+
\hphantom{\tau}\arr{0}{0}{0}{\One} \otimes \arr{\One}{0}{0}{0}\nonumber \\
&+& \tau \arr{\One}{0}{0}{0} \otimes \arr{\One}{0}{0}{0}+
\tau \arr{0}{0}{0}{\One} \otimes \arr{0}{0}{0}{\One} \,.
\eqnx
The rest of the calculation follows the Ginibre-Girko case, with the resolvents
for the elliptic case as in  section 4.1 
\eqn
\GG_1 &=& \frac{1}{1-\tau^2}
\left( \begin{array}{cc} \zb-\tau z &g_z\\g_z&z-\tau \zb \end{array}\right) \\
\GG_2 &=&\frac{1}{1-\tau^2}
\left( \begin{array}{cc} \bar{w}-\tau w &g_w\\g_w&w-\tau \bar{w}
	\end{array}\right) 
\eqnx
with $g_z^2=|z|^2(1-\tau)^2-\tau(z+\zb)^2-(1-\tau^2)^2$.
After defining $c_{\tau}= (1-\tau^2)^{-2}$, 
$z_{\tau}=z-\tau\bar{z}$, $w_{\tau}=w-\tau\bar{w}$,  we have
\eq
N^2 G_c(z,\bar{w})=-\partial_z \partial_{\bar{w}}
\log {\rm det\,} {\cal L}
\eqx
where ${\cal L}$ is given explicitly by
\eq
{\cal L}=\frac{1}{c_{\tau}}
\left(\begin{array}{cccc}
c_{\tau} -\tau \bar{z}_{\tau}\bar{w}_{\tau}&
-\zb_{\tau}g_w  & -g_z \bar{w}_{\tau} & -\tau g_z g_w \\
-\tau \zb_{\tau}g_w &c_{\tau}-\zb_{\tau}w_{\tau}
 &- g_zg_w  &-\tau g_z w_{\tau} \\
-\tau g_z \bar{w}_{\tau}&- g_z g_w   & c_{\tau}-z_{\tau}\bar{w}_{\tau}
&-\tau z_{\tau}g_w\\
-\tau g_z g_w& -g_z w_{\tau} w & -z_{\tau}g_w&
c_{\tau}-\tau z_{\tau}w_{\tau}
\end{array}\right) \,.
\eqx
After some remarkable cancelations, the determinant ${\cal L}$
is again simply $|z-w|^2$, giving the correlators~(\ref{93}--\ref{64}).

\begin{figure}[htbp]
\centerline{\epsfysize=8truecm \epsfbox{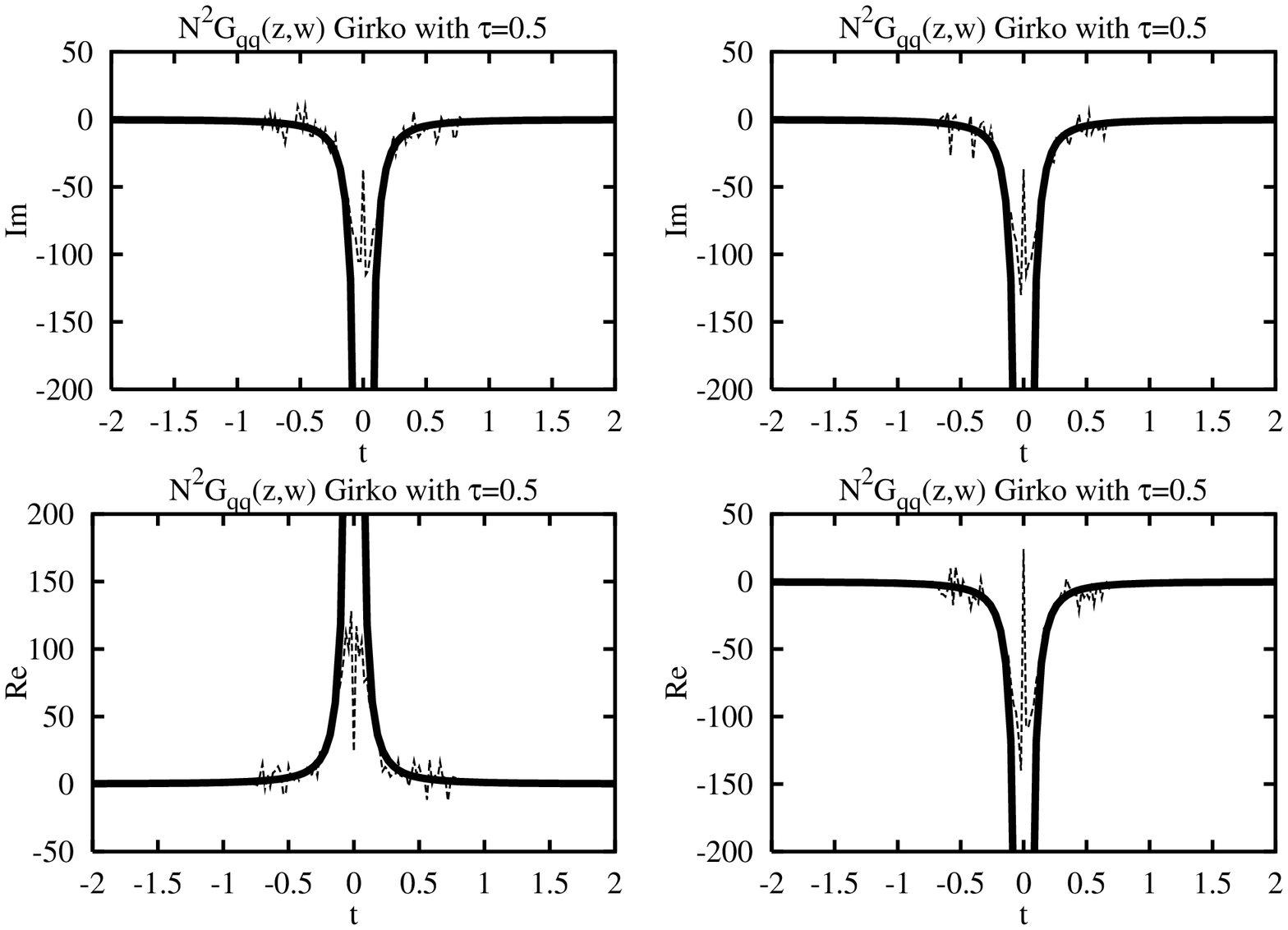}}
\vspace*{-5mm}
\caption{Numerical (dashed line) versus analytical (solid line) results for 
the elliptic correlators from GUE (see text).}
\label{fig.numgirko}
\end{figure}

\begin{figure}[htbp]
\centerline{\epsfysize=5.5truecm \epsfbox{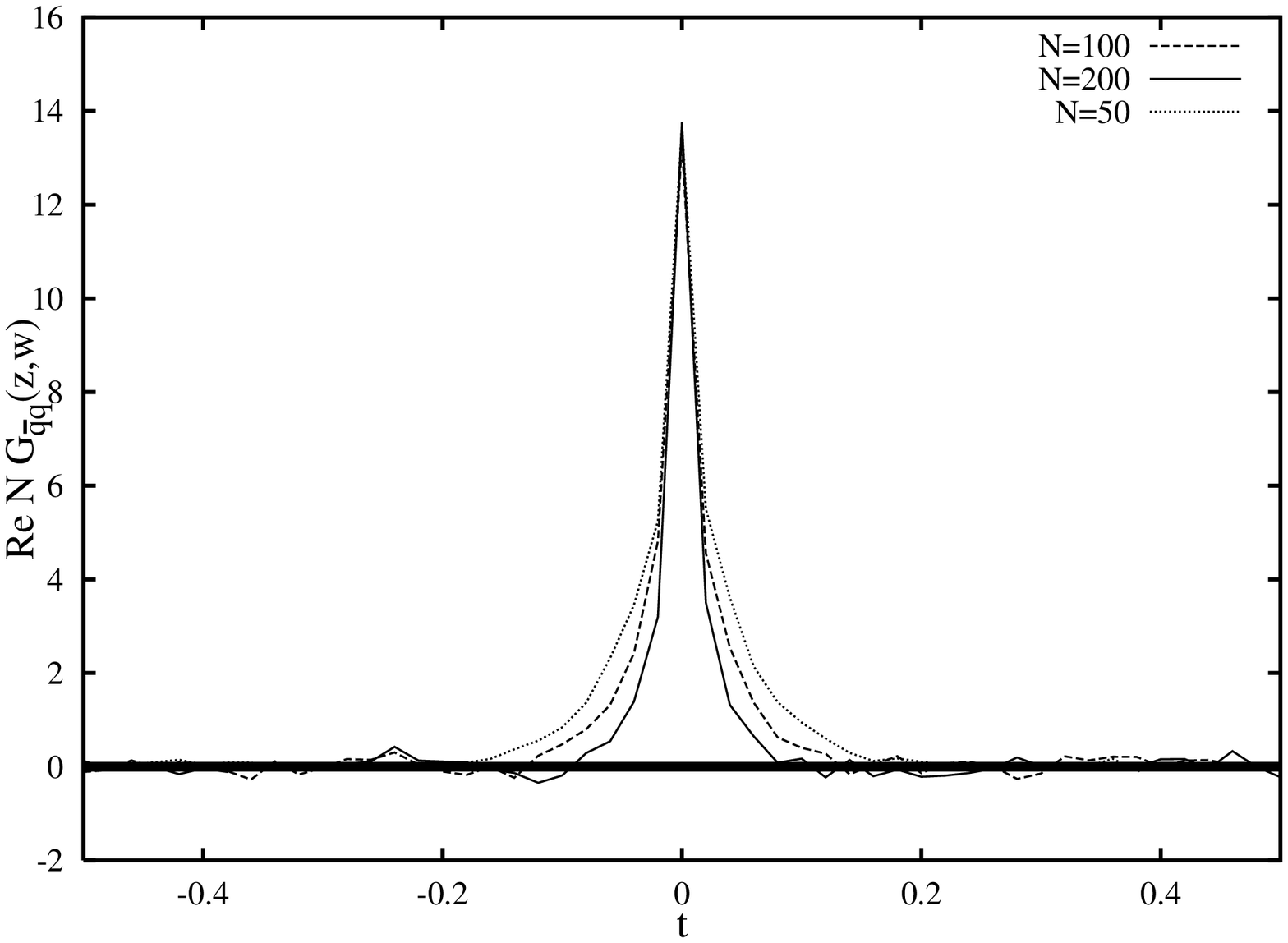}}
\caption{Same as Fig.~\protect\ref{fig.numgirko} but for the
``quark-conjugate-quark" correlator divided by $N$.
The analytical result~(\protect\ref{64}) is zero in this region.}
\label{fig.gircb}
\end{figure}

Figure~\ref{fig.numgirko} compares the analytical results (\ref{93})
(solid lines) for the ``quark-quark" correlator, to the numerical calculations 
(dashed lines) using an ensemble of 50000 matrices of size
100$\times$100 for $\tau=0.5$ (real/major axis is 1.5, imaginary/minor is 0.5).
The figures on the left are parameterized by $z=t$,
$w=t/\protect\sqrt{2} (1+i)$, while those on the right are parameterized by
$z=t/\protect\sqrt{2} (1+i)$, $w=i t$. In Fig.~\ref{fig.gircb} we show the
analytical result (thick solid line) for the ``quark-conjugate-quark"
correlator~(\ref{64}) versus
the numerical calculations for different matrix sizes $N=50,100,200$. The peak
around zero in the numerical calculations is proportional to $N$, while its 
width decreases with larger matrices, approaching zero in the large $N$ limit.
Our analytical evaluation of the correlators disregard delta-functions.

Let us now investigate the case of {\it real} asymmetric matrices 
defined by~(\ref{23}). When the 'twisted' propagator
\eq
\corr{M_{ij}M_{ij}}= \frac 1N
\eqx
couples at both ends to the same ``quark" (as is the case for the
self-energy) it gives a subleading contribution in $1/N$ (see 
Fig.~\ref{fig.twistedpastur}). But in the
calculation of the correlator it may couple to two different ``quark" loops
giving a planar contribution. Now it joins ``quarks" of the same species.
Similar arguments apply for the `$\tau$'-propagator which joins ``quarks"
and ``conjugate-quarks". Diagrammatically we observe
that the 'twisted' propagators and the ones considered earlier cannot mix
in the planar limit so we must add the two resummed contributions separately
--- the
one calculated earlier and the twisted one given by 
\eq
-\partial_1\partial_2 \log \det(\One-\GG_1(z)\otimes
	\GG_2^T(w)\cdot\Gamma_{twisted}) 
\eqx
where $\Gamma_{twisted}$ is given by the expression obtained from
(\ref{e.gamma}) after interchanging $1$ and $\tau$:
\eqn
\Gamma_{twisted}&=&\hphantom{\tau} \arr{\One}{0}{0}{0} \otimes
	\arr{\One}{0}{0}{0}+ 
\hphantom{\tau}\arr{0}{0}{0}{\One} \otimes \arr{0}{0}{0}{\One}\nonumber \\
&+& \tau \arr{\One}{0}{0}{0} \otimes \arr{0}{0}{0}{\One}+
\tau \arr{0}{0}{0}{\One} \otimes \arr{\One}{0}{0}{0} \,.
\eqnx
Here the terms correspond respectively to the ``gluonic" amplitudes in
Fig.~\ref{fig.eightglue}a, \ref{fig.eightglue}d, \ref{fig.eightglue}f
and \ref{fig.eightglue}g.
The determinant is now given by 
\eq
const\cdot(w-\br{z})(\br{w}-z) \,.
\eqx
This leads to correlators of the form:
\eqn
N^2 G_{qq}(z,w)&=&\f{-1}{(w-z)^2} \,, \\ 
N^2 G_{q\qb}(z,\br{w}) &=& \f{-1}{(\br{w}-z)^2} \,.
\label{64GOE}
\eqnx
Note that these are different expressions than in the
complex  case, although the Green's functions are identical
in both cases. 
%
%

\subsection{Chiral Correlator I (Holomorphic Region)}

In the nonhermitean (chiral) case the correlator in the {\it outside} region 
is calculated using the same arguments as above. In the outside, the resolvents 
are holomorphic, hence we may consider the derivatives as acting on 
analytical functions. The amplitude $\Gamma$ reads in this case 
\eq
\Gamma=\arr{0}{\One}{0}{0}_{ab} \otimes \arr{0}{\One}{0}{0}_{cd}+
\arr{0}{0}{\One}{0}_{ab} \otimes \arr{0}{0}{\One}{0}_{cd}
\label{66}
\eqx
where $ab$ are the ``quark" indices and $cd$ are the ``conjugate-quark"
indices. Note that 
the kernel is rewritten in the chiral  basis, i.e. the indices
$a$,$b$,$c$
 and $d$ run
from $1$ to $2N$. 
The correlator reads
\eq
N^2G_c(z,\zb)=-\frac{1}{4}\partial_z\partial_{\zb} \log \det(1-\GG_{qq}(z)
 \otimes \GG^T_{\qb\qb}(z)\Gamma)
\label{l.formm}
\eqx
{}From the definition of $\GG$ in~(\ref{44}) we see, that in the
holomorphic case ($\Sg_2=\Sg_3=0$ and $\Sg_1=G$) the $\GG_{qq}$
component (after performing the trivial tracing of the diagonal $N\times
N$ blocks) is simply
\be
  \GG_{qq} = \arr{z-\Sg_1}{\mu}{-\mu}{z-\Sg_1}^{-1}
\ee
where we used the explicit chiral structure~(\ref{43}). Performing the
inversion 
we arrive at
\eqn
\GG_{qq} &=& \f{1}{det}\arr{z-G}{-\mu}{\mu}{z-G} \,, \\
\GG^T_{\qb\qb} &=&
\f{1}{\overline{det}}\arr{\zb-\Gb}{-\mu}{\mu}{\zb-\Gb} 
\eqnx
with the determinants given by
\be
det&=&(z-G)^2+\mu^2 \quad ,\quad \overline{det} = (\zb-\Gb)^2+\mu^2
\ee
with $G$ being the solution of Pastur
equation~(\ref{cubicpasturmux}) and 
$\Gb$ is just the complex conjugate of $G$.
Now the operator $\GT\Gamma$ reads
\eq
\f{1}{|det|^2}\left[\arr{0}{z\!-\!G}{0}{\mu}
	\otimes\arr{0}{\zb\!-\!\Gb}{0}{\mu} \!+\!
	\arr{-\mu}{0}{z\!-\!G}{0}
	\otimes\arr{-\mu}{0}{\zb\!-\!\Gb}{0}\right] \,.
\eqx
or in $4\times 4$ form
\eq
\f{1}{D}\cdot
\left(\begin{array}{cccc}
\mu^2 & 0 & 0 & |z-G|^2 \\
-\mu(\zb-\Gb) & 0 & 0 & \mu(z-G) \\
-\mu(z-G) & 0 & 0 & \mu(\zb-\Gb) \\
|z-G|^2 & 0 & 0 & \mu^2
\end{array}\right) \,,
\label{deterchir}
\eqx
with
\eq
D\equiv |det|^2=\f{|z-G|^2}{|G|^2}\,.
\eqx
where holomorphic G is the appropriate branch of
(\ref{cubicpasturmux}) rewritten in the form $G [(z-G)^2+\mu^2] = z-G$.
The determinant of $1-\GT\Gamma$ gives
\eq
\frac{(D-\mu^2)^2-|z-G|^4}{D^2}
\label{detmodul}
\eqx
The zero of the determinant in (\ref{l.formm})
occurs for $(D-\mu^2)=|z-G|^2$, that is
\eq
|z-G|^2(1-|G|^2)-\mu^2 |G|^2=0
\label{e.ch1z}
\eqx
as quoted in \cite{USMUX}. In the case $\mu=0$ (and $\zb=w$), the determinant
in (\ref{l.formm}) is simply $(1-G^2(z) G^2(w))$ (chiral) as opposed to
$(1-G(z)G(w))$ (non-chiral). As a result, for $w=z$ and $\mu=0$,
 (\ref{l.formm}) is
\eq
N^2 G(z,z)=1/(z^2 (z^2-4)^2)
\eqx
which coincides with (5.5) in \cite{MAKEENKO}. For small $z$, the behavior
$G(z, z) \sim 1/z^2$ reflects  the exchange of two ``massless" modes for
$z=im \rightarrow 0$.

\begin{figure}[htbp]
\centerline{\epsfysize=5.5truecm \epsfbox{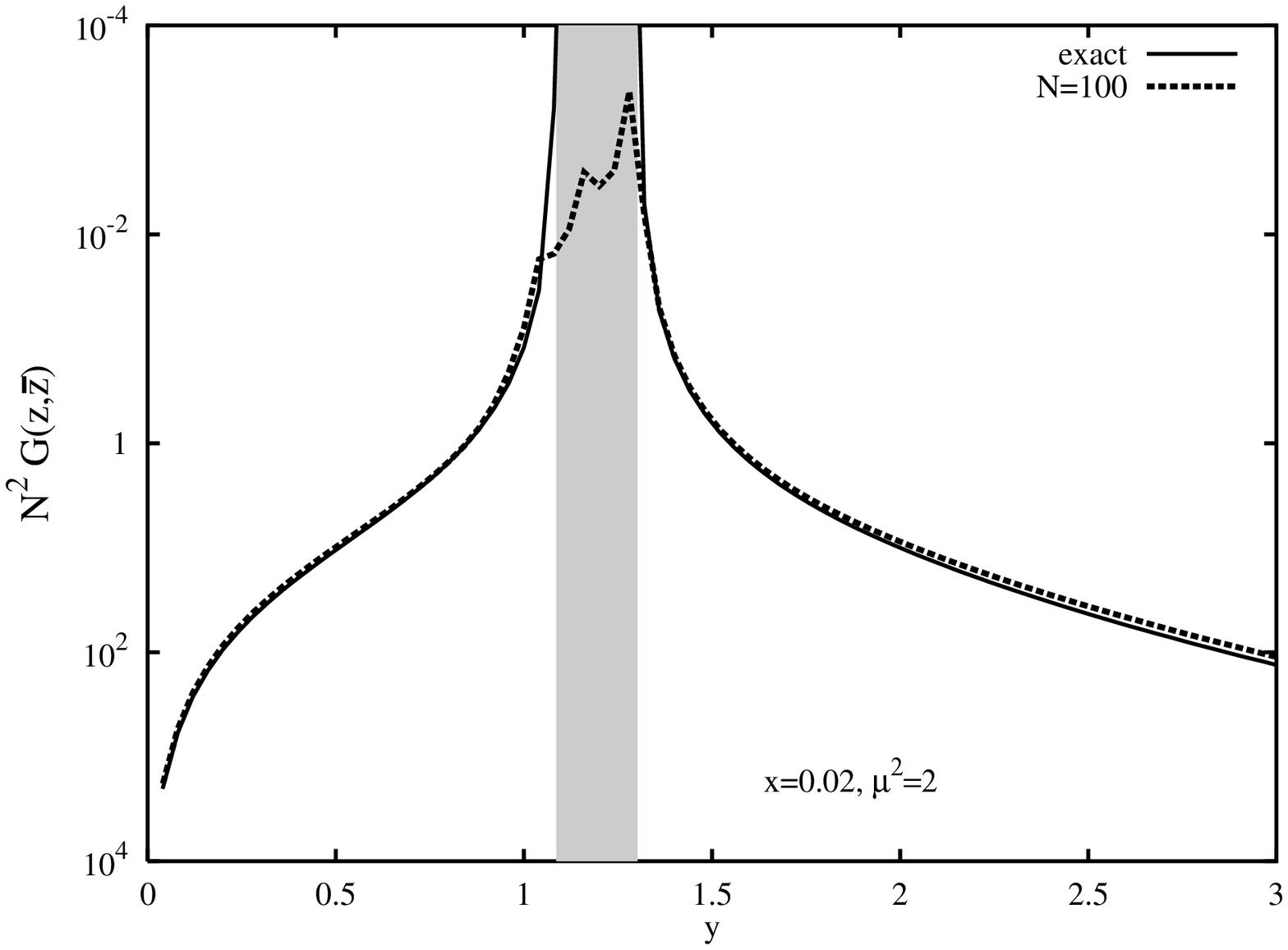}}
\caption{Numerical (dashed) versus analytical (solid line)
results for the chiral correlator I. The nonholomorphic region is shown 
dashed.}
\label{fig.ch1}
\end{figure}

The numerical
results for the chiral correlator versus the present analytical results 
are shown in Fig.~\ref{fig.ch1}. 
The dashed line is obtained from a numerical simulation of 200
100$\times$100 matrices at $\mu^2=2$, while the solid line
follows from~(\protect\ref{l.formm}). The dashed region indicates the
extent of the nonholomorphic region for this choice of $\mu^2=2.$, over
which the result (\ref{l.formm}) no longer holds.

\subsection{Chiral Correlator II (Non-Holomorphic Region)} 

As an extension of this construction we consider the case
of a nonhermitean deterministic matrix plus a hermitean chiral
random matrix. The extension amounts to adding an ``isospin" index
to the resolvents which are now $4\times 4$  matrices (2 ``isospin"+ 2 chirality).
The original $4\times 4$  matrix
$\GG_{qq}(z)\otimes\GG_{\qb\qb}^T(w)\Gamma$  is now
changed into a $16\times 16$ matrix (each factor in the tensor product is
now $4\times 4$ so the tensor product has dimension 16).
The matrix $\GG(z)\otimes\GG^T(w)\Gamma$ is  of the form
\eq
\left(\begin{array}{cccc}
\ggd{qq}{qq} & \ggd{qq}{q\qb} & \ggd{q \qb}{qq} & \ggd{q \qb}{q \qb}\\
\ggd{qq}{\qb q} & \ggd{qq}{\qb\qb} & \ggd{q\qb}{\qb q} & \ggd{q\qb}{\qb\qb}\\
\ggd{\qb q}{qq} & \ggd{\qb q}{q\qb} & \ggd{\qb \qb}{q q} & 
  \ggd{\qb \qb}{q\qb}\\
\ggd{\qb q}{\qb q} & \ggd{\qb q}{\qb \qb} & \ggd{\qb\qb}{\qb q} & 
  \ggd{\qb\qb}{\qb\qb}
\end{array}\right)
\label{67}
\eqx
The complexity in this case stems from the fact that in the 
nonholomorphic region ``quarks" may turn to ``conjugate-quarks" and vice-versa, with 
all ``quark" species interacting with themselves.

\begin{figure}[htbp]
\centerline{\epsfysize=5.5truecm \epsfbox{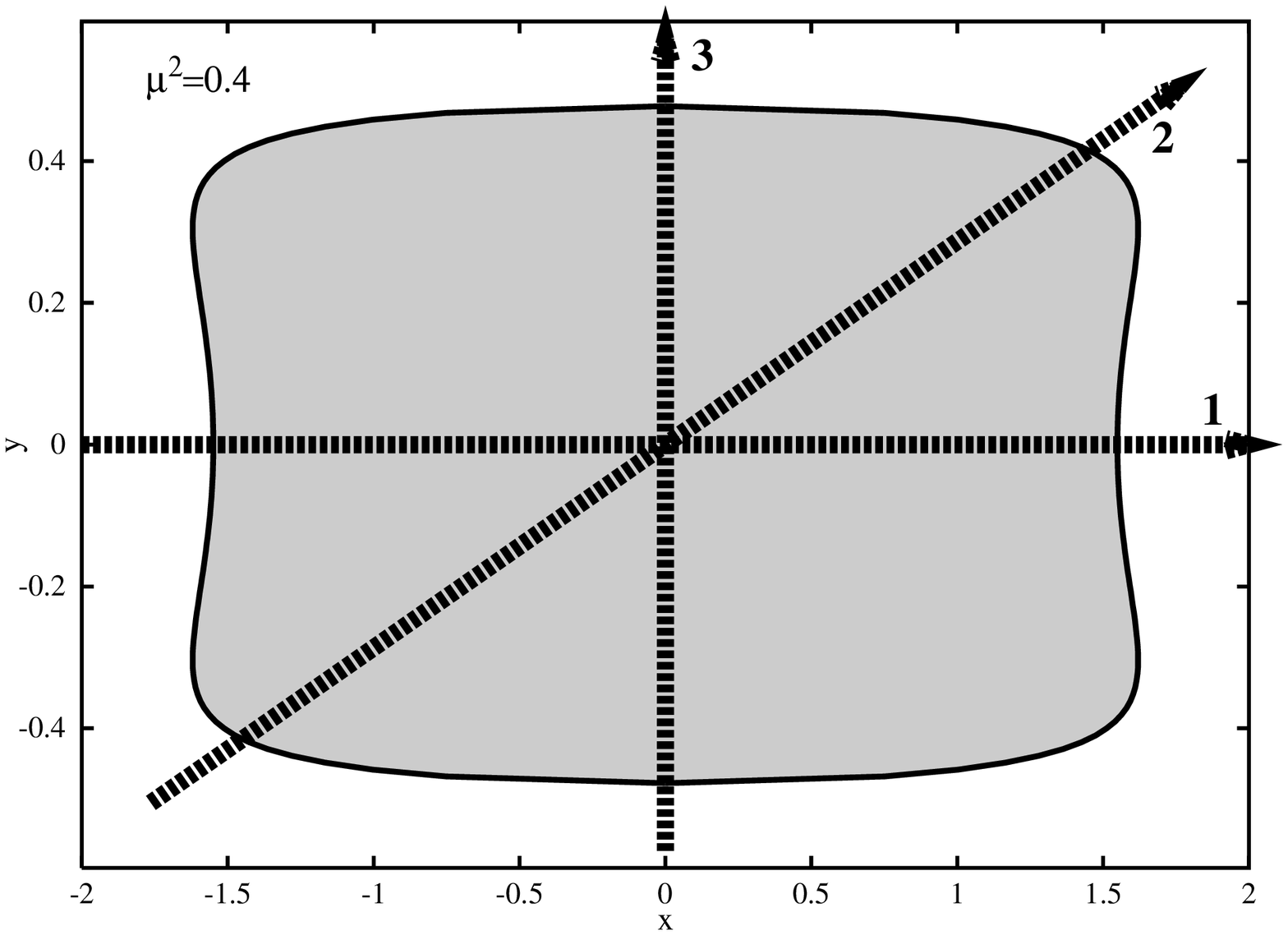}}
\caption{Extent of the non-holomorphic region (shadowed) for the chiral
ensemble at $\mu^2=0.4$. The thick dashed lines indicate the paths along
which the numerical correlator was evaluated.}
\label{fig.stpath}
\end{figure}  

\begin{figure}[htbp]
\centerline{\epsfysize=8truecm \epsfbox{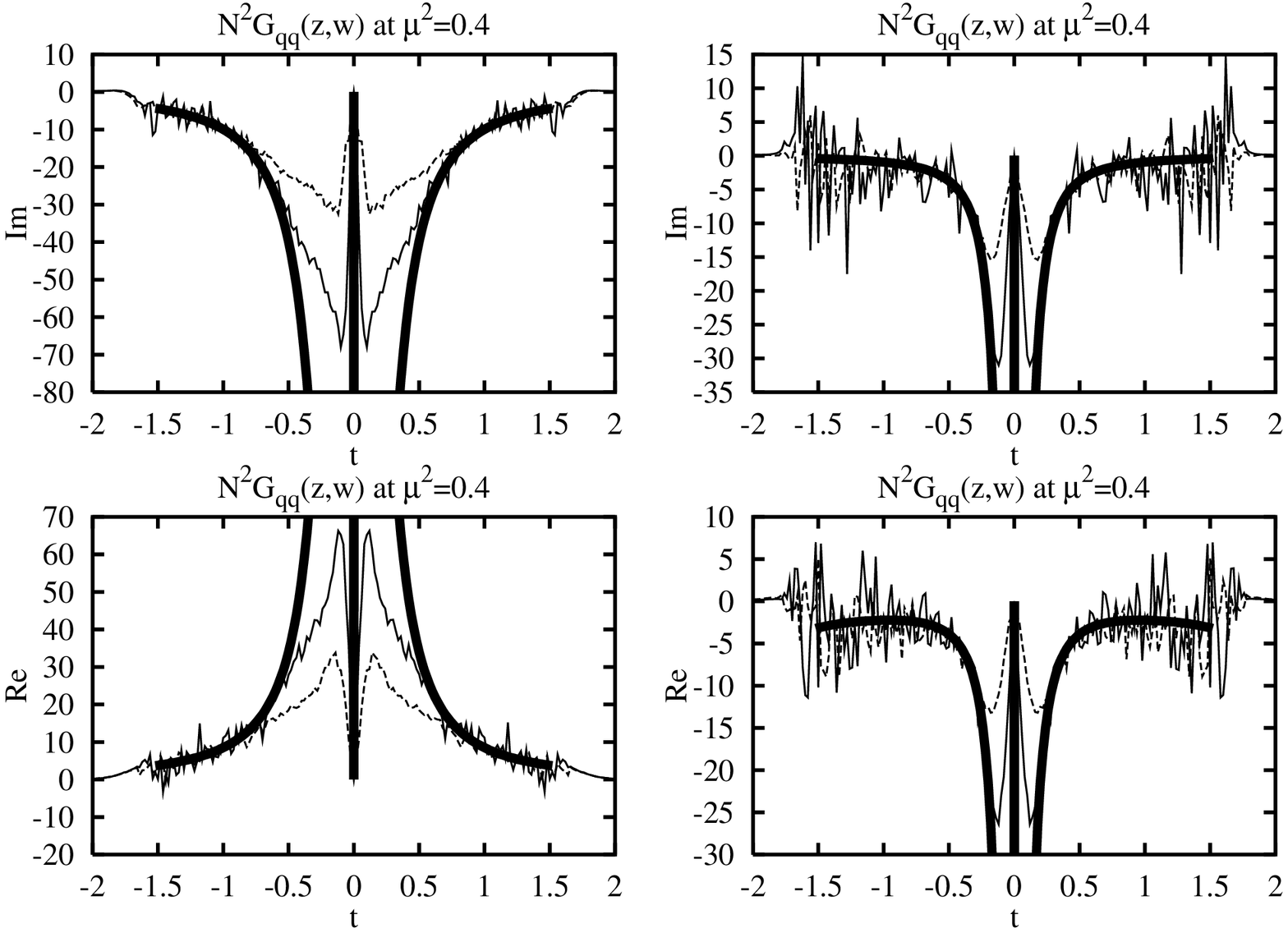}}
\vspace*{-5mm}
\caption{Numerical (thin and dashed lines) versus analytical (thick
solid lines) results for the chiral ``quark-quark" correlators II. The
correlators were evaluated along the paths shown in
Fig.~(\protect\ref{fig.stpath}): $z\!-\!w$=1--2 (left), 2--3 (right).} 
\label{fig.numchirII}
\end{figure}

\begin{figure}[htbp]
\centerline{\epsfysize=5.5truecm \epsfbox{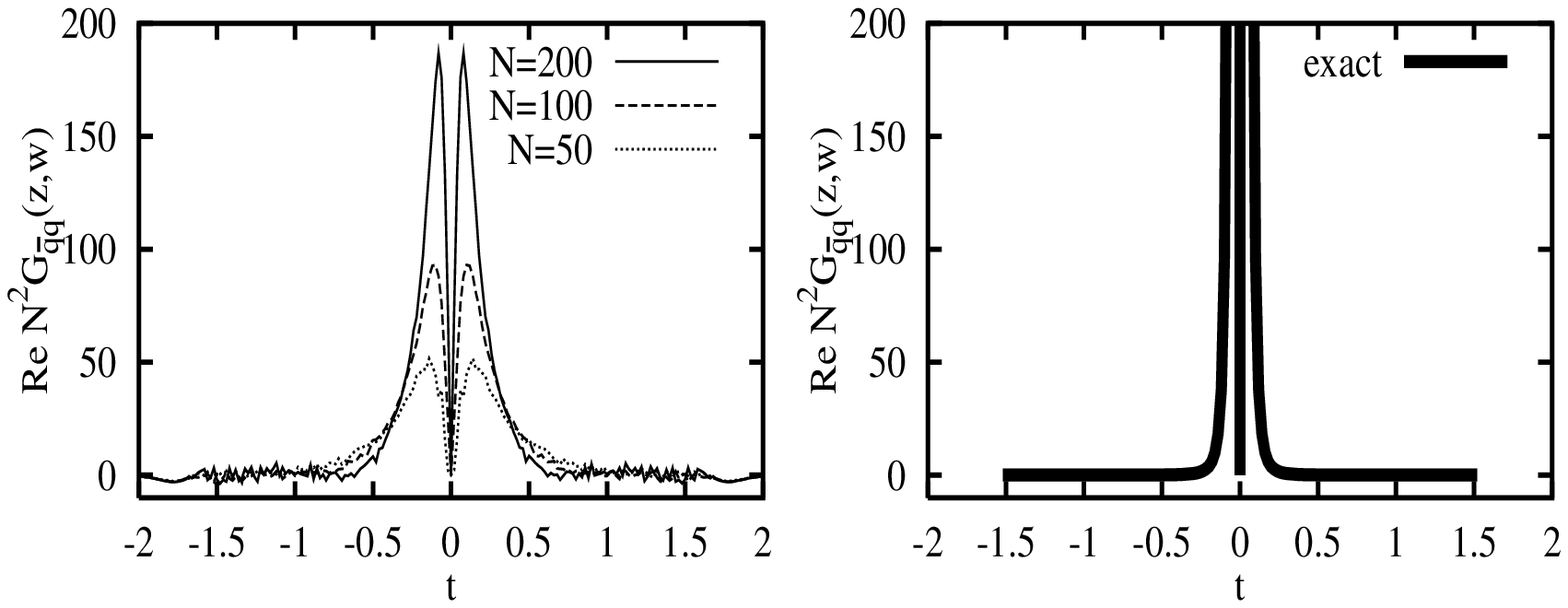}}
\caption{Same as Fig.~\protect\ref{fig.numchirII} for
``quark-conjugate-quark" chiral correlator II.}
\label{fig.stbc}
\end{figure}

The resolvents in (\ref{67}) follow from the solutions presented
in section \ref{s.chiral}. Defining $z=x+iy, w=u+iv$, we have
\eqn
\GG_{qq} &=& (\zb+\mu\gamma-\Sg_4)\cdot \Delta^{-1}=
    \arr{\Sg_1}{(\zb-\Sg_4){iy}/{\mu}-\mu}{-(\zb-\Sg_4){iy}/{\mu}+\mu}
     {\Sg_1} \,, \nonumber\\
\GG_{\qb\qb} &=& (z-\mu\gamma-\Sg_1)\cdot \Delta^{-1}=
    \arr{\Sg_4}{(z-\Sg_1){iy}/{\mu}+\mu}{-(z-\Sg_1 ){iy}/{\mu}-\mu}
     {\Sg_4} \,,\nonumber\\
\GG_{q\qb} &=& R\cdot \Delta^{-1}= R\arr{1}{{iy}/{\mu}}{-{iy}/{\mu}}{1}
\label{70}
\eqnx
where
\eq
(z-\Sg_1)(\zb-\Sg_4)+\mu^2-R^2=det=\f{-\mu^2}{y^2-\mu^2}
\label{71}
\eqx
and 
\eqn
\Sg_1 &=& \f{x}{2}+iy+\f{1}{2}\f{iy}{y^2-\mu^2} \,,\\
\Sg_4 &=& \f{x}{2}-iy-\f{1}{2}\f{iy}{y^2-\mu^2} \,.
\label{72}
\eqnx

\noindent
Using (\ref{70}-\ref{72}), the determinant of (\ref{67}) viewed as
a $16\times 16$ matrix simplifies to
\be
\hspace*{-8mm}
\det(1\!-\!\GG(z)\otimes\GG^T(w)\Gamma) = 
	|z\!-\!w|^2|z\!+\!w|^2
	\f{(\mu^2\!-\!(\mu^2\!-\!y^2)(\mu^2\!-\!v^2))^2\!-\!v^2y^2}{\mu^4}
	\,. 
\ee

We have evaluated the  correlator numerically inside the nonholomorphic
region. Figure~\ref{fig.stpath} shows the island for $\mu^2=0.4$, and
indicates the paths $z$ and $w$ taken during the numerical
evaluation of the correlator. In 
Fig.~\ref{fig.numchirII} the analytical results (\ref{67}) are compared to
the numerical simulations. 
The thin solid line is obtained from a numerical simulation of 73000
200$\times$200 matrices at $\mu^2=0.4$, the dashed line represents 75000
events of 100$\times$100 matrices, while the thick solid line is
deduced from the analytical prediction~(\protect\ref{67}). The numerical
results converge slowly to the analytical one. In Fig.~\ref{fig.stbc},
the thick solid line on the right corresponds to the analytical
prediction~(\protect\ref{67}), while the thin lines on the left are the
numerical simulation using 15000 50$\times$50 (dotted), 15000
100$\times$100 (dashed) and 10000 200$\times$200 matrices (solid
line). The height of the numerical peak is proportional to $N$,
diverging in the large $N$ limit. Its width is slowly approaching the 
analytical one.

\subsection{Scattering Correlator}

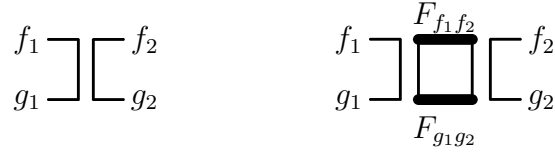
\begin{figure}[t]
\centerline{%
\parbox{10mm}{\raisebox{6mm}{%
  \begin{fmfgraph*}(10,8)
	\fmftop{f1,u1,u2,f2}
	\fmfbottom{g1,l1,l2,g2}
	\fmffreeze
	\fmfforce{(0,h)}{f1}
	\fmfforce{(0,0)}{g1}
	\fmfforce{(w,h)}{f2}
	\fmfforce{(w,0)}{g2}
	\fmfforce{(.4w,h)}{u1}
	\fmfforce{(.6w,h)}{u2}
	\fmfforce{(.4w,0)}{l1}
	\fmfforce{(.6w,0)}{l2}
	\fmf{plain}{f1,u1,l1,g1}
	\fmf{plain}{f2,u2,l2,g2}
	\fmfv{d.si=0,l=$f_1$,l.a=180,l.d=1mm}{f1}
	\fmfv{d.si=0,l=$f_2$,l.a=0,l.d=1mm}{f2}
	\fmfv{d.si=0,l=$g_1$,l.a=180,l.d=1mm}{g1}
	\fmfv{d.si=0,l=$g_2$,l.a=0,l.d=1mm}{g2}
  \end{fmfgraph*}
}}
\hspace*{30mm}
\parbox{20mm}{%
  \begin{fmfgraph*}(20,8)
	\fmftop{f1,u1,u2,u3,u4,f2}
	\fmfbottom{g1,l1,l2,l3,l4,g2}
	\fmffreeze
	\fmfforce{(0,h)}{f1}
	\fmfforce{(0,0)}{g1}
	\fmfforce{(w,h)}{f2}
	\fmfforce{(w,0)}{g2}
	\fmfforce{(.2w,h)}{u1}
	\fmfforce{(.32w,h)}{u2}
	\fmfforce{(.68w,h)}{u3}
	\fmfforce{(.8w,h)}{u4}
	\fmfforce{(.2w,0)}{l1}
	\fmfforce{(.32w,0)}{l2}
	\fmfforce{(.68w,0)}{l3}
	\fmfforce{(.8w,0)}{l4}
	\fmf{plain}{f1,u1,l1,g1}
	\fmf{plain}{f2,u4,l4,g2}
	\fmf{plain}{u2,l2}
	\fmf{plain}{u3,l3}
	\fmf{plain,wi=2thick,label=$F_{f_1f_2}$,l.side=left}{u2,u3}
	\fmf{plain,wi=2thick,label=$F_{g_1g_2}$,l.side=right}{l2,l3}
	\fmfv{d.si=0,l=$f_1$,l.a=180,l.d=1mm}{f1}
	\fmfv{d.si=0,l=$f_2$,l.a=0,l.d=1mm}{f2}
	\fmfv{d.si=0,l=$g_1$,l.a=180,l.d=1mm}{g1}
	\fmfv{d.si=0,l=$g_2$,l.a=0,l.d=1mm}{g2}
  \end{fmfgraph*} \vspace*{1mm}
}} 
\caption{a) Graph contributing to $\Gamma^{VV^{\dagger}}$ (left) and b)
graph contributing to $\Gamma^H$ (right).}
\label{fig.Gdif}
\end{figure}

Here we consider the correlator for the model discussed in section 3.4, that is
\eq
H-i\gamma VV^{\dagger}=H-\dl VV^{\dagger}
\eqx
where $H$ is random Gaussian Hermitean. The precedent discussion can be easily 
adapted to this case by assessing $\Gamma$ the ``quark-quark" scattering 
amplitude. In this case, the latter splits into a sum $\Gamma^H$ and 
$\Gamma^{VV^{\dagger}}$. The first one is just
\eq
\Gamma^H_{f_1f_2;g_1g_2}=\dl_{f_1 f_2}\dl_{g_1 g_2}
\eqx
where $f_1,f_2$ are the ingoing and respectively outgoing ``quark isospin"
on the upper line, while the $g_i$'s are the corresponding ones for the
lower lines (see Fig.~\ref{fig.Gdif}a). The graphs $\Gamma^{VV^{\dagger}}$ 
are considerably more involved (see Fig.~\ref{fig.Gdif}b) due to the fact 
that the $V$ vertices always appear in pairs. Therefore they always contain 
an exchange of two $F$ propagators between the ``quark" lines.
Note that the 1PI ``quark" scattering amplitude
$\Gamma^{VV^{\dagger}}$ flips ``quarks" into ``conjugate-quarks" and
vice-versa on the upper and the lower lines. With this in mind,
\eq
\Gamma^{VV^{\dagger}}_{f_1f_2;g_1g_2}=m\cdot F_{f_1f_2}\cdot F_{g_1 g_2}
\eqx
where the $F$'s are defined by Fig.~\ref{fig.LS} and the factor $m$ comes 
from the interior loop of Fig.~\ref{fig.Gdif}b. Hence
($z=x+iy, w=u+iv$)
\eqn
F_{qq}&=&\dl(1+\dl \br{G})/den\\
F_{\qb\qb}&=&-\dl(1-\dl G)/den \\
F_{q\qb}&=&-\dl^2 G_{q\qb}/den
\eqnx
where
\eq
den=\f{-\gm m}{y}
\eqx
and
\eq
G_{q\qb}=\sqrt{G\br{G}-\f{1}{1-\gm y}}
\eqx

In the present model the ``quark-quark" correlator is now
\eq
N^2 G_{qq}(z,w)=-\partial_z\partial_w \log\det(1-\GG(z)\otimes\GG^T(w)\cdot 
[\Gamma^H+\Gamma^{VV^{\dagger}}])
\label{corrrandom}
\eqx
where the dot stands for ordinary matrix multiplication. 
The determinant in (\ref{corrrandom}) may be explicitly evaluated to give
\eq
\det\,(\ldots)=|z-w|^2\f{(1-\gm y)(1-\gm v)(m+yv)-\gm^2 yv}{m(1-\gm y)^2(1-\gm v)^2}
\,.
\label{ddd}
\eqx
As in the case of the elliptic ensemble, the correlators differ
depending on whether the matrices $\Gamma$ and $V$ stem from GUE or GOE 
ensembles. In the last case, a new set of diagrams have to be added. They are 
leading in GOE and subleading in GUE. We have checked that the additional 
contribution is the same as (\ref{ddd}), resulting into an 
extra factor of 2 for the GOE correlators.

\begin{figure}[htbp]
\centerline{\epsfysize=5.5truecm \epsfbox{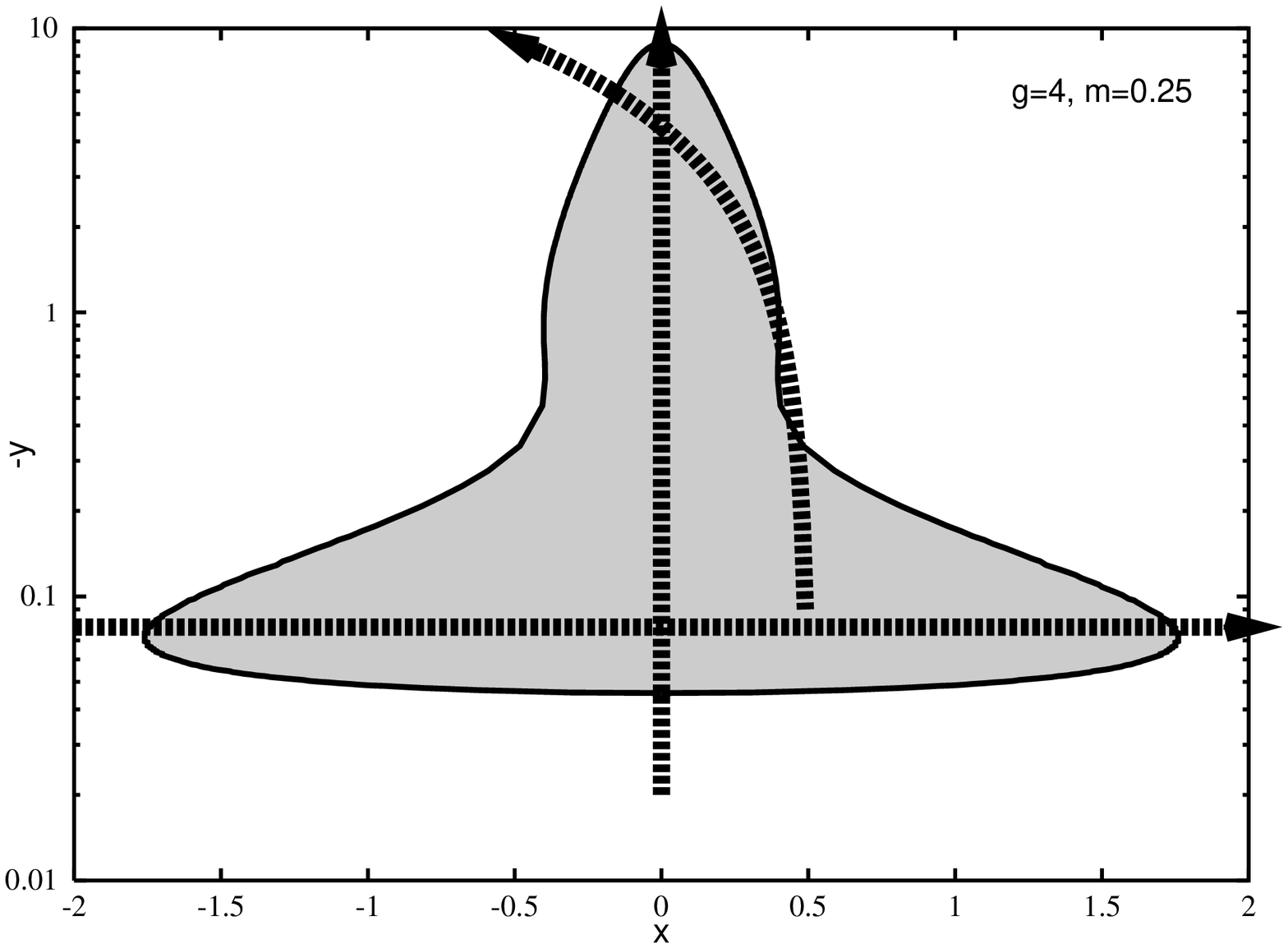}}
\caption{Extent of the nonholomorphic region (shadowed) for the scattering
problem at $g=4, m=0.25$. The thick dashed lines indicate the paths along
which the numerical correlator was evaluated (see below).}
\label{fig.hakpath}
\end{figure}

\begin{figure}[htbp]
\centerline{\epsfysize=5.5truecm \epsfbox{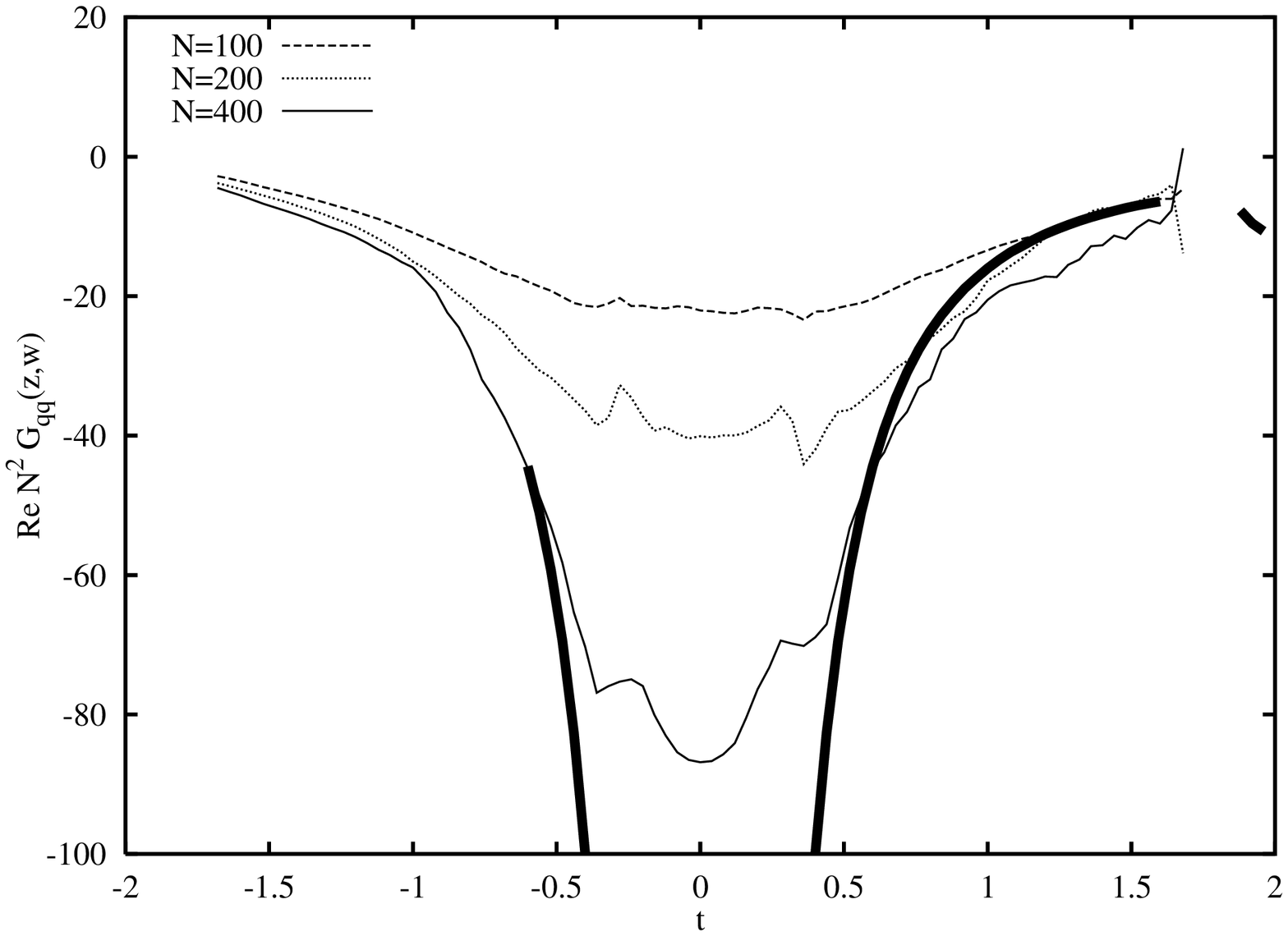}}
\caption{Numerical versus analytical (thick solid lines) 
for the ``quark-quark" correlator of the random scattering problem.}
\label{fig.numrandom}
\end{figure}  

\begin{figure}[htbp]
\centerline{\epsfysize=5.5truecm \epsfbox{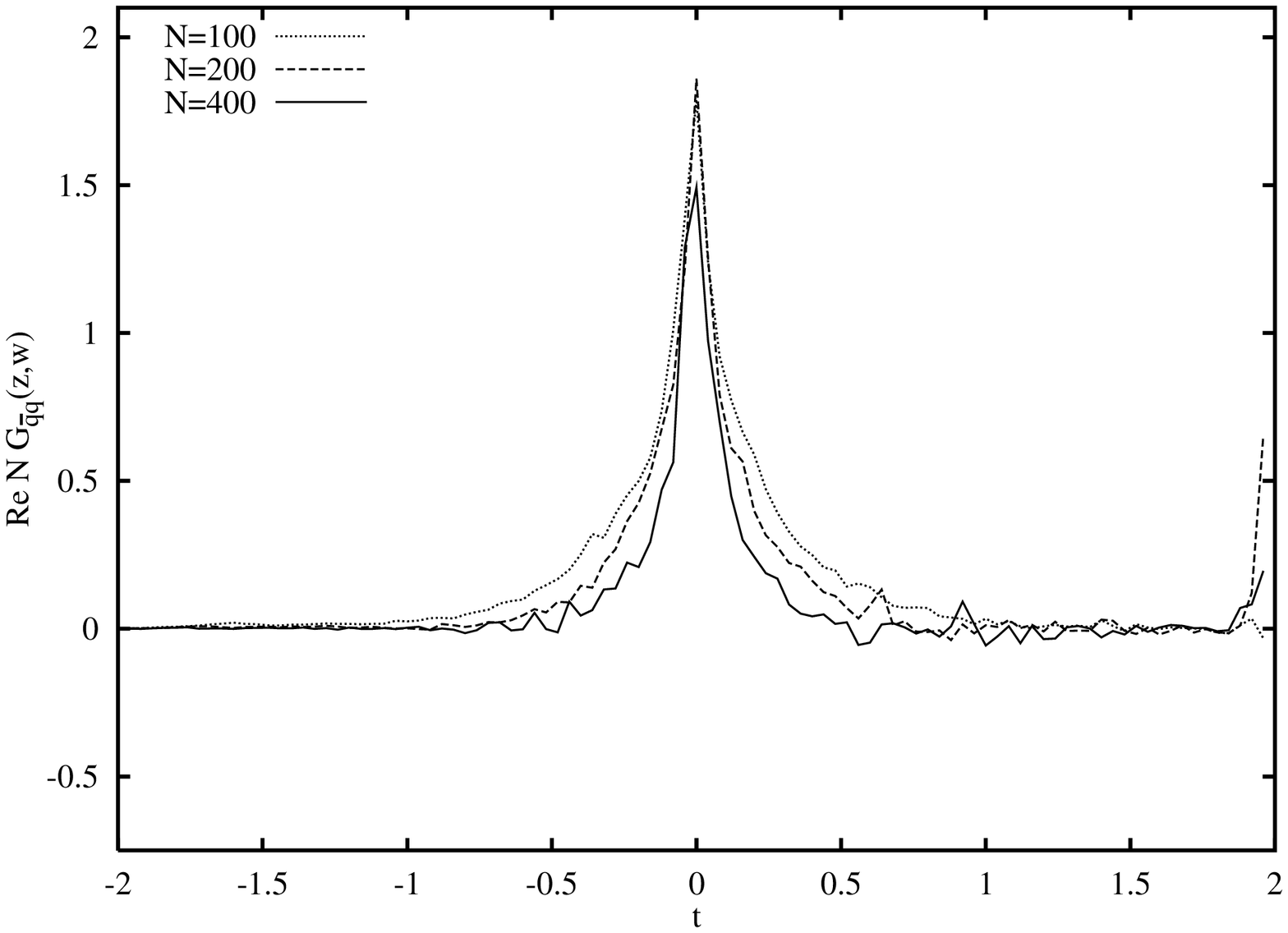}}
\caption{Same as Fig.~\protect\ref{fig.numrandom} but 
for the ``quark-conjugate-quark" correlation function divided by $N$. The analytical
result~(\protect\ref{corrrandom}) is not distinguishable from zero on
the plot (see 
text).}
\label{fig.hkcqqb}
\end{figure}

Figure~\ref{fig.numrandom} compares the numerical versus analytical 
results for random scattering correlator using GUE ensembles along the
paths indicated on Fig.~\ref{fig.hakpath}. The thick solid line is
the analytical result~(\protect\ref{corrrandom}), while the thin lines
follow from numerical simulation of 10000 100$\times$100
(dashed), 3000 200$\times$200 (dotted) and 2000 400$\times$400 matrices
(solid) at $\gamma=4$, $m=0.25$ along the trajectory $z=0.777 t
(2+0.2865 i)$, $w=0.2865 i t$. Part of the path goes outside the
non-holomorphic region as shown in Fig.~\ref{fig.hakpath}
 and the analytical curve 
is missing for that values. The convergence of the numerical results to 
the analytical one is slow. The numerical results for the ``quark-conjugate-quark" 
correlators are shown in Fig.~\ref{fig.hkcqqb}. The analytical result
(\ref{corrrandom}) is not distinguishable from zero  in the region
shown.
 The large central
peak is a finite size effect, as it starts to shrink for $N\sim 400$.

\section{Partition Functions}

\newcommand{\sg}{\sigma}
\newcommand{\al}{\alpha}
\newcommand{\PP}{{\cal P}}
\newcommand{\dz}{\partial_z}
\newcommand{\dzb}{\partial_{\zb}}
\renewcommand{\gg}{\GG\otimes \GG^T \cdot \Gamma}
\newcommand{\ggb}{\GG\otimes \br{\GG}^T \cdot \Gamma}
\newcommand{\gbgb}{\br{\GG}\otimes \br{\GG}^T \cdot \Gamma}
\newcommand{\ggg}{\det(1-\GG\otimes \br{\GG}^T \cdot \Gamma)}

{}From a ``thermodynamical" point of view, the information carried 
by the one- and two-point functions is sufficient to
specify the ``thermodynamical"  potential to order ${\cal O}(1/N)$ in the
entire $z$-plane modulo isolated singularities, as we now discuss. 
Similar ideas were used either in the 
context of two-dimensional models in matter \cite{STEELE}, or 
as a way to calculate next-to-leading corrections to hermitean
correlators~\cite{ITOI}, exploiting the functional formalism
discussed in~\cite{AMBJORNETAL}.

\begin{figure}[htbp]
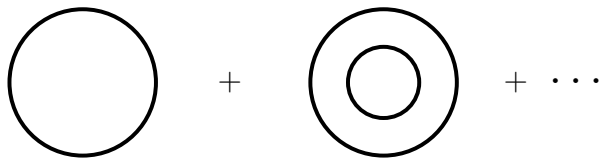

\centerline{%
\psset{unit=1truecm}
\pspicture(0,0)(8,2)
  \pscircle[linewidth=1.5pt](1,1){1}
  \uput{0}[0](2.8,1){+}
  \pscircle[linewidth=1.5pt](5,1){1}
  \pscircle[linewidth=1.5pt](5,1){0.5}
  \uput{0}[0](6.6,1){+}
  \uput{0}[0](7.2,1){\Large $\cdots$}
\endpspicture
}
\caption{Leading contributions to $Z_N$ in (\ref{X106}).}
\label{fig.E2}
\end{figure}

Let $Z_N$ be the partition function in the presence of an external parameter
$z$. In the $1/N$ approximation, the diagrammatic contributions to the
 partition function $Z_N$ are shown in Fig.~\ref{fig.E2}, corresponding to
\be
\log\, Z_N = NE_0 + E_1 + {\cal O} (\frac 1N )
\label{YL1}
\ee
$E_0$ is the contribution of the ``quark" or ``conjugate quark" loop
in the planar approximation, and $E_1$ is contribution of the 
``quark-quark" loop, and so on, in the same approximation. Throughout
this section, 
we will restrict our attention to nonhermitean matrices with unitary 
randomness, in which case the non-planar corrections to $E_0$ are
of order $1/N^2$.\footnote{A similar observation was made
in~\protect\cite{ITOI},
while evaluating the correction to the resolvent for hermitean GUE ensembles.}
Hence, $E_0$ is determined by the one-point function and $E_1$ by the 
two-point functions. 

For $z$ such that  (\ref{YL1}) is real, continuous and 
nondecreasing function of the extensive parameters \cite{HUANG}, 
$\log\, Z_N/N$ may be identified with the ``pressure" of the random matrix 
model. As a
result, the isolated singularities in the ``pressure" are just the ``phase"
boundaries provided that the expansion is uniform. Below we give 
examples where the ``phase" boundary is either mean-field-driven or 
fluctuation-driven.

\subsection{Holomorphic $Z$}

To illustrate the above arguments, consider the partition function
\eq
Z_N =\corr{\det (z-\MM) }=\left<\int d\psi d\psib 
e^{-\psib(z-\MM)\psi}\right>
\label{X106}
\eqx
In contrast to the one- and two-point correlators discussed above, the 
determinant in (\ref{X106}) is not singular in the $z$-plane configuration 
by configuration. Hence, (\ref{X106}) is {\it a priori} holomorphic in 
$z$ (minus isolated singularities).

The one- and two-point functions on their holomorphic support may be
obtained from  
$\mbox{log}\,Z_N$ by differentiation with respect to $z$. Therefore, from 
(\ref{YL1})
\eq
E_0=\int^z\! dz'\ G(z') + const
\label{eq.E0}
\eqx
or equivalently
\be
E_0 =zG - \int\! dG\ z(G) + const
\label{trick}
\ee
after integrating by part. Note that $z(G) =B(G)$ is just the Blue's 
function~\cite{ZEE} of $G$ (where the Blue's function is the functional inverse
of the resolvent, i.e. $B[G(z)]=z$). The constant in $E_0$ is fixed by the 
asymptotic behavior of (\ref{X106}), that is $Z_N\sim z^N$. 
The planar contribution to $E_1$ in (\ref{YL1}) follows from the
``quark-quark" wheel  of Fig.~\ref{fig.E2}, that is
\eq
\quad E_1 = -\frac 12 \mbox{log}\, \det(1-\gg)
\label{e.e1l}
\eqx
which can be understood by 
opening one of the lines in the two-point correlator (\ref{logxx}):
 the factor $1/2$ is
combinatorial (two identical ``quarks") and  the minus sign corresponds 
to decreasing the number of fermion loops by one. The derivative amputates
the left out propagator after ``opening'' the fermionic loop. We note that the
factor $-1/2$ in (\ref{e.e1l}) causes the ``quark-quark" contribution to be 
overall ``bosonic"
\footnote{A similar result can be obtained using a bozonized 
version to (\ref{X106}) in the Gaussian approximation.}. Hence
\eq
Z_N =e^{N E_0}\cdot \left(\left\{ \det(1-\gg)\right\}^{-\frac 12} + 
{\cal O} \left( \frac 1N \right) \right)
\label{X107}
\eqx
In contrast to the ``quark" contribution, the ``bosonic" contribution is not 
extensive in $N$, since in random matrix models $N$ is like ``color" not 
``space". The ``bosonic" contribution is dwarfed by the ``quark" contribution
($1:N$) \cite{USNJL}.
Both $E_0$ and $E_1$ are simple functions of the resolvent on a specific
branch as we will show in the examples below. 
When the ``quark-quark" contribution (pre-exponent) to (\ref{X107})
does not vanish in the $z$-plane, the
singularities of $Z_N$ are those of the ``quark" contribution in $E_0$ in large 
$N$. Such singularities are mean-field-driven. Otherwise, they are 
fluctuation-driven. 

We note that the partition function 
$Z_N$ through (\ref{X107}) exhibits an essential singularity in $1/N$ as 
expected from thermodynamical arguments. Assuming that the expansion 
for $\log Z_N /N$ is uniform, then $\log Z_N/N$ follows from (\ref{X107}) using
the holomorphic resolvent $G(z)$ for large $z$. The small $z$ region
follows by 
analytical continuation. However, since $G(z)$ is multi-valued (already the
simple case of the Ginibre-Girko ensemble yields two branches for the
resolvent in~(\ref{31})), the analytical
continuation is ambiguous. The ambiguity may be removed by identifying
$\log |Z_N|/N$ with some {\it generalized}
``pressure" and taking $G(z)$ so that $\log |Z_N|/N$ is maximum 
\footnote{This condition is equivalent to the saddle point condition.
It does not necessarily fulfill the conditions discussed in \cite{HUANG}.}.  
As a result, $\log \, Z_N/N$ is piece-wise 
analytic in leading order in $1/N$ with ``cusps" at
\eq
F^{(ij)}(x,y) \equiv V_N^{(i)}(x,y) - V_N^{(j)}(x,y) = 0\,,
\label{CUSPS}
\eqx
following the transition from branch $i$ to branch $j$ of $G$
with $V_N=\log |Z_N|/N$. 

The character of the transition in the $1/N$ approximation
can be highlighted by noting that
for any finite $N$, the partition function (\ref{X106}) 
is a complex polynomial in $z$ of degree $N$ with random coefficients. 
In large $N$,
\eq
V_N=\frac 1N \log\, |Z_N| = \frac 12 \int \, dv\,d\overline{v}\,\varrho (v, 
\overline{v})\, \log |z-v|^2 \,.
\label{INTEGRAL}
\eqx
To leading order, the distribution of 
singularities along the ``cusps" (\ref{CUSPS}) is 
\be
\varrho (z, \zb ) = \frac 1{2\pi} | \dz F |^2 \,\, \delta ( F (z, \zb) )
\label{densdel}
\ee
which is normalized to 1 in the $z$-plane. Redefining the density of 
singularities by unit length along the curve $F(z,\zb)=0$, we may rewrite
(~\ref{densdel}) as
\be
\left. \varrho(z,\zb)\right|_{F=0} = \frac 1{2\pi} |\dz F | \equiv
	\frac 1{4\pi} |G^{(i)}-G^{(j)}| \,.
\label{densun}
\ee
For $\varrho \neq 0$, the integrand
in (\ref{INTEGRAL}) is singular at $z=v$ which results into different forms
for $V_N$, hence a cusp. For $\varrho =0$, that is $\dz F =0$,
$V_N$ is differentiable. For physical 
$V_N$ (real and monotonically increasing), the points $\varrho =0$ are 
multi-critical points. At these points all $n$-point ($n\geq 2$)
functions diverge.
This observation also holds for Ising models with complex external
parameters \cite{ROBERT}. Assuming
macroscopic universality~\cite{AMBJORN} for all $n$-points ($n\geq 2$), 
we conclude that $\partial_z F =0$ means a branching point for the
resolvents, hence $\dz G =\infty$ 
or $B' (G) =0$ \cite{ZEE}. 
 For hermitean matrices, these conditions coincide with 
the end-points of the eigenvalue distributions \cite{ZEE,USBLUE}.

\renewcommand{\dz}{\partial_z}
\renewcommand{\dzb}{\partial_{\zb}}
\newcommand{\ddx}{\partial_x}
\newcommand{\ddy}{\partial_y}

\vskip .5cm
$\bullet$ \,\,{\it Example}
\vskip .25cm

To illustrate the above concepts, consider the Ginibre-Girko ensemble of 
section~4.1. The resolvent in the holomorphic region satisfies (\ref{31}), so 
\be
z = \tau G+\frac{1}{G}\, .
\label{defgir}
\ee
The integration~(\ref{trick}) in $E_0$ is straightforward, and after
fixing the asymptotic behavior we obtain 
\be
Z_N  = G^{-N} e^{\frac{\tau}{2}N G^2} \left( 
(1-G^2(z)\tau)^{-\frac{1}{2}}
 + {\cal O} \left(\frac 1N\right) \right)\,.
\label{girhol}
\ee
Here $G$ is the solution of (\ref{defgir}). The pre-exponent in (\ref{girhol})
follows from (\ref{X107}) with the matrix $\GG$ replaced 
by $G$ and $\Gamma=\tau$, as seen in the 
``quark-quark" component of the vertex matrix in (\ref{24}).
Using (\ref{defgir}) we observe that the pre-exponent diverges at two 
points in the $z$-plane, $z^2 =4\tau$. At these points there is a ``phase" 
change as we now show. 

Given (\ref{girhol}), the generalized ``pressure" in leading order is
\be
V_\pm = -\frac{1}{2}\mbox{log\,}(G_\pm\overline{G}_\pm) + \frac{\tau}{4}
	(G_\pm^2+\overline{G}_\pm^2) +{\cal O} \left( \frac 1N\right)\,.
\ee
$V_\pm$ define two intersecting surfaces valued in the $z$-plane, 
for two branches $G_{\pm}$ of the solutions to (\ref{defgir}). 
The parametric equation for the intersecting curve is  
\be
F (z , \zb ) = V_+ -V_- =0
\label{CUTX}
\ee
As indicated above, $V_N$ is piece-wise differentiable.
Note that $F=0$ on the cut along the real axis, 
$-2\sqrt{\tau}<z<+2\sqrt{\tau}$, and from~(\ref{densun}) the density of
singularities per unit length is
\eq
\left. \varrho (z,\zb)\right|_{F=0} = \frac 1\pi 
	\frac{\sqrt{4\tau-z^2}}{2\tau} \,. 
\label{charge1}
\eqx
Along $F$, the density of singularities is semi-circle. The density
(\ref{charge1}) vanishes at the end-points $z=\pm 2\sqrt{\tau}$. This is
easily seen to be the same as $\dz G = \infty$, or $dB(G)/dG =0$ with 
$B (G) =\tau G + 1/ G$.  As noted above the term in bracket in
Eq.~(\ref{girhol}) vanishes at these points, with a diverging
``quark-quark" contribution. The transition is fluctuation-driven. 
These points may signal the onset of scaling regions with possible universal
microscopic behavior for nonhermitean random matrix models. This issue
will be pursued elsewhere. At these points the $1/N$ expansion we have used 
breaks down. 

Note that the case of the circular ensemble with $\tau=0$ is very specific. 
{}From our analysis, we see that the support of $F$
shrinks to a single point $z=0$. In fact $Z_N=z^N$ to
all orders in $1/N$. This can be understood by noting that since in
\eq
Z_N=\corr{\det(z-\MM)}
\eqx
the determinant is just
\eq
\det(z-\MM)=\varepsilon_{i_1 i_2\ldots i_N} (z-\MM)^1_{i_1}
(z-\MM)^2_{i_2} \ldots (z-\MM)^N_{i_N}
\eqx
{\em all} expectation values vanish (exactly to all orders in $1/N$)
for random gaussian complex matrices, {\it i.e.}
\eq
\corr{\MM^a_b \MM^c_d \ldots \MM^x_y}=0 \,.
\eqx
Hence our claim. 

The present construction for the partition function for the
Ginibre-Girko ensemble extends to the other cases of nonhermitean ensembles
with unitary randomness. We briefly mention the case of chiral 
random ensembles  plus deterministic nonhermitean matrix of
(off-diagonal) strength $\zeta$. 
The relevant one- and two-point correlators in the holomorphic and
nonholomorphic domains were explicitly constructed in sections
4.3, 5.4 and 5.5. 
Using these results  and in analogy with (\ref{X107}), 
elementary integration for this case leads to 
\eq
Z_N (z, \zeta) =e^{N E_0}\cdot \left(\left\{ D^{-2} [(D+\zeta^2)^2 -(z-G)^4]
\right\}^{-\frac 12} + 
{\cal O} \left( \frac 1N \right) \right)\,,
\label{X107ch}
\eqx
where $D=(z-G)^2/G^2$, and
\eq
E_0 (z, \zeta ) = G^2 +\log \f{z-G}{G}
\eqx
with the appropriate branch of holomorphic $G$ solution to 
(\ref{cubicpasturmux}),
with 
$\mu=\zeta$.
Note that for $z=0$ and $G^2=-1-\zeta^2$, the pre-exponent in (\ref{X107ch})
diverges. It also diverges at $z=z_*$ which are the zeros of
(\ref{densun}) for the present case (two zeros for small $\zeta$ and
four zeros for large $\zeta$). Again, at these points, the $1/N$ expansion 
breaks down marking  the onset of scaling regions and the possibility of 
microscopic universality. The $z=0$ divergence is just the notorious 
``Goldstone" mode in chiral models, illustrating the noncommutativity of 
$N\rightarrow \infty$ and $z\rightarrow 0$. The rest of the arguments 
follow easily from the preceding example, in agreement with the 
``thermodynamics" discussed in~\cite{USNJL}.

\vskip .5cm
$\bullet$\,\,\,{\it Replicas}
\vskip .25cm
We note that in the circular case, the partition function with $N_f$
``replicas"  
\eq
Z_{N,N_f} = \left<{\det}^{N_f} (z- \MM )\right>
\label{Z1NF}
\eqx
yields simply 
\eq
Z_{N,N_f} =z^{N \,N_f}
\label{ZNF}
\eqx
For general $\MM$ and 
$N_f$ noninteger the determinant in (\ref{Z1NF}) is multivalued, and a 
determination is in general needed. Here, it will be assumed by 
{\it interpolating} the integer result \cite{USNJL} \footnote
{Another way is to use the $\xi$-function in combination with the
heat-kernel construction, through $det^{N_f} A = e^{-N_f \xi_A' (0)}$ 
and $\xi_A (s) = {\rm Tr} (1/A^s)$. We have checked that the leading term
is in agreement with the interpolation for smooth operators.}. While the 
limits $N\rightarrow\infty$ and $N_f\rightarrow 0$ are compatible for 
$\ln Z_{N, N_f}/ N$ through the diagrammatic analysis, it is not necessarily the
case for the connected functions ($z$-derivatives) since singularities for
$z\sim \MM$ are generated.

In the $1/N$ expansion, the ``quark" insertions are suppressed by 
powers of $N_f/N$ for singularity-free operators. In general, this
implies that random matrix models are self-quenched \cite{USNJL}, and 
holomorphy in the external parameter $z$ is preserved as $N\rightarrow \infty$
(modulo cusps). In the presence of singularities, 
holomorphic symmetry is usually upset as $N\rightarrow\infty$ for a certain 
range of $N_f$ and $z\sim \MM$, and arguments similar to those of section 3
should be used. This confirms the shortcomings of 
(\ref{Z1NF}) to yield the quenched one-point functions \cite{STEPHANOV}, 
and two-point functions \cite{USMUX} for small $z$.

\subsection{Nonholomorphic $Z$}

The above analysis for the holomorphic thermodynamical potential may also be 
extended to nonholomorphic partition functions, again for Gaussian unitary
randomness. For example,
\eq
Z_N [z, \zb ] =\corr{\det |z-\MM|^2}=\left<\int d\psi d\psib d\phi d\overline{\phi}
e^{-\psib(z-\MM)\psi -\phib(\zb-\MM^{\dagger})\phi}\right>
\label{X100}
\eqx
which is the random matrix analogue of the partition function discussed 
in \cite{GOCKSCH} for unquenched QCD. In these partition functions the phase
of the ($N_f=2$) determinant is set to one. Again, the one- and two-point
functions may be  
obtained from $\log \,Z_N$ by differentiation with respect to $z$ and $\zb$ 
(modulo delta functions).

{}From (\ref{YL1}) we 
have
\be
G(z, \zb)= \frac{\partial E_0(z,\zb)}{\partial z} \qquad
\overline{G}(z, \zb)= \frac{\partial E_0(z,\zb)}{\partial \zb} \qquad
\ee
These expressions may be integrated
and the integration constant is set by the matching
condition
\eq
E_0 \sim \log |z|^2 \quad \quad\quad\quad \mbox{\rm for } z\to \infty \,.
\eqx
For the Ginibre-Girko ensemble the nonholomorphic resolvent~(\ref{25})
yields
\be
E_0(z, \zb) = \frac 1{1-\tau^2} \left[ |z|^2-\frac 12 \tau (z^2+\zb^2) 
	\right] + c_{nh}
\ee
while the holomorphic resolvent~(\ref{31}) gives
\be
E_0(z, \zb) = \frac z{4\tau} \left( z\mp \sqrt{z^2-4\tau} \right)
	\pm \mbox{log} \left[ \frac 12 (z+\sqrt{z^2-4\tau}) \right] + h.c. + c_h
\ee
with the limit
\be
\lim_{z\to\infty} E_0(z, \zb) =
	1+\mbox{log}\ |z|^2 +c_h \,. 
\ee
The constant $c_h$ is $-1$, and $c_{nh}$ is determined from the continuity
condition of $E_0$ at the boundary~(\ref{34}). At $\tau=0$
\be
E_0 = \left\{ \begin{array}{r@{\quad:\quad}l}
	\mbox{log}\ |z|^2 & |z|^2 > 1 \\
	|z|^2-1 & |z|^2 <1
	\end{array} \right. \,.
\label{FREEN}
\ee
The ``phase'' change at $|z|=1$ is fluctuation-driven.

\noindent
The expression for the correlator in the holomorphic phase follows by
differentiating the term $E_1$, that is
\eq
\dz \dzb E_1= \corr{\tr\left|\f{1}{z-\MM}\right|^2}_c
\eqx
Hence
\eq
E_1=-\ln \ggg+ \Phi
\eqx
where $\Phi$ is a harmonic function. To determine $\Phi$ we note that
\eq
\dz E_1 = G_1(z)\, ,
\label{essential}
\eqx
$G_1(z)$ is the $1/N$ correction to the one-point function in the planar 
approximation. It follows from the insertion of ``quark" or ``conjugate-quark" 
loops in $G(z)$. Specifically
\eq
G_1\!\!=\!\!-\!\tr[\f{1}{1\!\!-\!\ggb}\dz (\ggb)
	\!+\!\f{1}{2}\f{1}{1\!\!-\!\gg}\dz (\gg)] ,
\eqx
which can be rewritten as
\eq
\dz \left\{ \ln \det(1-\ggb)+\f{1}{2}\ln \det(1-\gg) \right\} \,.
\eqx
A similar expression follows from $\br{G}$. Hence
\eqn
E_1&=&-\frac 12 \log \det(1-\ggb) -\frac 12 \log \det(1-\ggb)\nonumber\\&&
      -\f{1}{2} \log \det(1\!-\!\gg)-\frac 12 \log \det(1\!-\!\gbgb) \,.
\label{dwakol}
\eqnx
As a result, the partition function (\ref{X100}) reads
\eq
Z_N [z, \zb ]\!\!=\!\!e^{N E_0}\cdot\left(\!\left\{ 
	\det(1\!\!-\!\ggb)\, |\!\det(1\!\!-\!\gg)| \right\}^{-1}
	\!\!\!+\!{\cal O}\!\left(\!\frac 1N\!\right)\!\right)\,.
\label{X101}
\eqx
Note that the contributions from the wheel-diagrams
are of the form $1/\sqrt{{\rm det}}$ and hence ``bosonic" in character.
 In (\ref{dwakol}), there are two 
contributions from the ``quark-conjugate-quark" wheels 
(upper line in (\ref{dwakol})), one contribution from the ``quark-quark"  wheel
 and one contribution from the ``conjugate-quark-conjugate-quark" wheel, 
 thereby explaining all contributions in (\ref{X101}) at next
to leading order. Like in the holomorphic case, again
similar results can be obtained using a bozonized 
version to (\ref{X100}) in the Gaussian approximation.

Again, the partition function $Z_N$ has an essential singularity in
 $1/N$, but $\log Z_N/N$ does not. For any finite $N$, the latter 
diverges for 
\eq
\det (1-\ggb) =0
\label{LINESING}
\eqx
The divergence condition from the second pre-exponent in (\ref{X101}),
{\it i.e.} 
\eq
\det (1-\gg) =0
\label{DISPOINTS}
\eqx
is not essential, since
the line of singularities following from (\ref{LINESING}) defines the
boundary of a surface that in general contains the cusps at
$\partial_z G =\infty$. In particular, the set of discrete
points (\ref{DISPOINTS}).  In this case, the transition in the $z$-plane is 
fluctuation-driven. For the Ginibre-Girko example, the nonholomorphic partition
function diverges according to (\ref{LINESING}), that is for
\eq
1-|G|^2 =0
\label{ellipse}
\eqx
Indeed in the holomorphic region, we could replace $\GG$ by $G$, 
the modulus follows from the ``conjugate-quark" resolvent, and the vertex 
corresponds to the ``quark-conjugate-quark" element of the kernel
(\ref{24}) set to  1. The line of singularities
(\ref{ellipse}) reproduces in this case the ellipse
(\ref{34}). The ellipse includes the points of the ``phase"
change, i.e. the focal points $z^2=4\tau$, connected by the interval
(\ref{CUTX}), {\it i.e.} $F=0$. We recall that these 
focal points were obtained as a divergence condition for (\ref{girhol}), 
a special case of the general condition (\ref{DISPOINTS}).     

The same results were also discussed in~\cite{USMUX} for the case of
a chiral random matrix model with a chemical potential $\mu=\zeta$
in the quenched approximation. In  particular, it was shown
that the condition (\ref{LINESING}) exactly reproduces the islands of
``mixed-condensate'', obtained using the replica methods~\cite{STEPHANOV}. 
In this case the condition (\ref{LINESING}) reads~\cite{USMUX}
\eq
 D^{-2} [(D-\zeta^2)^2 -|z-G|^4] =0
\label{X107cch}
\eqx
with $D=|(z-G)/G|^2$, therefore exactly the condition (\ref{e.ch1z}). 
Comparing to the ``quark-quark'' correlator in (\ref{X107ch}), 
we notice the appearance of the modulus and the flip in the sign of
$\zeta$ due to the fact that now we measure
the correlation between the ``quark'' and the ``conjugate-quark''.
Equation (\ref{X107cch}) defines the boundaries of the fore mentioned
``islands''. Inside (\ref{X107cch}), the resolvents in (\ref{X101})
are nonholomorphic, while outside (\ref{X107cch}) they are holomorphic. 
The holomorphic and nonholomorphic resolvents match on (\ref{X107cch}).

\section{Conclusions}

We have shown that chiral and hermitean random matrix models are amenable
to diagrammatic techniques in the context of the $1/N$ approximation. Since 
the matrices are hermitean, the eigenvalue distributions and their macroscopic 
correlations follow from one- and two-point functions that are analytic in the
complex $z$-plane modulo isolated singularities. Although our results were 
carried using ``quenched" measures, we expect them to hold for
``unquenched" ones as well  since the ``quark" effects are $1/N$ suppressed
\cite{USNJL}.

We have shown how to extend the diagrammatic arguments of Br\'{e}zin,
Hikami and Zee to a large variety
of nonhermitean random matrix models. For nonhermitean matrices, the one- and
two-point correlation functions are analytic in the complex $z$-plane modulo
surface singularities. These singularities originate from an accumulation of 
eigenvalues in the large $N$ limit, and cause the one- and two-point functions 
to be nonholomorphic in finite domains. The breakdown condition is set by the 
divergence of the correlations between pairs of eigenvalues, conjugate of each 
other and a distance $N$ apart in the spectrum \cite{USMUX}. 
We have explicitly 
constructed the one- and two-point functions in the planar approximation and 
shown that the analytical results are reproduced by direct numerical 
calculations using large samples of random and nonhermitean matrices. 

We have explicitly shown how the partition functions of nonhermitean
matrix models with unitary randomness
can be related to the one- and two-point functions for holomorphic 
as well as nonholomorphic kernels to order ${\cal O} (1/N)$, thereby providing
a simple physical interpretation to several preceding results.
The partition functions to the order calculated are solely 
a function of the resolvent $G$. We expect this property to hold to higher 
orders, and even suspect a generalized form of macroscopic universality when 
extended to non-Gaussian weights \cite{AMBJORN}. Generic equations for the 
singularities of the partition functions were found both for the holomorphic 
as well as the nonholomorphic cases to order ${\cal O} (1/N)$. 
These singularities are related to 
structural changes in the random matrix model, such as changes in the 
eigenvalue distributions and correlations in  models with
\cite{STEPHANOV,USMUX} or without dissipation \cite{USBLUE,USNJL}. 
Our construction may be useful for investigating a number of 
related issues in random matrix and polynomial models \cite{BOHIGAS}.

\ack
This work was partially supported by the US DOE grant DE-FG-88ER40388,
by the Polish Government Project (KBN) grant 2P03004412
and by the Hungarian Research Foundation OTKA grants T022931 and
F019689. The authors would like to thank Jurek Jurkiewicz for discussions.
I.Z. thanks Robert Schrock for a discussion. G.P. thanks 
the Soros Foundation for financial support and the Nuclear
Theory Group at Stony Brook for hospitality.
R.A.J. would like to
thank the Nuclear Theory Group at Stony Brook and GSI, where part of
this work was done.

\vskip 2cm
{\bf Note Added}

After submission of this work, some of our results were
confirmed by other authors.  Our diagrammatic approach and part of our 
results have been confirmed in two interesting papers by Feinberg and Zee,
cond-mat/9701387 and cond-mat/9704191. The analytical results
of section~6 as illustrated in our second example and the discussion on
the replicas, account for the nature and location of all singularities
and their distribution as they appeared in an extensive 
numerical analysis (up to 500 digits accuracy) by Halasz et al.,
hep-lat/9611008 and hep-lat/9703006.


\end{fmffile}

\end{document}